\documentclass[twocolumn]{aastex631}
\shorttitle{Constraints on High-Redshift Gas Flows from {\it JWST}/NIRSpec}
\shortauthors{Kehoe et al.}
\graphicspath{{./}}

\begin{document}

\title{Unprecedented Constraints on Gas Flows at High Redshift Using Deep JWST/NIRSpec Observations from the LyC22, EXCELS, and AURORA Surveys}

\author[0000-0002-5657-9301]{Emily Kehoe}\affiliation{Department of Physics \& Astronomy, University of California, Los Angeles, 430 Portola Plaza, Los Angeles, CA 90095, USA}
\email{ekehoe@astro.ucla.edu}

\author[0000-0003-3509-4855]{Alice E. Shapley}\affiliation{Department of Physics \& Astronomy, University of California, Los Angeles, 430 Portola Plaza, Los Angeles, CA 90095, USA}

\author[0000-0002-1482-5818]{Adam C. Carnall}
\affiliation{Institute for Astronomy, University of Edinburgh, Royal Observatory, Edinburgh, EH9 3HJ, UK}

\author[0000-0002-3736-476X]{Fergus Cullen}\affiliation{Institute for Astronomy, University of Edinburgh, Royal Observatory, Edinburgh, EH9 3HJ, UK}

\author[0000-0002-0827-9769]{Thomas M. Stanton}\affiliation{Institute for Astronomy, University of Edinburgh, Royal Observatory, Edinburgh, EH9 3HJ, UK}

\author[0000-0001-7144-7182]{Daniel Schaerer}\affiliation{Observatoire de Gen\`eve, Universit\'e de Gen\`eve, Chemin Pegasi 51, 1290 Versoix, Switzerland}\affiliation{CNRS, IRAP, 14 Avenue E. Belin, 31400 Toulouse, France}

\author[0000-0001-8442-1846]{Rui Marques-Chaves}\affiliation{Geneva Observatory, Department of Astronomy, University of Geneva, Chemin Pegasi 51, CH-1290 Versoix, Switzerland}

\author[0000-0002-4834-7260]{Charles C. Steidel}\affiliation{Cahill Center for Astronomy and Astrophysics, California Institute of Technology, MS 249-17, Pasadena, CA 91125, USA}

\author[0000-0003-4792-9119]{Ryan L. Sanders}\affiliation{Department of Physics and Astronomy, University of Kentucky, 505 Rose Street, Lexington, KY 40506, USA}

\author[0000-0001-9489-3791]{Natalie Lam}\affiliation{Department of Physics \& Astronomy, University of California, Los Angeles, 430 Portola Plaza, Los Angeles, CA 90095, USA}

\author[0000-0002-2644-3518]{Karla Z. Arellano-Cordova}\affiliation{Centro de Estudios de Física del Cosmos de Aragón (CEFCA), Plaza San Juan 1, 44001 Teruel, Spain}

\author[0000-0003-0629-8074]{Ryan Begley}\affiliation{Armagh Observatory and Planetarium, College Hill, Armagh, BT61 9DG, N. Ireland, UK}

\author[0000-0002-0159-2613]{Sophia R. Flury}\affiliation{Institute for Astronomy, University of Edinburgh, Royal Observatory, Edinburgh, EH9 3HJ, UK}

\author{Natalia G. Guseva}\affiliation{Bogolyubov Institute for Theoretical Physics, National Academy of
Sciences of Ukraine, 14-b Metrolohichna str., Kyiv, 03143, Ukraine}

\author{Timothy Heckman}\affiliation{School of Earth \& Space Exploration, Arizona State University, 781 Terrace Mall, Tempe, AZ 85287, USA}\affiliation{Department of Physics \& Astronomy, Johns Hopkins University, Bloomberg Centre, 3400 N. Charles Street, Baltimore, MD 21218,
USA}

\author[0000-0002-6586-4446]{Alaina Henry}\affiliation{Space Telescope Science Institute, 3700 San Martin Drive, Baltimore, MD 21218, USA}

\author[0000-0002-7779-8677]{Akio K. Inoue}\affiliation{Waseda Research Institute for Science and Engineering, Faculty of Science and Engineering, Waseda University, 3-4-1 Okubo, Shinjuku, 169-8555 Tokyo, Japan}\affiliation{Department of Physics, School of Advanced Science and Engineering, Faculty of Science and Engineering, Waseda University, 3-4-1 Okubo, Shinjuku, 169-8555 Tokyo, Japan}

\author[0000-0002-1416-6082]{Yuri I. Izotov}\affiliation{Bogolyubov Institute for Theoretical Physics, National Academy of
Sciences of Ukraine, 14-b Metrolohichna str., Kyiv, 03143, Ukraine}

\author[0000-0003-0486-5178]{Ho-Hin Leung}\affiliation{Institute for Astronomy, University of Edinburgh, Royal Observatory, Edinburgh, EH9 3HJ, UK}

\author[0000-0003-4368-3326]{Derek J. McLeod}\affiliation{Institute for Astronomy, University of Edinburgh, Royal Observatory, Edinburgh, EH9 3HJ, UK}

\author[0000-0001-7883-8434]{Kate Rowlands}\affiliation{AURA for ESA, Space Telescope Science Institute, 3700 San Martin Drive, Baltimore, MD 21218, USA}\affiliation{William H. Miller III Department of Physics and Astronomy, Johns Hopkins University, Baltimore, MD 21218, USA}

\author[0000-0001-8419-3062]{Alberto Saldana-Lopez}
\affiliation{Department of Astronomy, Oskar Klein Centre, Stockholm University, 106 91 Stockholm, Sweden}

\author[0000-0002-6867-1244]{Dirk Scholte}\affiliation{Institute for Astronomy, University of Edinburgh, Royal Observatory, Edinburgh, EH9 3HJ, UK}

\author[0009-0004-0844-0657]{Maya Skarbinski}
\affiliation{William H. Miller III Department of Physics and Astronomy, Johns Hopkins University, Baltimore, MD 21218, USA}

\author[0000-0001-5642-752X]{Struan D. Stevenson} \affiliation{Institute for Astronomy, University of Edinburgh, Royal Observatory, Edinburgh, EH9 3HJ, UK}

\author[0000-0001-8728-2984]{Elizabeth Taylor}\affiliation{Institute for Astronomy, University of Edinburgh, Royal Observatory, Edinburgh, EH9 3HJ, UK}

\author[0000-0001-9269-5046]{Bingjie Wang}
\thanks{NHFP Hubble Fellow}
\affiliation{Department of Astrophysical Sciences, Princeton University, Princeton, NJ 08544, USA}

\author[0000-0001-9687-4973]{Naveen A. Reddy}\affiliation{Department of Physics \& Astronomy, University of California, Riverside, 900 University Avenue, Riverside, CA 92521, USA}



\begin{abstract}
    We investigate how low-ionization gas flows in typical star-forming galaxies at $z\sim3$ depend on galaxy intrinsic properties and viewing angle. For this analysis we use {\it JWST}/NIRSpec observations of rest-frame near-UV Fe\,{\sc ii} and Mg\,{\sc ii} absorption, and rest-frame optical Na\,{\sc d} absorption. This study combines galaxies from the LyC22, EXCELS, and AURORA surveys and contains 176, 197, and 315 galaxies, respectively, with Fe\,{\sc ii}, Mg\,{\sc ii}, and Na\,{\sc d} coverage. Based on both individual and composite spectra, we find no statistically significant correlations between outflow velocity and galaxy properties. However, galaxies with detected outflows tend towards higher stellar masses, SFR, and $\Sigma_{\rm SFR}$ than those without outflows, suggesting that the two samples are not drawn from the same parent population. Finally, we additionally find that Mg\,{\sc ii} emission is preferentially detected in galaxies with lower stellar mass and $A_V$, and higher sSFR, consistent with conditions that favor the escape of resonantly scattered line and ionizing continuum radiation. We present the first evidence in $z\sim3$ star-forming galaxies that properties of the absorption lines depend on galaxy inclination, with more face-on systems showing stronger absorption and higher outflow velocities, while inflowing gas is more frequently detected in more highly inclined galaxies. These trends are consistent with observations at $z\lesssim1$ and predictions from cosmological simulations in which galactic winds are launched perpendicular to the galactic disks, while accretion occurs primarily along the disk plane.  
\end{abstract}



\section{Introduction}
    Gas flows play a fundamental role in galaxy evolution across cosmic time by governing the baryon cycle between galaxies and their surrounding environments. These flows regulate the supply of star-forming material, redistribute metals, and enrich the circumgalactic medium (CGM) and intergalactic medium (IGM) with material expelled from the interstellar medium (ISM) \citep{2005NDiMatteo, 2005Scannapieco, 2008Hopkins, 2008Somerville, 2015Erb, 2017Tumlinson}. Feedback-driven outflows powered by massive stars, supernovae, and active galactic nuclei (AGN) inject energy and momentum into the ISM and launch material into the CGM, regulating star formation \citep{1998Silk, 2005Veilleux, 2011Murray}. Gas inflows replenish galaxies with fresh material that sustains star formation, establishing a continuous exchange of baryons that influences how galaxies evolve over time \citep{2017Tumlinson, Kacprzak2017}.

    Galaxy gas flows have a multiphase structure, spanning cold neutral gas, warm ionized gas, and hot X-ray emitting plasma \citep{1990Heckman, 2005Martin, Rubin2014, 2017Chisholm}. These phases are traced through a wide range of diagnostics, including near-ultraviolet (NUV; 2200--3000\,\AA) absorption lines such as Fe\,{\sc ii} and Mg\,{\sc ii} \citep[e.g.,][]{2009Weiner, 2012Erb, 2014Bordoloi}, far-ultraviolet (FUV; below 2200\,\AA) absorption lines such as Si\,{\sc ii}, C\,{\sc ii}, Si\,{\sc iv}, C\,{\sc iv}, and O\,{\sc vi},  along with resonantly scattered Ly$\alpha$ emission \citep[e.g.,][]{2003Shapley, 2010Steidel, Tumlinson2011, 2022Calabro}, optical features including H$\alpha$ and [OIII]$\lambda5007$ emission and Na\,{\sc d} absorption \citep[e.g.,][]{2010Chen, 2019Schreiber, 2024Davies, 2024Carniani, 2025Saldana}, and X-ray observations \citep[e.g.,][]{2009Strickland, 2010Tombesi}. Among these tracers, low-ionization absorption lines are especially useful for studying cool gas kinematics because they directly trace material moving along the line of sight through blueshifted and redshifted absorption.

    The geometry of gas flows is essential for understanding how feedback and accretion operate within galaxies. Cosmological simulations predict that star formation-driven outflows are not isotropic, but instead preferentially emerge perpendicular to the galactic disks, while inflowing gas is funneled through the galactic disk along the major axis \citep[e.g.,][]{2011Brook, 2019Nelson, Peroux2020}. In this picture, the properties of the observed absorption lines depend on galaxy inclination. Observational studies across multiple emission- and absorption-line tracers spanning $0 \lesssim z \lesssim 2$ generally support this model. Outflows measured from absorption lines are often found to show stronger blueshifted signatures in more face-on systems, while inflow signatures are more frequently observed in more highly-inclined galaxies \citep{1990Heckman, 2000Heckman, 2010Chen, Stewart2011, 2012Kacprzak, 2012Newman, Rubin2012, 2019Roberts}. However, the strength and consistency of these trends vary across studies, with several higher redshift ($z\gtrsim2$) datasets suggesting no dependence of gas flow properties on inclination \citep{Law2012, 2019Schreiber, 2022Weldon}, and indicating that such trends are more difficult to detect at earlier times. 

    The launch of the James Webb Space Telescope ({\it JWST}) has enabled a new regime for studying galactic gas flows at high redshift. Prior to {\it JWST}, NUV absorption features at $z>2$ were largely inaccessible from the ground due to atmospheric absorption and limited near-infrared spectroscopic capabilities, restricting studies of NUV tracers to lower redshift galaxies. With {\it JWST}/NIRSpec, it is now possible to observe these features in galaxies during the peak epoch of cosmic star formation, allowing direct measurements of NUV absorption in galaxies at $z>2$ \citep{Kehoe2025, 2026Lyu}. These new probes of gas flows at $z>2$ enable direct comparisons with trends established at lower redshifts. At the same time, {\it JWST} provides high-resolution images, yielding robust measurements of galaxy morphological properties, such as inclination. These measurements of galaxy orientation offer the opportunity to connect NUV inflow and outflow kinematics with galaxy inclination and therefore place new constraints on the geometry of gas flows in high-redshift galaxies. 

    Utilizing these new capabilities, \citet{Kehoe2025} analyzed {\it JWST}/NIRSpec observations from the Assembly of Ultradeep Rest-optical Observations Revealing Astrophysics (AURORA) Cycle 1 {\it JWST}/NIRSpec program \citep{Shapley2025}. We used Fe\,{\sc ii} and Mg\,{\sc ii} absorption to study outflowing gas in star-forming galaxies at $z\gtrsim 2.5$, providing direct comparisons of trends between NUV absorption kinematics and outflow properties between low ($z\sim 1$) and high ($z>2.5$) redshift galaxies for the first time. \citet{Kehoe2025} also demonstrated the feasibility of detecting and analyzing Na\,{\sc d} in these high redshift galaxies. 
    
    Building on this framework, we expand the previous {\it JWST} sample by incorporating two additional surveys: the LyC22 Cycle 1 {\it JWST}/NIRSpec program \citep{Schaerer2026} and the Early eXtragalactic Continuum and Emission Line Science (EXCELS) Cycle 2 {\it JWST}/NIRSpec program \citep{Carnall2024}. Including two additional surveys increases the overall sample size by a factor of four, enabling a more statistically robust exploration of NUV absorption in high redshift galaxies. With this expanded sample at $z\sim 3$, we perform a more detailed analysis of Fe\,{\sc ii}, Mg\,{\sc ii}, and Na\,{\sc d} absorption. In particular, we investigate how outflow signatures relate to the galaxy stellar mass ($M_*$), star formation rate (SFR), specific star formation rate (sSFR), star formation rate surface density ($\Sigma_{\rm SFR}$), UV slope ($\beta$), dust attenuation ($A_V$), and nebular $E(B-V)$. These correlations allow us to provide constraints on the physical driving mechanisms of galactic-scale outflows. We also examine trends with galaxy inclination to investigate the geometry of the flowing gas. Notably, this study provides the first statistically robust analysis on how gas flows vary with galaxy orientation, providing new constraints on the geometry of gas flows for galaxies at $z\sim3$.

    The outline of this paper is as follows. Section \ref{sec:Observations} describes the {\it JWST} observations and sample selection. Section \ref{section:Measurements} outlines the methods used to measure galaxy properties and outflow velocities derived from Fe\,{\sc ii}, Mg\,{\sc ii}, and Na\,{\sc d} absorption features, and the construction of the composite spectra. Section \ref{sec:Results} presents the results of our analysis, including correlations between outflow and absorption properties and galaxy properties, and trends with inflowing gas and galaxies with significant Mg\,{\sc ii} emission. Section \ref{sec:Discussion} compares our results to previous outflow studies, discusses the geometry of inflowing and outflowing gas, and examines the connection between Mg\,{\sc ii} emission and LyC escape. Throughout this work, we assume a $\Lambda$CDM cosmology with $H_0=70\,\rm km\,s^{-1}$, $\Omega_m=0.3$, and $\Omega_{\Lambda}=0.7$, and the \citet{2003Chabrier} initial mass function (IMF).

\section{Observations}\label{sec:Observations}
In this work, we analyze a sample of galaxies  with observations drawn from
three different {\it JWST}/NIRSpec programs: the Cycle 1 LyC22 survey \citep[GO 1869, PI: Schaerer;][]{Schaerer2026}; the Cycle 2 EXCELS survey \citep[GO 3543; PIs: Carnall, Cullen;][]{Carnall2024}; and the Cycle 1 AURORA program \citep[GO 1914; PIs: Shapley, Sanders;][]{Shapley2025}. We now describe each of these programs in turn, and the combined sample that we construct from them at $z_{{\rm med}}=3.07$.

    \begin{figure*}
        \centering
        \includegraphics[width=\linewidth]{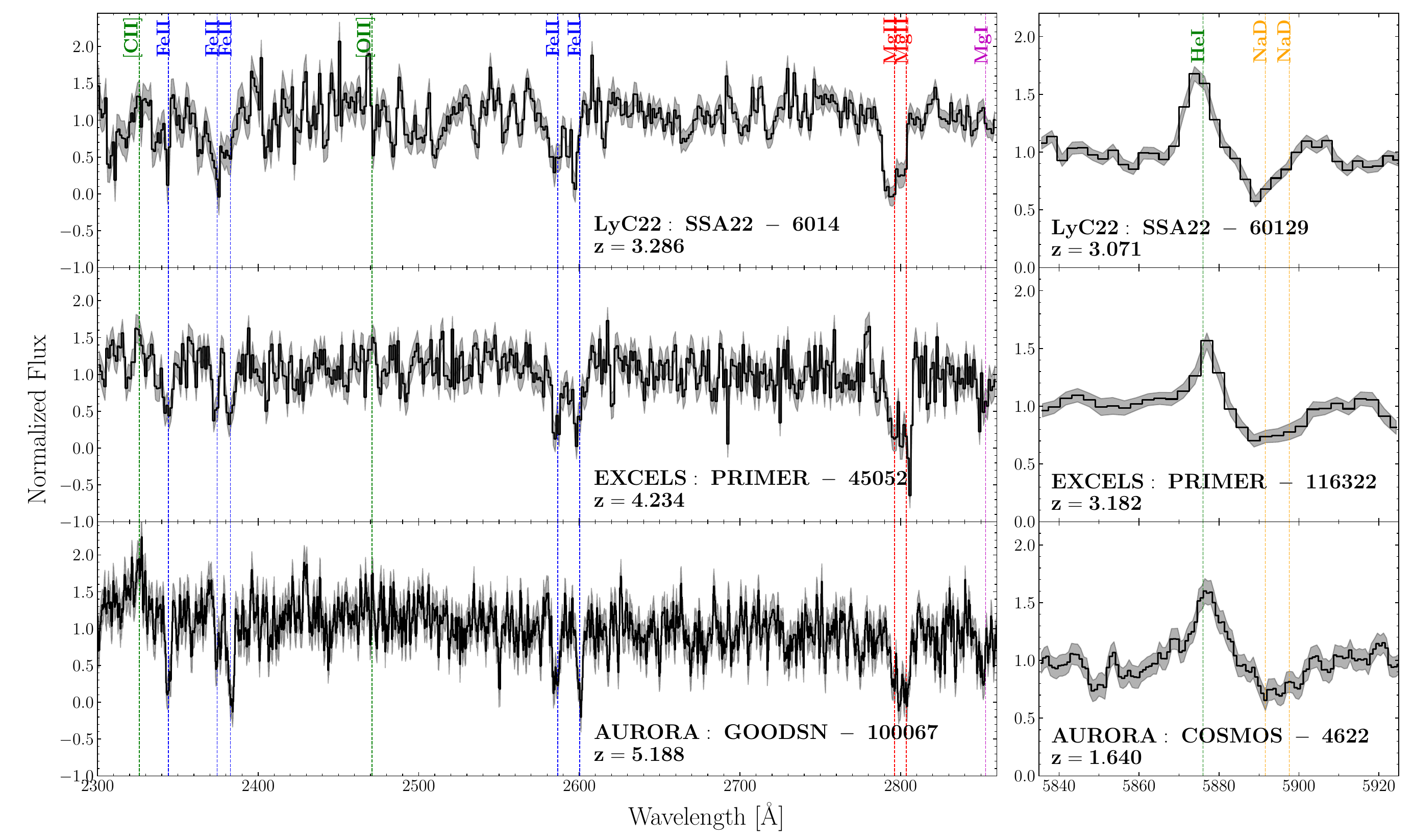}
        \caption{Rest-frame continuum-normalized spectra from the LyC22 survey (top), the EXCELS survey (middle), and the AURORA survey (bottom). Left panels show the NUV lines analyzed, with Fe\,{\sc ii} rest wavelengths indicated in blue and Mg\,{\sc ii} in red. Green lines mark emission lines not included in the analysis, and the magenta line marks the Mg\,{\sc i} $\lambda2850$\AA\,absorption line. Right panels show example Na\,{\sc d} profiles, with orange lines indicating the Na\,{\sc d} $\lambda\lambda5891,5897$\AA\,lines and the green line labeling the He\,{\sc i} $\lambda 5875$\AA\,emission line.}
        \label{fig:spectra}
    \end{figure*}
    \begin{figure*}
        \centering
        \includegraphics[width=\linewidth]{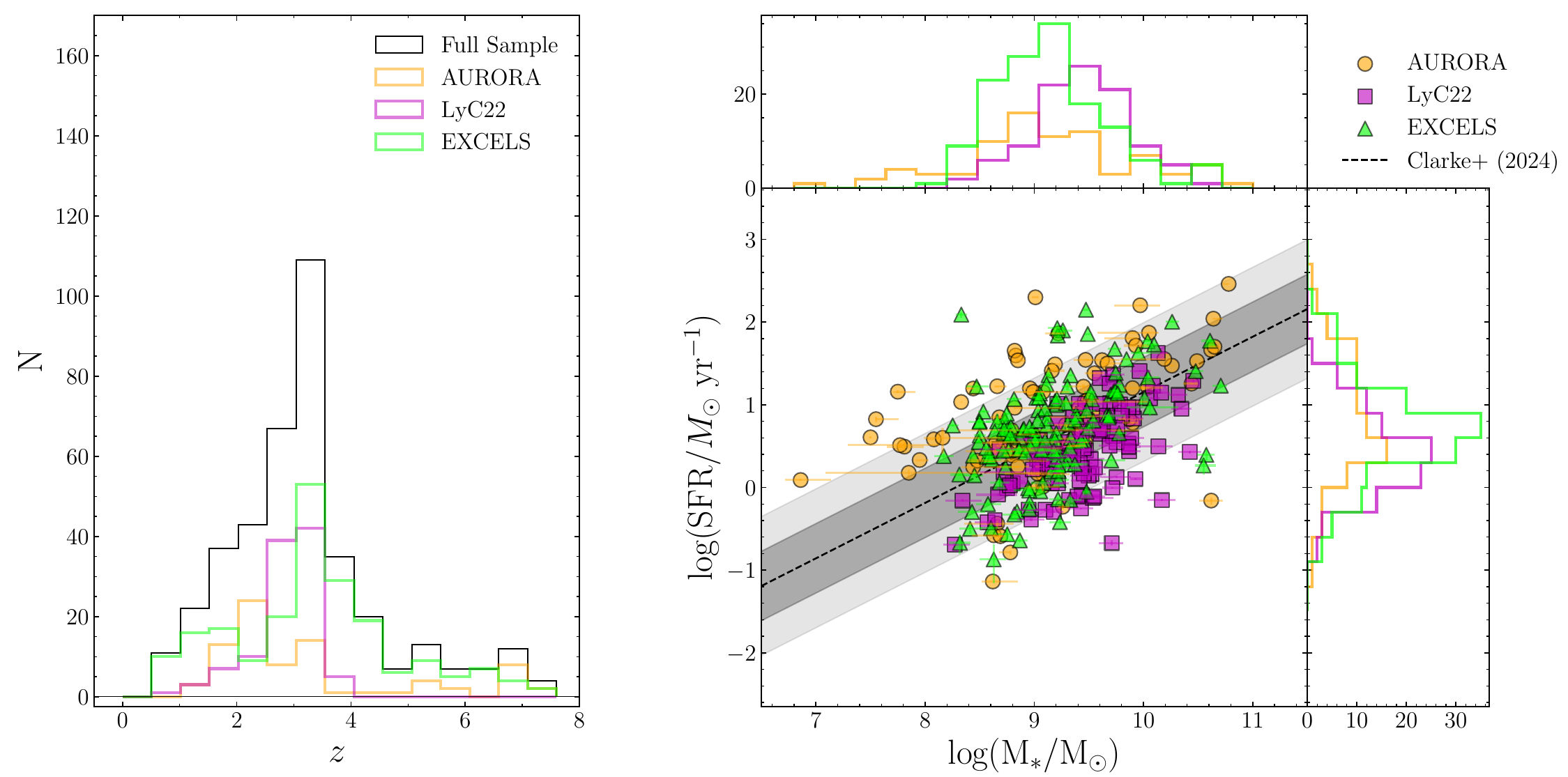}
        \caption{Properties of the \textit{JWST} sample analyzed in this work. Left: Histogram of H$\alpha$ redshifts for the AURORA survey (orange), the LyC22 survey (magenta), the EXCELS survey (green), and the full sample (black). Right: SFR (based on H$\alpha$) vs. stellar mass for the AURORA survey (orange circles), the LyC22 survey (pink squares), and the EXCELS survey (green triangles). The SFRs are derived from dust-corrected H$\alpha$ luminosities, and the stellar mass is derived from SED fitting. The dashed line represents the star-forming main sequence for $2.7<z\leq4$ galaxies from \cite{2024Clarke}, where the dark gray and light gray shaded regions represent the $1\sigma$ and $2\sigma$ intrinsic scatter intervals, respectively.}
        \label{fig:demographics}
    \end{figure*}
    
    \subsection{The LyC22 Survey}
    The LyC22 Survey is described in detail in \citet{Schaerer2026}. Targets were observed with the NIRSpec micro-shutter assembly (MSA) in three pointings, two of which were in the SSA22 field and one in the Westphal field. Each pointing was observed with both the G140M/F100LP and G235M/F170LP grating/filter combinations, providing wavelength coverage from 1.0 -- 3.1 $\mu$m. Exposure times were 9.2 hours in G140M and 8.8 hours in G235M, corresponding, respectively, to median 3$\sigma$ line flux sensitivities of $2.5\times 10^{-19} \: {\rm erg}\; {\rm cm}^{-2}\: {\rm s}^{-1}$ and $1.8 \times 10^{-19} \: {\rm erg}\; {\rm cm}^{-2}\: {\rm s}^{-1}$. The LyC22 survey aims to test indirect observational proxies for Lyman continuum escape in star-forming galaxies at $z\sim 2.5-3.5$ (both narrowband-selected Ly$\alpha$ emitters and color-selected Lyman break galaxies), which is critical for models of cosmic reionization \citep{Steidel2018}.  A total of 143 galaxies was targeted with LyC22, of which 117 yielded spectroscopic redshifts with NIRSpec and spanned the range $z=0.07-4.04$. LyC22 targets are also covered by extensive multi-wavelength photometry, including both ground-based data as well as imaging from the Hubble Space Telescope ({\it HST})/ACS and {\it HST}/WFC3, and {\it JWST}/NIRCam. LyC22 data reduction is described in detail in \citet{Schaerer2026}, including important corrections for slit-losses.

    \subsection{The EXCELS Survey}
        The EXCELS survey was first described in \citet{Carnall2024}. This NIRSpec MSA program consists of four pointings in the PRIMER UDS field, each of which was observed with the G140M/F100LP, G235M/F170LP, and G395M/F290LP grating/filter combinations. This set-up provides wavelength coverage from 1--5~$\mu$m. However, distinct MSA configurations were designed for each grating within each pointing in order to maximize the number of galaxies with coverage of key rest-UV and rest-optical features. Integration times were 4 hours in G140M and G395M and 5.5 hours in G235M, yielding slightly shallower emission-line sensitivities than LyC22 in G140M ($\times 0.65$) and G235M ($\times 0.79$) and a depth in G395M comparable to the depth of the LyC22 observations in G140M and G235M. The 401 EXCELS targets included a mixture of quiescent galaxies at $z=1-5$ and star-forming galaxies at $z\geq 0.5$, and a total of 341 secure spectroscopic redshifts were obtained with NIRSpec. EXCELS targets are covered by extensive photometric data, including both {\it HST}/ACS and {\it JWST}/NIRCam and {\it JWST}/MIRI F770W imaging \citep{Carnall2024}. We refer readers to \citet{Stanton2026} and \citet{Carnall2024} for a detailed description of the data reduction methods in the EXCELS survey, including the method for slit-loss corrections.
    
    \subsection{The AURORA Survey}
    The AURORA Survey is described in detail in e.g., \citet{Shapley2025} and \citet{Sanders2025}. Briefly, this survey includes NIRSpec MSA observations of two pointings, one in the COSMOS field, and one in GOODSN. As in the EXCELS survey, AURORA observations were designed to provide wavelength coverage from 1--5~$\mu$m using the G140M/F100LP, G235M/F170LP,
    and G395M/F290LP grating/filter combinations. Exposures in these gratings were, respectively, 12.3, 8.0, and 4.2 hours. Therefore, the AURORA G235M line-flux sensitivity is extremely comparable to that in LyC22, while the G140M depth is a factor of 1.14$\times$ greater.

These deep exposure times were designed to detect faint auroral emission lines and determine direct, $T_e$-based metallicities \citep{Sanders2026}. As described in detail in \cite{Shapley2025}, a total of 97 galaxies were targeted with AURORA, including 95 that yielded spectroscopic redshifts at $z=1.4-10.4$. As for LyC22 and EXCELS targets, AURORA galaxies are covered by extensive multi-wavelength photometry, including ground-based, {\it HST}/ACS and {\it HST}/WFC3, and {\it JWST}/NIRCam data. AURORA data reduction is described in \citet{Shapley2025} and \citet{Sanders2025}, including a detailed discussion of slit-loss corrections.

\subsection{Combined {\it JWST}/NIRSpec Sample}
        From the LyC22, EXCELS and AURORA surveys, we construct a combined sample of star-forming galaxies with \textit{JWST}/NIRSpec spectroscopy. To isolate star-forming galaxies, objects identified as AGN or quiescent are removed from the sample. In LyC22, AGN are identified from broad Balmer emission and high-ionization emission lines \citep{Schaerer2026}. In EXCELS, AGN are identified with broad Balmer emission, and quiescent galaxies are identified from best-fit SEDs showing low sSFRs and red rest-frame colors \citep{Carnall2024}. In AURORA, AGN are identified from broad Balmer lines and elevated [N\,{\sc ii}]/H$\alpha$ ratios ($> 0.5$), and quiescent galaxies are identified from SED fits showing suppressed star formation and spectra dominated by stellar absorption features with weak or absent nebular emission lines \citep{Shapley2025}. We further require coverage of rest-frame NUV Fe\,{\sc ii} and Mg\,{\sc ii} features, or the Na\,{\sc d} absorption doublet (Table \ref{tab:line_windows}; Figure \ref{fig:spectra}). The Fe\,{\sc ii}, Mg\,{\sc ii}, and Na\,{\sc d} absorption have lower redshift limits of $z =2.87$, $z=2.57$, and $z=0.70$, respectively, due to the wavelength coverage of NIRSpec.
        
        The resulting LyC22 sample contains 67 galaxies with Fe\,{\sc ii} coverage, 81 galaxies with Mg\,{\sc ii} coverage, and 103 galaxies with Na\,{\sc d} coverage. The EXCELS sample has 68 galaxies with Fe\,{\sc ii} coverage, 73 galaxies with Mg\,{\sc ii} coverage, and 130 galaxies with Na\,{\sc d} coverage. In AURORA, there are 41 galaxies with Fe\,{\sc ii} coverage, 43 galaxies with Mg\,{\sc ii} coverage, 82 galaxies with Na\,{\sc d} coverage. In total, the combined \textit{JWST}/NIRSpec sample contains 176, 197, and 315 galaxies with Fe\,{\sc ii}, Mg\,{\sc ii}, and Na\,{\sc d} coverage, respectively. The final sample spans $z=0.75$ -- 7.57 (median $z=3.07$), stellar masses  $\log(M_*/M_\odot)=$ 6.86 -- 10.78 (median $\log(M_*/M_\odot)=9.21$), and H$\alpha$-derived SFRs $\log(\rm SFR/M_\odot\rm yr^{-1})= -1.14$ -- 2.46 (median $\log(\rm SFR)=0.60$) (Figure \ref{fig:demographics}).

\section{Measurements}\label{section:Measurements}
\subsection{Galaxy Properties}
    \subsubsection{SED Fitting}\label{sec:SEDFitting}
        We determined the stellar masses ($M_*$) and dust attenuation values ($A_V$) of galaxies in our sample by fitting their spectral energy distributions (SEDs).         For the LyC22 and EXCELS surveys, details of the SED fitting methodology are provided in \cite{Stanton2026}, with a brief summary given here. The observed photometry was first corrected for contamination from nebular emission lines by constructing mock emission line spectra from Gaussian fits to detect lines and subtracting their contribution from each filter. We then fit the corrected photometry with the SED-fitting code \texttt{BAGPIPES} \citep{Carnall2018, Carnall2019} using Binary Population and Spectral Synthesis stellar population models (BPASS v2.21; \citealt{Eldridge_2017,2018Stanway}), which include the effects of binary stellar evolution and a \cite{Kroupa2001} initial mass function. Star-formation histories were modeled using a delayed-$\tau$ model (i.e., $\mathrm{SFR}(t)\sim(t-T_0)\times\exp(-(t-T_0)/\tau)$, where $T_0$ is the time at the onset of star formation, stellar metallicities and masses were varied with broad ranges, and modeled with the \citet{Salim2018} flexible attenuation law. To account for the remaining nebular continuum, we included a continuum component and varied its strength through the ionization parameter, which was allowed to vary freely in the fitting process. The photometric dataset for the EXCELS sample is described in detail in \citet{Carnall2024}. Photometry for the LyC22 sample consisted of a mixture of ground-based, {\it Spitzer}, {\it HST} and {\it JWST} bands, which was slightly different for the SSA22 and Westphal fields. The SSA22 filter set included ground-based $u$, $B$, $G$, $V$, $Rs$, $i$, $z$, $J$, $H$, and $Ks$ bands, \textit{HST}/WFC3 $F160W$, and \textit{JWST}/NIRCam F150W and F277W, whereas the Westphal photometry included ground-based $u$, $g$, $r$, $i$, $z$, $J$, $H$, $Ks$ bands, \textit{Spitzer}/IRAC channels 1, 2, 3, and 4, \textit{HST}/ACS F606W, \textit{HST}/WFC3 F125W and F160W, and \textit{JWST}/NIRCam F277W. 
        
        For the AURORA survey, \cite{Sanders2025} provide details of the SED modeling, including the photometric bands that were utilized. To summarize briefly, we first corrected the photometry for contributions from strong nebular emission lines and nebular continuum emission, and then performed SED fitting using this corrected photometry. The photometric measurements were drawn from the DAWN {\it JWST} Archive (DJA) \citep{Valentino2023}, which provided coverage from {\it JWST}/NIRCam \citep{Eisenstein2026} as well as {\it HST}/WFC3 and {\it HST}/ACS imaging, giving broad wavelength coverage of each galaxy's SED. We derived the stellar population properties and the stellar continuum by fitting flexible stellar population synthesis models \citep{2009Conroy} to the corrected photometry using Fitting and Assessment of Synthetic Templates (FAST) \citep{Kriek2009}, with a \cite{2003Chabrier} initial mass function and star-formation histories described by a delayed-$\tau$ model. Following earlier work \citep[e.g.,][]{2024Clarke}, we explored two sets of model assumptions, consisting of a $1.4Z_{\odot}$ model with the \cite{2000Calzetti} attenuation curve (the ``Calzetti" model) and a $0.2Z_{\odot}$ model with an SMC extinction curve from \cite{2003Gordon} (the ``SMC" model). For each galaxy, we adopted the solution corresponding to the minimum $\chi^2$ value. To ensure that no significant systematics were introduced by using the existing FAST population synthesis fits for AURORA, we compared the best-fit stellar masses that would result from instead using Bagpipes models. We found excellent agreement, with a median difference  of $\log(M_{*,{\rm FAST}})-\log(M_{*,{\rm Bagpipes}})=-0.03\pm 0.22$. The joint distributions of $A_V$ vs. $M_*$ are also consistent when using either FAST or Bagpipes.
    \subsubsection{UV $\beta$ Slopes}
        We estimated the UV continuum slope, $\beta$, which characterizes the spectral shape as $F_\lambda \propto \lambda^{\beta}$. $\beta$ was determined by performing linear regression on the rest-frame UV photometry of each galaxy \citep{2024Clarke}. Only filters covering the rest-frame wavelength range 1250--2600\,\AA \,were included to ensure robust measurements of $\beta$. To estimate the uncertainties on $\beta$, we generated $10{,}000$ Monte Carlo realizations of the photometry by perturbing each measurement according to its reported error. We then computed $\beta$ for each realization and adopted the 16th, 50th, and 84th percentiles of the resulting distribution as the lower bound, median, and upper bound, respectively.
    \subsubsection{Galaxy Inclination}\label{sec:measurementsInc}
\begin{figure}
            \centering
            \includegraphics[width = \linewidth]{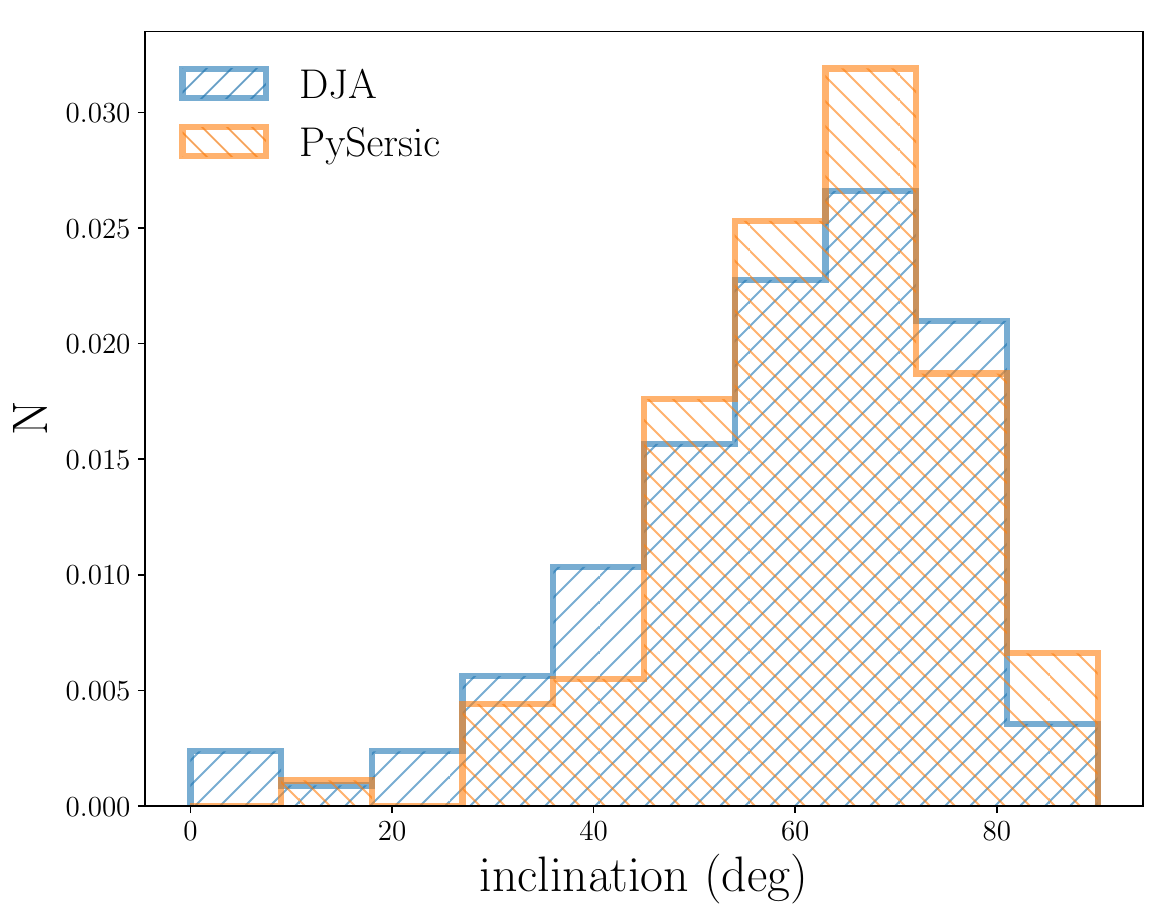}
            \caption{Normalized histograms of galaxy inclinations measured from the DJA catalogs (blue; AURORA and EXCELS) and from \texttt{PySersic} (orange; LyC22). Both types of measurements show similar distributions, demonstrating that the use of different methods does not introduce a significant bias.}
            \label{fig:inc_compare}
        \end{figure}

        We calculated the galaxy inclination using,
        \begin{equation}
            \cos(i)=\frac{q^2-\gamma^2}{1-\gamma^2}
        \end{equation}
        where $q$ is the axis ratio that ranges between 0 and 1, and $\gamma$ is the intrinsic disk thickness. We adopt $\gamma\sim0.2$, consistent with previous studies of distant star-forming galaxies \citep[e.g.,][]{2015Wisnioski,2019Wisnioski}. For the EXCELS and AURORA surveys, the axis ratios and uncertainties are a component of the S\'ersic profile fits reported in the DJA morphological catalogs \citep{2025Genin}. These fits were performed using \texttt{Sextractor++} \citep{Bertin2020}.
        Since galaxies from the LyC22 survey are not available in the DJA catalogs, we derived the axis ratios and uncertainties by modeling NIRCAM/F277W light distributions with PSF-convolved two-dimensional Gaussian fits using \texttt{PySersic} \citep{Pasha2023}, where the PSF was constructed from stacked bright stars in the F277W field using \texttt{PSFr} \citep{Birrer2021, Birrer2022}. As shown in Figure \ref{fig:inc_compare}, the inclinations obtained from the DJA and \texttt{PySersic} measurements show similar distributions. A Kolmogorov–Smirnov (KS) test indicates that the two distributions are drawn from the same parent population, suggesting that the different fitting methodologies do not introduce a significant systematic bias.

    \subsubsection{Spectroscopic Measurements}
        We used spectroscopic detections of Balmer lines (corrected for underlying stellar absorption) to calculate the systemic redshift, star-formation rate (SFR) and nebular dust reddening ($E(B-V)_{\mathrm{gas}}$) for galaxies in our sample. Dust attenuation was determined from the Balmer decrement, with $E(B-V)$ derived from the $\rm H\alpha/\rm H\beta$ flux ratio, assuming an intrinsic value of 2.79 for Case B recombination. We assumed $T_e=15{,}000\rm \,K$ and $n_e = 300\,\rm cm^{-3}$, typical for star-forming galaxies at these redshifts \citep{Topping2025,Sanders2026}, and applied the \cite{1989Cardelli} dust law. In cases where $\rm H\alpha$ was not covered by the wavelength range of the observations, the $\rm H\gamma/\rm H\beta$ flux ratio, with an intrinsic value of $0.475$, was used. When $\rm H\beta$ was not covered, the $\rm H\gamma/\rm H\alpha$ flux ratio, with an intrinsic value of $0.173$, was used. The dust-corrected $\rm H\alpha$ luminosities were then converted to SFRs using the following equation:
            \begin{equation}
            \log \left (\frac{\mathrm{SFR}}{M_{\odot}\mathrm{yr}^{-1}}\right) = \log \left (\frac{\mathrm{L}_{\mathrm{H}\alpha}}{\text{erg s}^{-1}}\right) + C
        \end{equation}
        where $C$ is a metallicity-dependent conversion factor, following \cite{2024Clarke}. The conversion factor was determined from a set of BPASS models with upper mass limits of $100\,M_{\odot}$ \citep{2018Stanway, 2022Reddy}. For the AURORA survey, galaxies fit with the ``Calzetti" model had a correction factor of $C=-41.37$, and galaxies fit with the ``SMC" model had a correction factor of $C=-41.59$. For the LyC22 and EXCELS surveys, which adopted different SED modeling assumptions, we applied the ``SMC" model conversion factor of $C=-41.59$, following \cite{Stanton2026}, who adopt this conversion and show consistency with other SFR conversion factors. 

        From the dust-corrected SFRs and stellar masses described above, we calculated the specific star-formation rate ($\mathrm{sSFR} = \mathrm{SFR}/M_*$). We also derived the star-formation rate surface density $\Sigma_{\rm SFR}=\mathrm{SFR}/(2\pi R_{e}^2)$ using the dust-corrected SFR and the galaxy's half-light radius. For the AURORA survey, we use the methodology from \cite{Pahl2022} to calculate $R_e$. Briefly, $R_e$ was measured using S\'ersic profile fits with \texttt{galfit}. For galaxies well described by a single S\'ersic component, a circularized radius is computed using the semimajor-axis half-light radius and the axis ratio. For galaxies needing multiple S\'ersic components, $R_e$ was determined from the area containing half of the total flux. For the EXCELS survey, effective radii were adopted directly from the DJA catalogs \citep{Valentino2023,2025Genin}. For the LyC22 survey, we adopt the PSF-deconvolved effective radii from the same \texttt{PySersic} morphological fits described in section \ref{sec:measurementsInc}
        
        Uncertainties in both $E(B-V)$ and SFR were quantified through Monte Carlo simulations with $10{,}000$ realizations of the observed fluxes. For sSFR and $\Sigma_{\rm SFR}$, we propagated the uncertainties by sampling the distributions of the underlying parameters. Specifically, sSFR uncertainties were estimated by drawing SFRs and stellar masses from Gaussian distributions defined by their measured values and associated uncertainties, computing the sSFR for each realization. Similarly, the uncertainties in $\Sigma_{\rm SFR}$ were obtained by sampling the SFRs and effective radii confidence intervals and recalculating $\Sigma_{\rm SFR}$ for each iteration.

    \subsection{Outflow Velocities}\label{sec:vel}
        \begin{deluxetable}{lccc}
            \tablecaption{Fitted Absorption Lines \label{tab:line_windows}}
            \tablehead{
              \colhead{Line} & \colhead{$\lambda\textsubscript{rest}$ (\AA)} & \colhead{Blue Window (\AA)\tablenotemark{a}} & \colhead{Red Window (\AA)\tablenotemark{a}}
              }
              \startdata
                  Fe\,{\sc ii} & 2249.88 & $2200-2220$ & $2300-2330$ \\
                  Fe\,{\sc ii} & 2260.78 & $2200-2220$ & $2300-2330$ \\
                  Fe\,{\sc ii} & 2344.21 & $2300-2330$ & $2405-2425$ \\
                  Fe\,{\sc ii} & 2374.46 & $2300-2330$ & $2405-2425$ \\
                  Fe\,{\sc ii} & 2382.77 & $2300-2330$ & $2405-2425$ \\
                  Fe\,{\sc ii} & 2586.65 & $2540-2560$ & $2640-2660$ \\
                  Fe\,{\sc ii} & 2600.17 & $2540-2560$ & $2640-2660$ \\
                  Mg\,{\sc ii} & 2796.35 & $2705-2730$ & $2820-2840$ \\
                  Mg\,{\sc ii} & 2803.53 & $2705-2730$ & $2820-2840$ \\
                  Na\,{\sc d} & 5891.59 & $5810-5830$ & $6000-6020$ \\
                  Na\,{\sc d} & 5897.56 & $5810-5830$ & $6000-6020$ \\
              \enddata
              \tablenotetext{a}{The blue and red regions represent the wavelength ranges used for local continuum fitting for each feature.}
          \end{deluxetable}

        To characterize the kinematics of the absorbing gas, we measure the velocity offsets and equivalent widths (EWs) from Fe\,{\sc ii}, Mg\,{\sc ii}, and Na\,{\sc d} absorption features (Table \ref{tab:line_windows}). These tracers probe progressively more complex line profiles, reflecting different combinations of pure absorption, emission filling, and stellar and nebular contamination, which require different modeling approaches tailored to each transition. For example, both Mg\,{\sc ii}, and Na\,{\sc d} can be significantly affected by emission-line infilling, which is less important for Fe\,{\sc ii} \citep{Martin2012, 2015Zhu}. For each tracer, we adopt the approach that most robustly reproduces the observed spectra and yields well-constrained centroid measurements.
        
        \begin{figure}
            \centering
            \includegraphics[width=\linewidth]{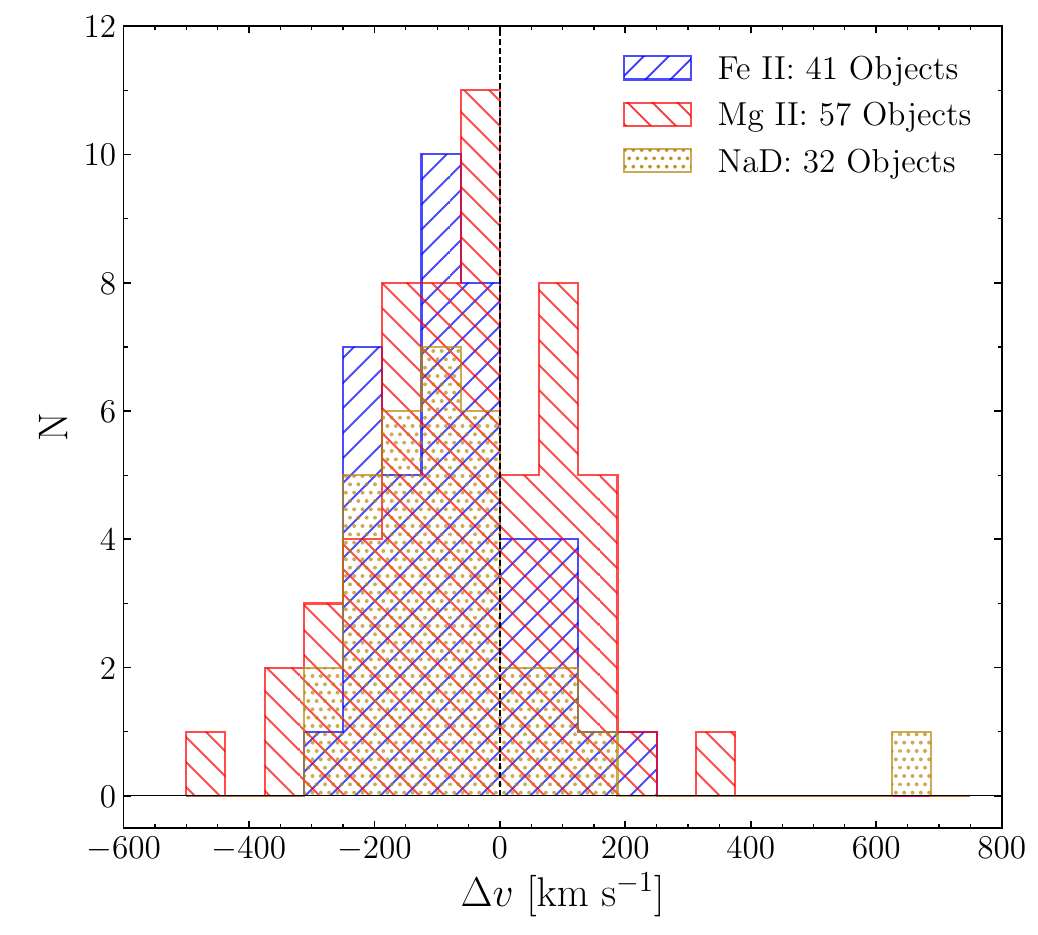}
            \caption{Histogram of the measured velocity offsets of Fe\,{\sc ii} (blue), Mg\,{\sc ii} (red), and Na\,{\sc d} (gold) relative to the galaxy redshift from the H$\alpha$ centroid. Velocity offsets were measured for 41 galaxies using Fe\,{\sc ii}, with an average velocity of $\langle\Delta v_{\rm Fe\,II}\rangle=-70\pm 18 \rm \, km\,s^{-1}$, 57 galaxies using Mg\,{\sc ii}, with $\langle\Delta v_{\rm Mg\,II}\rangle=-50\pm 20 \rm \, km\,s^{-1}$, and 32 galaxies using Na\,{\sc d}, with $\langle\Delta v_{\rm Na\,I\,D}\rangle=-79\pm 29 \rm \, km\,s^{-1}$.}
            \label{fig:vel_hist}
        \end{figure}
        
        \subsubsection{Fe\,{\sc ii}}
            Doppler shifts in Fe\,{\sc ii} absorption lines (Table \ref{tab:line_windows}), provide a direct probe of large-scale galactic flows. We identified significantly detected absorption lines using a nonparametric method to estimate the line flux following the method describe in detail in \cite{Kehoe2024}. In summary, we first determined the local continuum using line-free regions on the blue and red sides around each absorption feature (Table \ref{tab:line_windows}) determined from a high S/N composite spectrum of the full sample of galaxies with Fe\,{\sc ii} coverage (see Section \ref{sec:comp} for details of composite spectra). We defined the continuum level as the average flux density of the two regions and used the continuum level to calculate the equivalent width (EW) of each line. The EW was measured by integrating over a wavelength interval extending $\pm6$\,\AA\,from the line center. This integration window was selected after testing a range of widths, with $\pm6$\,\AA\,providing the most robust measurements while minimizing contamination from neighboring spectral features. We estimated uncertainties on the flux through Monte Carlo simulations, perturbing each spectrum by its error spectrum for 1000 realizations. A line was considered significantly detected if the flux was greater than 4$\sigma$. This threshold was determined by evaluating a range of significance levels and choosing the smallest value above 3$\sigma$ that consistently produced robust results upon visual inspection.
    
            For all detected lines, we measured the observed centroid of each line by fitting a Gaussian profile in the observed frame using a non-linear least squares fitting routine. Uncertainties were determined using the same Monte Carlo approach, where the measured centroid was taken as the mean of the realizations and the uncertainty as the standard deviation.
    
            We calculated velocity shifts by comparing the measured centroid wavelength ($\lambda_{\rm obs}$) to the rest wavelength of each feature ($\lambda_{\rm rest}$): 
            \begin{equation}
                \Delta v_{\rm} = \frac{\lambda_{\rm obs}-\lambda_{\rm rest}(1+z)}{\lambda_{\rm rest}(1+z)}\times c
                \label{equation:delv}
            \end{equation}
           where $z$ is the galaxy's systemic redshift, determined from H$\alpha$. For galaxies with multiple detected Fe\,{\sc ii} features, we computed an inverse-variance-weighted average of the individual measurements to obtain a single velocity shift. Negative velocity offsets correspond to blueshifted absorption and may be indicative of outflowing gas, and positive velocity offsets correspond to redshifted absorption and are indicative of inflowing gas. We see instances of both outflow (blueshifted absorption) and inflow (redshifted absorption) and will discuss this in Section \ref{sec:in_geo}. In our sample, we measured Fe\,{\sc ii} velocities for 41 galaxies out of a total of 176 galaxies (23\%) with Fe\,{\sc ii} coverage. The average velocity was $\langle\Delta v_{\rm Fe\,II}\rangle=-70\pm 18 \rm \, km\,s^{-1}$ (where the uncertainty listed is the standard error on the mean), with velocities ranging from $-281$ to $+213\,\rm km\,s^{-1}$. The median uncertainty on individual Fe\,{\sc ii} velocity measurements was $22\,\rm km\,s^{-1}$.

        \subsubsection{Mg\,{\sc ii}}
            The Mg\,{\sc ii} doublet $\lambda\lambda2796,2803$ often shows a mixture of absorption and emission, making the centroid more challenging to measure than for Fe\,{\sc ii}. To capture this complexity, we used the Markov Chain Monte Carlo (MCMC) fitting approach, \texttt{emcee} \citep{2013Foreman}, applying the same procedure and priors as in \citet{Kehoe2025}, using three different parameterizations. We first modeled the profile as an absorption doublet, with one Gaussian for each transition, then as an emission doublet, and lastly as a four-component model that included both an emission and an absorption component for each transition \citep{Martin2013}. This method allowed us to capture the variety of Mg \,{\sc ii} profiles observed in the sample.

            For each model, the spectra were restricted to the wavelength range surrounding the Mg\,{\sc ii} doublet and normalized by the continuum level, estimated from line-free regions on the blue and red sides of the line profile (Table \ref{tab:line_windows}). The free parameters in the fit included the centroids and widths of the Gaussian components, the integrated fluxes of each line, and the continuum level. Physically motivated priors were applied to guide the fit. For the absorption components, we constrained the flux ratio between the $\lambda2796\rm\,\AA$ and $\lambda2803\rm\,\AA$ lines to fall between 1 and 2, where a ratio of 2 corresponded to optically thin lines and a ratio of 1 indicates saturated absorption. For the emission components, the allowed flux ratio ranged from 0.8 to 2.7, as the dominant excitation mechanism sets the relative strengths of the emission lines. The flux ratio reaches values near 2 when collisional excitation dominates and decreases toward 1 when radiative excitation dominates, as both lines emitted photons with roughly equal probability. Observed line ratios often deviate from these theoretical predictions, indicating that the intervening medium consists of a mixture of uniform and porous gas that absorbs and scatters light \citep{Martin2012,Martin2013, Kornei2013, 2018Henry, Chisholm2020}. To maintain a consistent velocity shift, we tied the centroids of the two doublet members. The absorption amplitude was also constrained, so that the lowest point of the absorption could only go $1\sigma$ below a flux density of zero. Lastly, we enforced limits on the line widths to prevent nonphysical results, requiring the Gaussian width to be at least as large as the galaxy's velocity dispersion traced by H$\alpha$, or H$\beta$ if H$\alpha$ was not detected. When modeling Mg\,{\sc ii} as a combination of absorption and emission, the absorption and emission line widths were inferred independently.

            We explored the posterior distributions using MCMC sampling, taking the median values as the best estimates of the parameters and the standard deviations as the uncertainties. The best model was selected based on $\chi^2$ comparisons, where the two-component model (emission+absorption) was selected over the one-component model if it provided a statistically significant  reduction in $\chi^2$ ($p\rm{-value}< 0.01$), with the $p$-value calculated from the difference in $\chi^2$ based on the change in degrees of freedom between models. We set the minimum total absorption flux for Mg\,{\sc ii} to be significantly detected at 5$\sigma$. Similar to Fe\,{\sc ii}, we determined this threshold by exploring a range of significance levels and adopting the smallest value above 3$\sigma$ that produced robust results upon visual inspection. We found that 46 galaxies were best described by an absorption only fit, 2 galaxies by an emission only fit, and 11 galaxies by a combined emission and absorption fit (see Figure 3 of \citealt{Kehoe2025}). In total, we successfully fit Mg\,{\sc ii} profiles for 59 of the 197 galaxies (31\%) with available Mg\,{\sc ii} coverage. The velocity shifts were calculated from the absorption centroids using equation \ref{equation:delv}. The average velocity for these detections was $\langle\Delta v_{\rm Mg\, II}\rangle=-50\pm 20 \rm \, km\,s^{-1}$, over a range of $-464$ to $+334\,\rm km\, s^{-1}$. The median uncertainty on individual Mg\,{\sc ii} velocity measurements was $23\,\rm km\,s^{-1}$.
            
        \subsubsection{Na\,{\sc d}}\label{sec:MeasurementsNaD}
            Measuring the centroids from the Na\,{\sc d} $\lambda\lambda5891,5897$ doublet is more complex than for the Fe\,{\sc ii} lines because the profile is shaped by both ISM and systemic stellar absorption components. To remove the stellar contribution, each spectrum was divided by its best-fit SED. Both the spectrum and SED were shifted into the rest frame, and the SED was interpolated onto the wavelength grid of the spectrum and locally scaled to match the continuum around the Na\,{\sc d} doublet using continuum windows on either side of the lines (Table \ref{tab:line_windows}). The observed flux was then divided by the scaled SED to produce a residual spectrum dominated by ISM absorption. 
            
            The Na\,{\sc d} profile shape is further complicated by the nearby He\,{\sc i} $\lambda5876\,$\AA \,emission line. Therefore, both the He\,{\sc i} emission line and Na\,{\sc d} absorption lines must be fit simultaneously. The absorption was fit with a partial covering model from \cite{2005Rupke}, where the residual flux, $F_{\rm Na\,D}(\lambda_i)$, is given by
            \begin{equation}
                F_{\rm Na\,D}(\lambda_i) = 1 - C_f+C_f\exp\left[-(\tau_b(\lambda_i)+\tau_r(\lambda_i)\right]
            \end{equation}
            where $C_f$ is the gas covering fraction and $\tau_b$ and $\tau_r$ are the optical depth profiles of the 5891 and 5897\, absorption lines:
            \begin{equation}
                \tau_b(\lambda_i)=\tau_{0,b}\exp\left[-\frac{(\lambda_i-\lambda_c)^2}{2\sigma^2}\right],
            \end{equation}
            \begin{equation}
                \tau_r(\lambda_i)=\tau_{0,r}\exp\left[-\frac{(\lambda_i-\lambda_c-6)^2}{2\sigma^2}\right]
            \end{equation}
            where $\tau_{0,b}$ and $\tau_{0,r}$ are the central optical depths of the Na\,{\sc d} 5891 and 5897\,\AA \,lines, $\lambda_c$ is the central wavelength of the 5891\,\AA \,absorption line, and $\sigma$ is the Gaussian width of the profile and corresponds to the velocity dispersion of the absorbing gas. Prior to our fitting, we account for the NIRSpec instrumental resolution by defining the observed line width, $\sigma_{\rm tot}$, as $\sigma_{\rm tot}=\sqrt{\sigma^2_{\rm Na\,D}+\sigma^2_{\rm inst}}$, where $\sigma^2_{\rm Na\,D}$ is the intrinsic velocity width of the  Na\,{\sc d} absorption feature, and $\sigma^2_{\rm inst}$ represents the instrumental broadening. In our model, we allow the ratio of $\tau_{0,b}/\tau_{0,r}$ to vary between 1 and 2. This range encompasses both the optically thick regime ($\tau_{0,b}/\tau_{0,r}=1$) and the optically thin regime ($\tau_{0,b}/\tau_{0,r}=2$). The centroids of the two lines are fixed relative to each other to ensure a consistent velocity shift across the doublet.
            
            MCMC was used to sample the posterior distributions, where the medians were adopted as the best-fit parameters and the standard deviations as their uncertainties. Lines were considered significantly detected if the integral of the absorption profile exceeded 3$\sigma$. The threshold was determined following the same approach used for Fe\,{\sc ii} and Mg\,{\sc ii}. Furthermore, each spectrum was visually inspected due to the complexity and noisy nature of the profiles in our sample. Out of the 315 galaxies in our sample with Na\,{\sc d} coverage, 32 had significantly detected absorption (10\%). The velocity shift of Na\,{\sc d} was calculated using the absorption centroid and equation \ref{equation:delv}. The average velocity of the sample was $\langle\Delta v_{\rm Na\, D}\rangle=-79\pm 29 \rm \, km\,s^{-1}$, with a range of $-261$ to $+664\,\rm km\,s^{-1}$. The median uncertainty on individual Na\,{\sc d} velocity measurements was $93\,\rm km\,s^{-1}$. We summarize the velocity offsets for Fe\,{\sc ii}, Mg\,{\sc ii} and Na\,{\sc d} in Figure \ref{fig:vel_hist}.
        
    \subsection{Composite Spectra}\label{sec:comp}
        The individual spectra in our sample spanned a continuum S/N per pixel range of $\sim$0.4--24 near the features of interest, with a median value of 3.5, while galaxies with reliable velocity measurements had a higher median continuum S/N of 5.6. Given the limited S/N of many individual spectra, we constructed composite spectra to maximize the use of our full sample and explore the dependence of outflow velocity on galaxy properties through the average absorption line profiles. We focused on the Fe\,{\sc ii} $\lambda\lambda2587, 2600$\AA, Mg\,{\sc ii} $\lambda\lambda 2796, 2803$\AA\,lines, as well as Na\,{\sc d} absorption.
        
        Only galaxies with valid spectral coverage over the region of interest were included in the stacking procedure. For the Fe\,{\sc ii} and Mg\,{\sc ii} stacks, we limited the sample to galaxies at $z<4$ to ensure comparable redshift distributions across all bins. We then divided the sample into bins based on individual galaxy properties. For Fe\,{\sc ii} and Mg\,{\sc ii}, we used three bins per property. For $M_*$, stellar $A_V$ and $\beta$, and inclination composites, the bins for Fe\,{\sc ii} and Mg\,{\sc ii} contain 46 galaxies. In our sample, 30 galaxies lack at least two detected Balmer lines. Therefore, for the SFR, $E(B-V)$, sSFR, and $\Sigma_{\rm SFR}$ composites, the number of galaxies in each bin for the Fe\,{\sc ii} and Mg\,{\sc ii} stacks was reduced to 36 galaxies per bin. The properties of the NUV composite spectra are presented in Table \ref{tab:comp_NUV}.

        For the Na\,{\sc d} composites, before stacking, we removed the stellar continuum contribution from each spectrum by dividing by each galaxy's best-fit SED as described in Section \ref{sec:MeasurementsNaD}. We initially constructed composites using the full sample, but the resulting spectra did not yield Na\,{\sc d} absorption that was strong enough for reliable line fitting. This is likely due to a combination of low S/N in many individual spectra and dilution of the absorption when stacking spectra that did not show clear Na\,{\sc d} detections. To better recover the Na\,{\sc d} absorption in the stacks, we limited the Na\,{\sc d} composites to galaxies with significant Na\,{\sc d} absorption in their individual spectra, yielding a subsample of 32 galaxies. We then divided this subsample into two bins with 16 galaxies each for all Na\,{\sc d} composites. The Na\,{\sc d}\ composite properties are listed in Table \ref{tab:comp_NaD}.

        To create the composites, we shifted each spectrum to the rest frame and interpolated it onto a common wavelength grid. We normalized the spectra by scaling them to a common flux level, determined from the median flux within a line-free region, 3100--3150\,\AA\,for Fe\,{\sc ii} and Mg\,{\sc ii} and 5920--5940\,\AA\,for Na\,{\sc d}. We combined the spectra by taking the median flux at each wavelength. The error spectrum of each composite was calculated by summing the squared individual flux uncertainties at each wavelength, dividing by the number of spectra contributing to that wavelength, and taking the square root. 

\section{Results}\label{sec:Results}
    In this section, we examine how velocity offsets correlate with galaxy properties. This work builds on our previous analysis \citep{Kehoe2025}, which used only the AURORA survey, by combining the sample with the LyC22 and EXCELS surveys, significantly increasing the number of galaxies to be analyzed. This expanded dataset allows for a more robust investigation of trends between outflow kinematics and galaxy properties, as well as a direct comparison across multiple tracers, including Fe\,{\sc ii}, Mg\,{\sc ii}, and Na\,{\sc d}. In addition to outflow velocity trends, we examine subsamples of galaxies exhibiting inflows and Mg\,{\sc ii} emission to explore how these populations differ from the overall sample.
    
    \subsection{Correlations with Galaxy Properties}
        \begin{figure*}
            \centering
            \includegraphics[width=\linewidth]{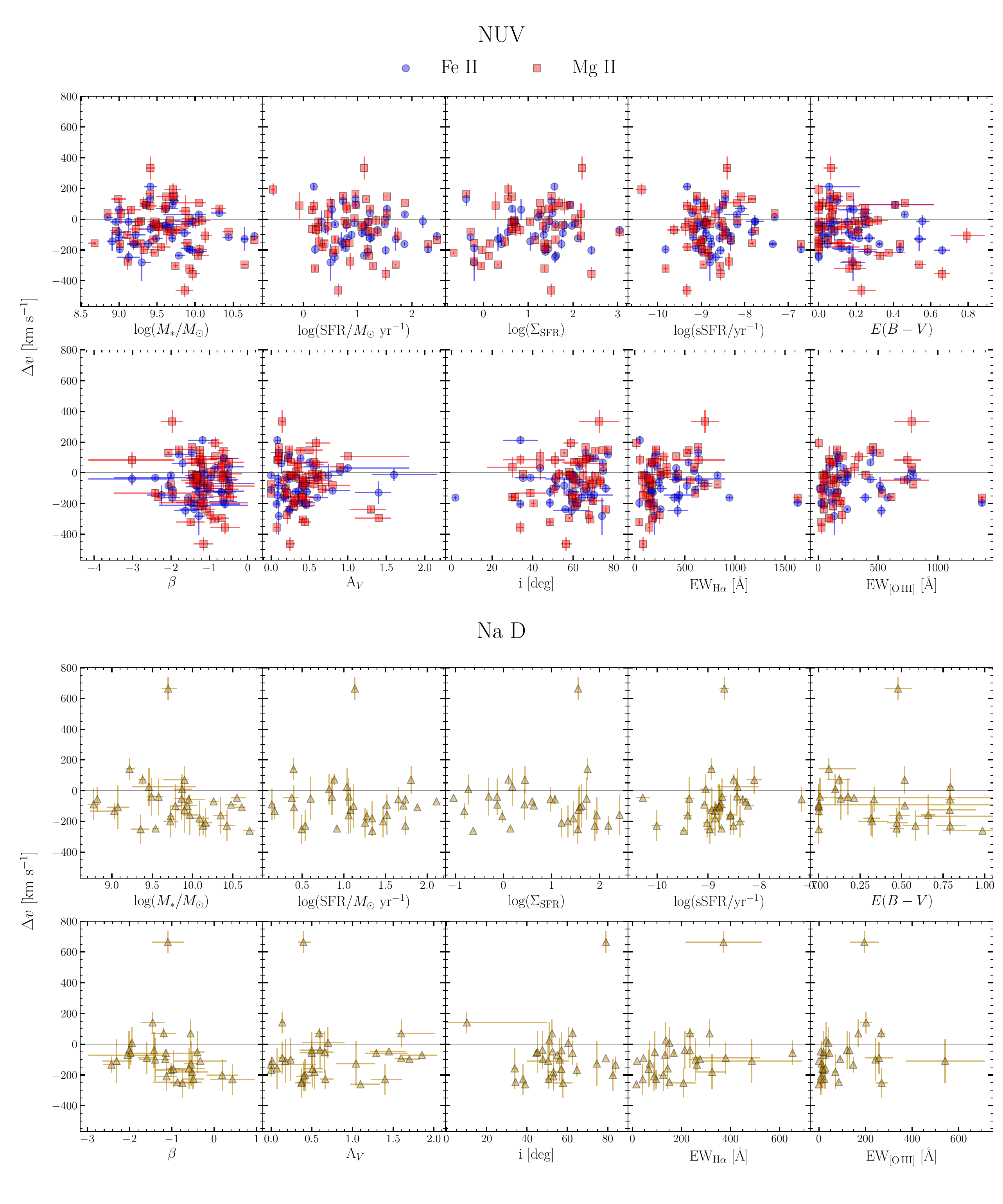}
            \caption{Gas kinematics traced by NUV absorption lines (top; Fe\,{\sc ii}, blue circles and Mg\,{\sc ii}, red squares), and optical absorption line (bottom; Na\,{\sc d}, gold triangles) absorption line centroids versus galaxy properties. These individual measurements show no correlation between outflow velocity and any galaxy property.}
            \label{fig:scatter_vel}
        \end{figure*}

        \begin{figure}
            \centering
            \includegraphics[width = \linewidth]{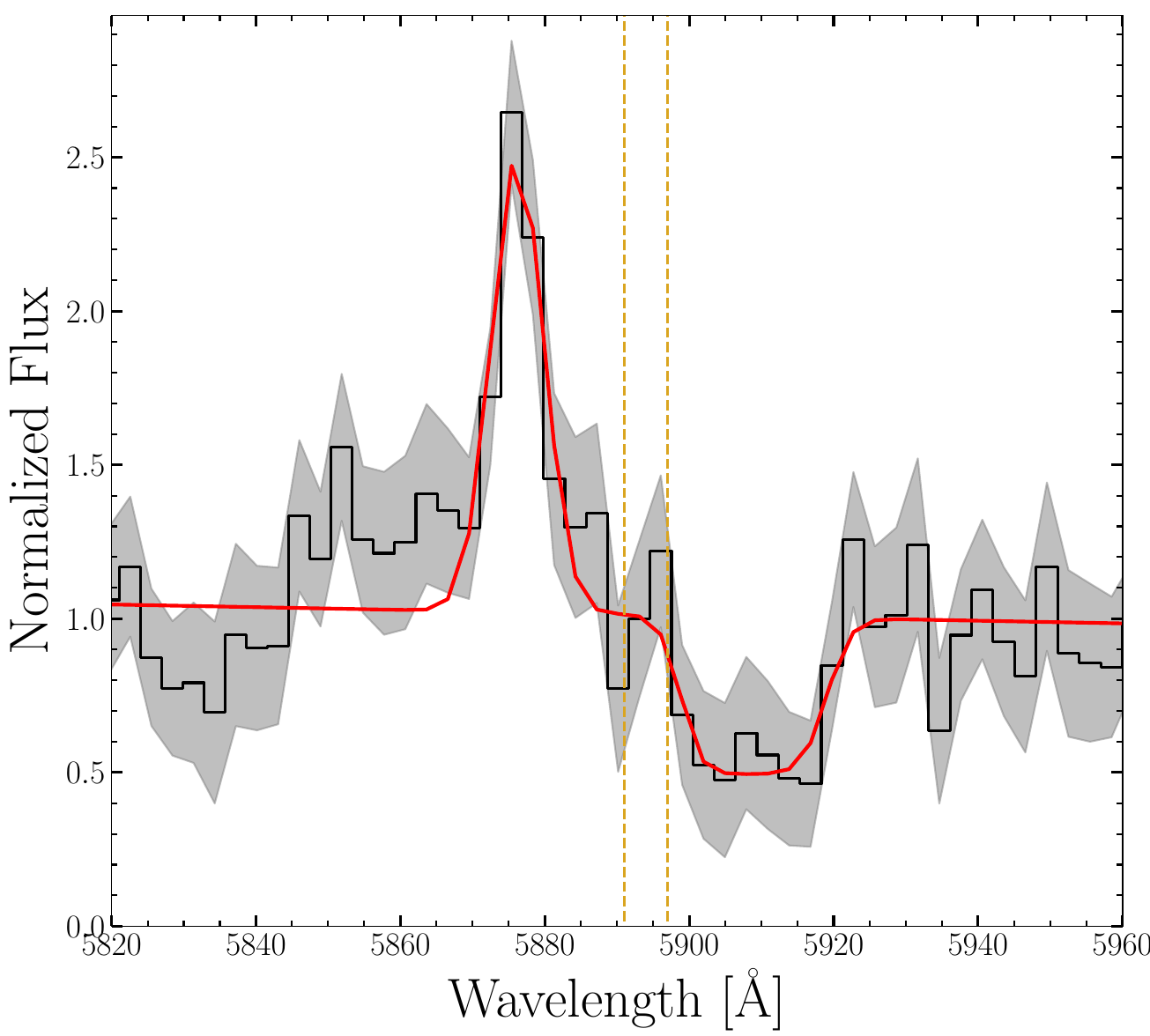}
            \caption{Spectrum in the Na\,{\sc d} region for SSA22-20013 from the LyC22 survey at $z=3.0495$. The best-fitting model is shown in red, and the vertical dashed orange lines mark the rest wavelengths of the Na\,{\sc d} $\lambda\lambda 5891,5897$ absorption doublet. The absorption feature is clearly redshifted relative to the expected rest-frame wavelengths, consistent with the large positive velocity offset identified in Figure \ref{fig:scatter_vel}.}
            \label{fig:inflow_outlier}
        \end{figure}

        \begin{figure*}
            \centering
            \includegraphics[width=\linewidth]{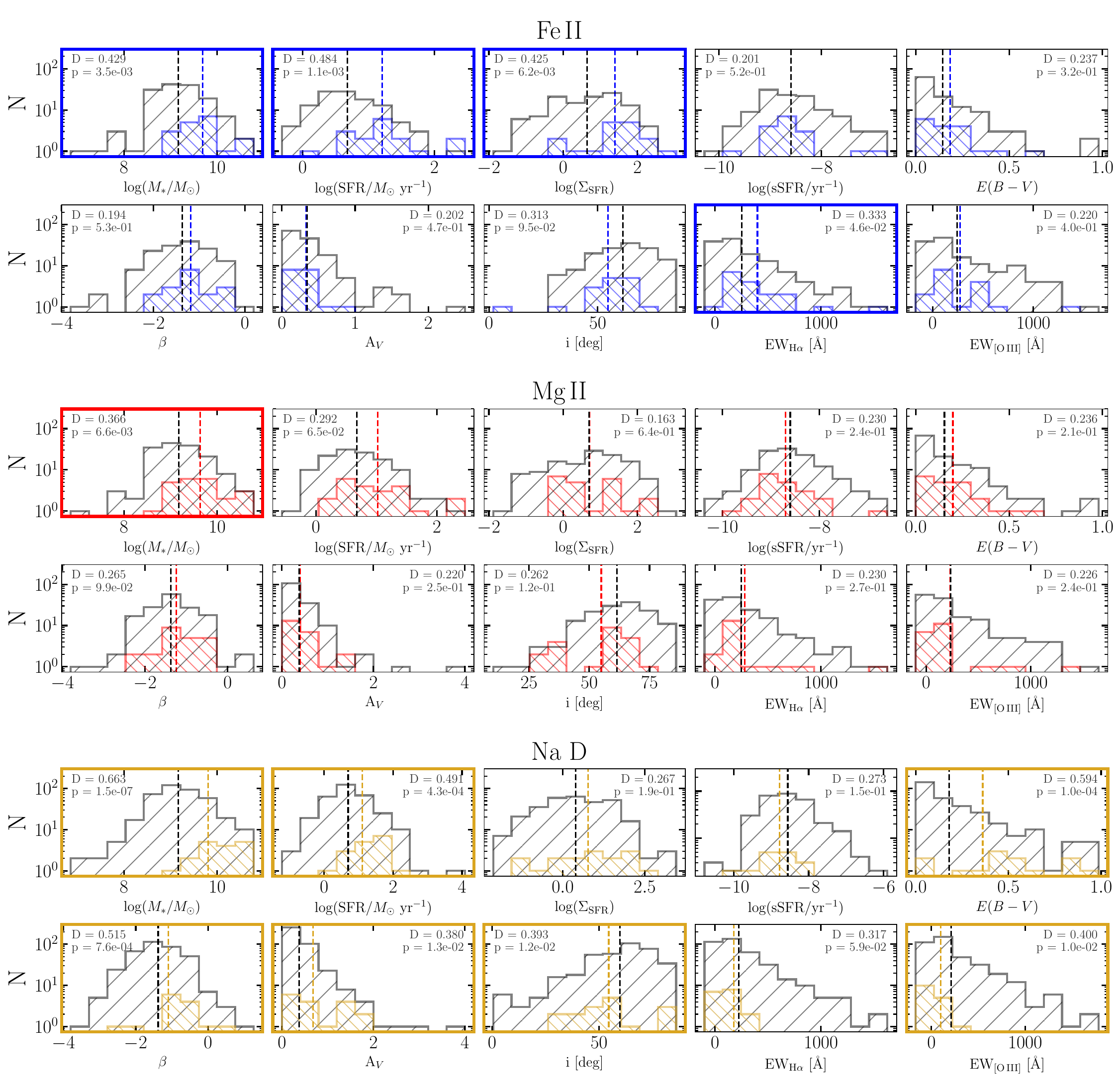}
            \caption{Distributions of galaxy properties for those with significant Fe\,{\sc ii} (top; blue), Mg\,{\sc ii} (middle; red), and Na\,{\sc d} (bottom; gold) outflow detections ($\Delta v < 0$ and $\left|\Delta v\right| > \sigma_{\Delta v}$) and galaxies with no outflow detections (gray). Vertical dashed lines indicate mean values for each sample. The KS $D$ statistic and $p$-values are reported in the upper corners of each panel. Significant results are highlighted in blue (Fe\,{\sc ii}), red (Mg\,{\sc ii}), and gold (Na\,{\sc d}). Galaxies with Fe\,{\sc ii}, Mg\,{\sc ii} and Na\,{\sc d} outflows exhibit significantly higher $M_*$,  galaxies with Fe\,{\sc ii} outflows have significantly higher SFR and $\Sigma_{\rm SFR}$, and galaxies with Na\,{\sc d} outflows have significantly higher SFR, nebular $E(B-V)$, and stellar $A_V$ and $\beta$, and lower inclination compared to those without detected outflows.}
            \label{fig:prop_hist_all}
        \end{figure*}
        
        Analyzing correlations between outflow velocity and galaxy properties provides insight into the physical processes that drive galactic outflows. We explore how outflow velocities relate to stellar and dust properties, as well as galaxy inclination. Stellar properties, including $M_*$, SFR, sSFR, and $\Sigma_{\rm SFR}$, trace the overall growth, star formation activity, and energetic output of galaxies. In addition to these stellar properties, we included the rest-frame EW of [O\,{\sc iii}] and H$\alpha$ emission lines as indirect tracers of the specific star-formation rate. Dust properties including nebular $E(B-V)$, and stellar $A_V$ and $\beta$ (though $\beta$ also reflects the underlying stellar population and nebular continuum), describe the amount and distribution of dust, which have been shown to correlate with the absorption EW of outflowing gas \citep{2021Du}. Correlations with galaxy inclination are sensitive to the geometry of the outflow. 
        
        Outflow velocities from individual galaxies, measured using Fe\,{\sc ii}, Mg\,{\sc ii}, and Na\,{\sc d} absorption lines, are compared against these galaxy properties in Figure \ref{fig:scatter_vel}. One notable outlier is present in the Na\,{\sc d} measurements, where a single galaxy (SSA22-20013; $z=3.0495$) exhibits a significantly redshifted absorption velocity traced by the line centroids of $+664\,\rm km\,s^{-1}$. This inflow velocity is greater in magnitude than others recently reported for $z\geq 2$ galaxies \citep{Weldon2023, 2024Davies}. We inspected this spectrum in detail (shown in Figure \ref{fig:inflow_outlier}) and confirm that the absorption signal is robust and not the result of noise or an unreliable fit. A detailed analysis of this source is warranted, but outside the scope the current work.
        
        We tested for correlations using the Spearman rank coefficient and found no significant trend between gas kinematics and any galaxy property. For each tracer, we also separated the sample into galaxies with and without significant outflow detections. Significant outflow detections are defined as $\left|\Delta v\right| > \sigma_{\Delta v}$, where $\sigma_{\Delta v}$ is the uncertainty on the velocity measurement, and $\Delta v< 0$. The non-outflow sample includes both galaxies without line detections and galaxies where $\left|\Delta v\right| \leq \sigma_{\Delta v}$. Using the KS test to assess whether these two populations come from the same parent distribution (Figure \ref{fig:prop_hist_all}), we found that galaxies with detected outflows in any tracer had statistically higher masses. Galaxies with Fe\,{\sc ii} outflows also showed higher SFR, $\Sigma_{\rm SFR}$, and H$\alpha$ EW and galaxies with outflows detected from Na\,{\sc d} also had higher SFR, nebular $E(B-V)$, stellar $A_V$ and $\beta$, and lower [O\,{\sc iii}] EW and inclination (i.e., more face-on).

        \begin{figure*}
            \centering
            \includegraphics[height=\linewidth]{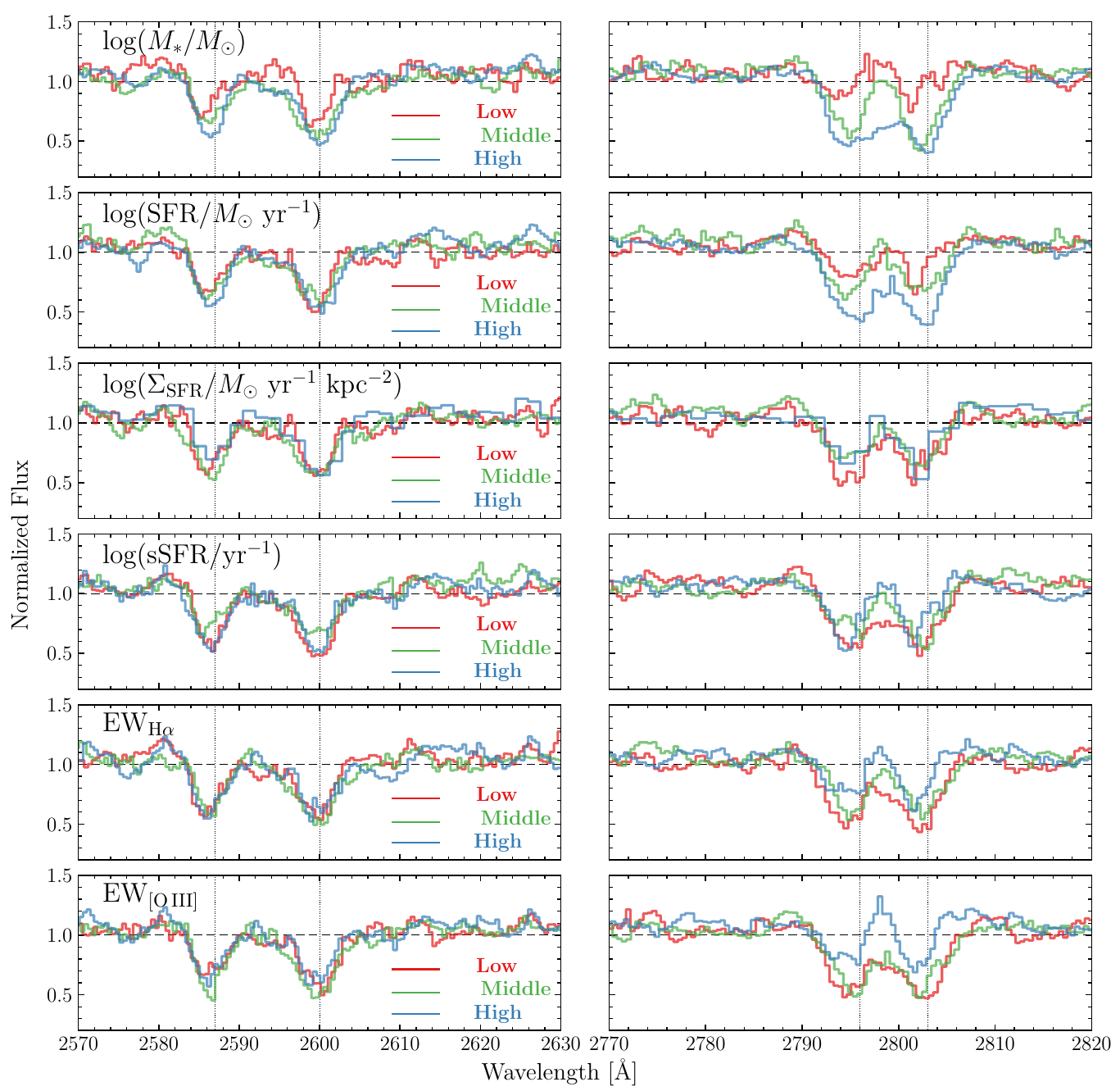}
            \caption{Composite spectra of galaxies binned by stellar properties, with Low, Middle, and High bins shown in red, green, and blue, respectively. The left panels show Fe\,{\sc ii} $\lambda2587$ and $\lambda2600$, while the right panels show the Mg\,{\sc ii} $\lambda\lambda2796,2803$ doublet. Vertical dashed lines indicate the rest wavelengths of the absorption features. Emission filling is most evident in Mg\,{\sc ii} profiles, particularly for stellar mass, SFR and [O\,{\sc iii}] EW.}
            \label{fig:comp_stellar_NUV}
        \end{figure*}

        \begin{figure*}
            \centering
            \includegraphics[width=\linewidth]{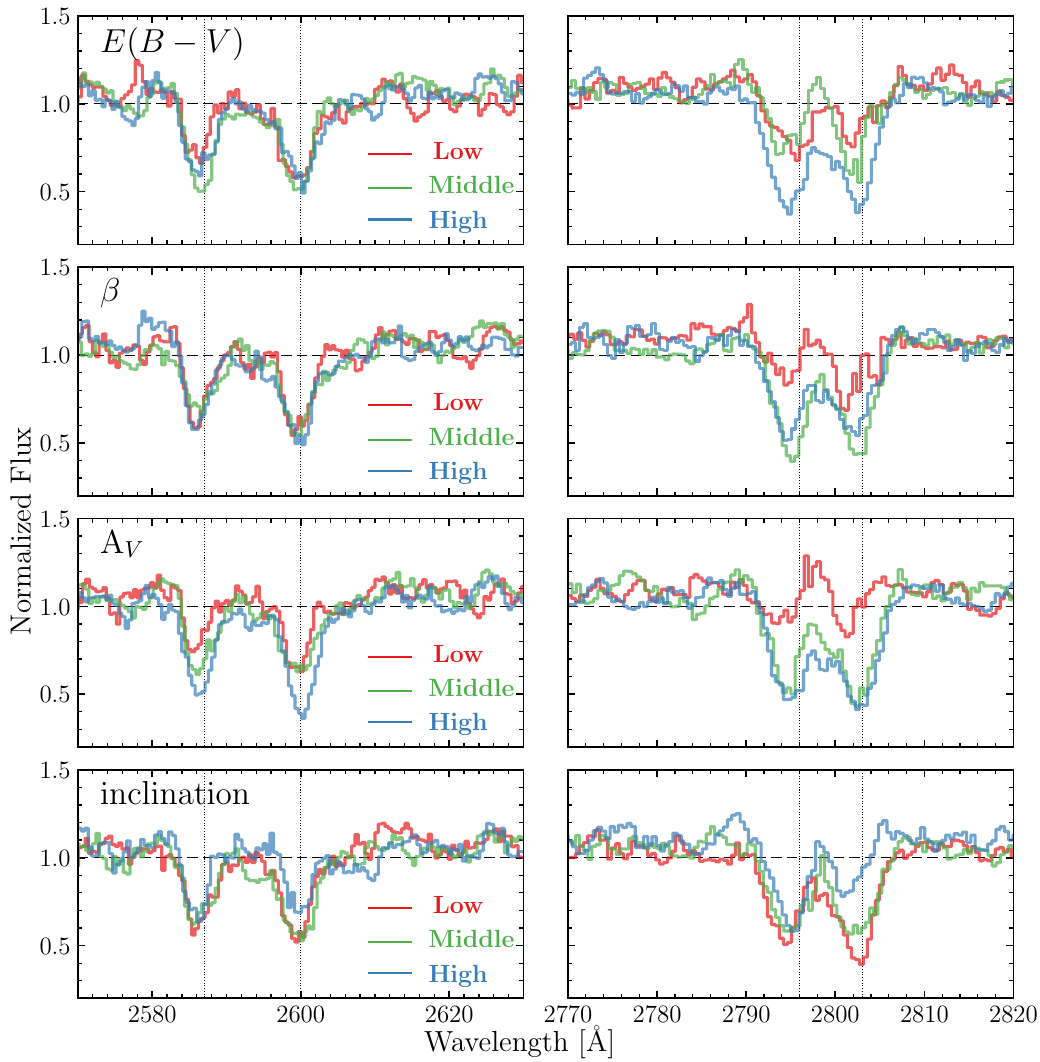}
            \caption{Same as Figure \ref{fig:comp_stellar_NUV}, but for galaxies binned by dust properties and inclination. Low, Middle, and High bins are shown in red, green, and blue, respectively. Emission filling is most pronounced in the Mg\,{\sc ii} profiles for nebular $E(B-V)$, and stellar $A_V$ and $\beta$.}
            \label{fig:comp_dust_NUV}
        \end{figure*}

        \begin{figure*}
            \centering
            \includegraphics[width=\linewidth]{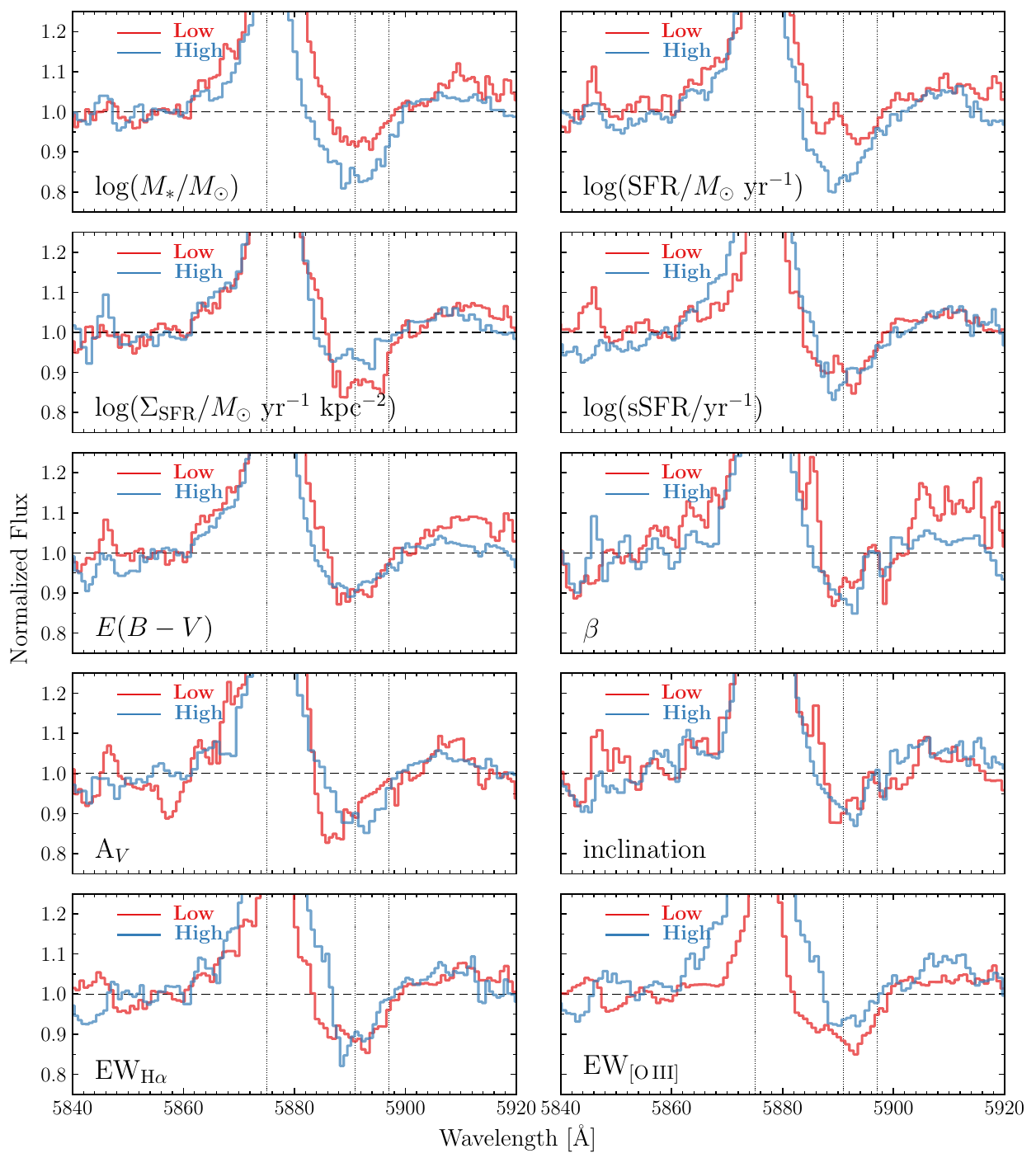}
            \caption{Continuum-divided composite spectra of galaxies for Na\,{\sc d} $\lambda\lambda5891,5897$ with He\,{\sc i} $\lambda5875$ emission, binned by galaxy properties. Low and High bins are shown in red and blue, respectively. Vertical dashed lines indicate the rest wavelengths of the line features.}
            \label{fig:comp_NaD}
        \end{figure*}

   \begin{deluxetable*}{llcccccccc}
        \tablecaption{Composite Spectra Measurements for Fe\,II and Mg\,II}
        \label{tab:comp_NUV}
        \tablehead{
        \colhead{Property} & \colhead{Composite} & \colhead{N\tablenotemark{a}} & \colhead{N$_{\rm Mg\,II, em}$\tablenotemark{b}} & \colhead{S/N}  & \colhead{Average Property Value} & \colhead{$v_{80,\,\text{FeII}}$\tablenotemark{c}} & \colhead{$v_{80,\,\text{MgII}}$\tablenotemark{c}} & \colhead{EW$_{\rm FeII}$\tablenotemark{d}} & \colhead{EW$_{\rm MgII}$\tablenotemark{d}} \\
         & & & & [pixel$^{-1}$] & \colhead{and Range} & \colhead{$\left[\text{km\,s}^{-1}\right]$} & \colhead{$\left[\text{km\,s}^{-1}\right]$} & [\AA] & [\AA]
        } 
            \startdata
             & Low & 46 & 7 & 15 & $8.85\,(8.43$--$9.08)$ & $-215 \pm 51$ & $-325 \pm 259$ & $0.9 \pm 0.2$ & $2.6 \pm 0.2$  \\
             $\log(M_{*})$ & Middle & 46 & 2 & 26 & $9.30\,(9.10$--$9.49)$ & $-347 \pm 32$ & $-398 \pm 14$ & $4.5 \pm 0.1$ & $4.0 \pm 0.1$\\
             & High & 46 & 2 & 14 & $9.86\,(9.50$--$10.78)$ & $-278 \pm 32$ & $-442 \pm 20$ & $6.5 \pm 0.2$ & $4.8 \pm 0.2$ \\
             \hline
             & Low & 36 & 2 & 20 & $0.22\,((-0.31)$--$0.49)$ & $-274 \pm 39$ & $-213 \pm 91$ & $2.1 \pm 0.2$ & $3.4 \pm 0.2$  \\
             log(SFR) & Middle & 36 & 4 & 33 & $0.66\,(0.49$--$0.87)$ & $-241 \pm 50$ & $-388 \pm 56$ & $3.7 \pm 0.1$ & $4.0 \pm 0.2$ \\
             & High & 36 & 4 & 15 & $1.30\,(0.88$--$2.46)$ & $-266 \pm 37$ & $-443 \pm 26$ & $6.5 \pm 0.2$ & $4.7 \pm 0.2$ \\
             \hline
             & Low & 36 & 1 & 30 & $-9.26\,((-10.28)$--$(-8.95))$ & $-312 \pm 22$ & $-438 \pm 33$ & $ 5.0 \pm 0.1$ & $4.1 \pm 0.1$ \\
             log(sSFR) & Middle & 36 & 5 & 24 & $-8.81\,((-8.95)$--$(-8.59))$ & $-259\pm 27$ & $-170 \pm 61$ & $3.7 \pm 0.1$ & $3.8 \pm 0.1$\\
             & High & 36 & 4 & 13 & $-8.22\,((-8.58)$--$(-7.23))$ & $-256 \pm 36$ & $-357 \pm 27$ & $2.9 \pm 0.2$ &  $4.5 \pm 0.2$\\
             \hline
             & Low & 36 & 3 & 17 & $-0.03\,((-1.03)$--$0.54)$ & $-250 \pm 43$ & $-396 \pm 22$ &  $4.6 \pm 0.2$ & $3.5 \pm 0.2$\\
             $\log(\Sigma_{\rm SFR})$ & Middle & 36 & 4 & 12 & $1.14\,(0.60$--$1.39)$ & $-195 \pm 60$ & $-259 \pm 146$ &  $3.8 \pm 0.2$ & $4.6 \pm 0.2$\\
             & High & 36 & 4 & 38 & $1.82\,(1.39$--$3.04)$ & $-319 \pm 35$ & $-423 \pm 83$ & $3.3 \pm 0.1$ & $3.7 \pm 0.1$ \\
             \hline
             & Low & 36 & 3 & 24 & $0.004\,(0.00$--$0.03)$ & $-338 \pm 37$ & $-196 \pm 51$ &  $2.8 \pm 0.1$ & $3.6 \pm 0.2$ \\
             $E(B-V) $ & Middle & 36 & 6 & 12 & $0.09\,(0.03-0.16)$ & $-253\pm 41$ & $-292 \pm 121$  & $2.8 \pm 0.2$ & $4.4 \pm 0.2$\\
             & High & 36 & 1 & 24 & $0.32\,(0.16$--$0.98)$ & $-254 \pm 47$ & $-495 \pm 27$ & $6.4 \pm 0.1$ & $4.3 \pm 0.2$ \\
             \hline
             & Low & 46 & 6 & 12 & $0.12\,(0.00$--$0.19)$ & $-189 \pm 56$ & $-247 \pm 348$ & $0.7 \pm 0.2$ & $2.8 \pm 0.2$ \\
             $A_{\rm V}$ & Middle & 46 & 4 & 17 & $0.28\,(0.20$--$0.40)$ & $-247 \pm 48$ & $-392 \pm 28$ & $5.1 \pm 0.1$ & $3.8 \pm 0.2$ \\
             & High & 46 & 1 & 27 & $0.63\,(0.40$--$1.64)$ & $-296 \pm 22$ & $-437 \pm 21$ & $6.0 \pm 0.1$ & $5.0 \pm 0.1$ \\
             \hline
             & Low & 46 & 8 & 13 & $-1.82\,((-3.02)
             $--$(-1.46))$ & $-248 \pm 42$ & $-157 \pm 165$ & $2 \pm 0.2$ & $3.3 \pm 0.2$ \\
             $\beta$ & Middle & 46 & 1 & 26 & $-1.21\,((-1.46)$--$(-1.06))$ & $-313 \pm 32$ & $-428 \pm 15$ & $5.4 \pm 0.1$ & $4.2 \pm 0.1$\\
             & High & 46 & 2 & 21 & $-0.60\,((-1.05)
             $--$0.20)$ & $-281 \pm 23$ & $-426 \pm 32$ & $4.8 \pm 0.1$  & $4.0\pm0.1$\\
             \hline
             & Low & 46 & 4 & 23 & $46\,(2$--$58)$ & $-225.3 \pm 32$ & $-466 \pm 245$ & $5.3 \pm 0.1$ & $4.4 \pm 0.1$ \\
             inclination & Middle & 46 & 2 & 26 & $63\,(58$--$68)$ & $-241 \pm 42$ & $-423 \pm 34$ & $4.9 \pm 0.1$ & $4.2 \pm 0.1$ \\
             & High & 46 & 5 & 15 & $76\,(68$--$86)$ & $-88 \pm 51$ & $-323 \pm 38$ & $2.9 \pm 0.1$ & $2.3 \pm 0.2$\\
             \hline
             & Low & 41 & 1 & 25 & $116\,(38$--$170)$ & $ -225\pm 45$ & $-516\pm 28$ & $ 3.7\pm 0.2$ & $ 5.2\pm 0.1$ \\
             EW$_{\rm H\alpha}$ & Middle & 41 & 1 & 20 & $232\,(170$--$343)$ & $ -230\pm 34$ & $ -411\pm 197$ & $ 4.6\pm 0.2$ & $ 4.3\pm 0.1$ \\
             & High & 41 & 7 & 12 & $644\,(365$--$1530)$ & $-135\pm55 $ & $ -138\pm 56$ & $ 3.8\pm 0.2$ & $ 4.0\pm 0.2$\\
             \hline
             & Low & 41 & 0 & 27 & $72\,(4$--$128)$ & $ -215\pm 56$ & $-451\pm 23$ & $ 3.8\pm 0.1$ & $ 5.6\pm 0.1$ \\
             EW$_{[\rm O\,III]}$ & Middle & 41 & 1 & 26 & $181\,(130$--$261)$ & $ -277\pm 35$ & $ -380\pm 242$ & $ 4.8\pm 0.1$ & $ 5.3\pm 0.1$ \\
             & High & 41 & 8 & 12 & $680\,(269$--$1648)$ & $-106\pm52 $ & $ -127\pm 116$ & $3.6 \pm 0.2$ & $ 1.6\pm 0.2$\\
            \enddata
           \tablenotetext{a}{Number of galaxies included in each composite spectrum.} 
          \tablenotetext{b}{Number of Mg\,{\sc ii} emitters (Section \ref{sec:result_mgiiem}) included in each composite spectrum.}
          \tablenotetext{c}{$v_{80}$ is defined as the velocity where the absorption reaches 80\% of the continuum on the blue side of the profile.}
          \tablenotetext{d}{Rest-frame absorption equivalent width.}
        \end{deluxetable*}

           \begin{deluxetable*}{llccccc}
        \tablecaption{Composite Spectra Measurements for Na\,{\sc d}}
        \label{tab:comp_NaD}
        \tablehead{
        \colhead{Property} & \colhead{Composite} & \colhead{N\tablenotemark{a}} & \colhead{S/N} & \colhead{Average Property Value} & \colhead{$\Delta v_{\text{NaD}}$\tablenotemark{b}} & \colhead{EW$_{\rm NaD}$\tablenotemark{c}}  \\
         & & & [pixel$^{-1}$]& \colhead{and Range} & \colhead{$\left[\text{km\,s}^{-1}\right]$} & [\AA] 
        } 
        \startdata
        $\log(M_{*})$ & Low & 16 & 26 & $9.4\,(8.77$--$9.87)$ & $-77 \pm 37$ & $2.4 \pm 0.2$ \\
         & High & 16 & 59 & $10.2\,(9.87$--$10.71)$ & $-142 \pm 9$ & $2.6 \pm 0.1$ \\
         \hline
        log(SFR) & Low & 16 & 27 & $0.68\,(0.13$--$1.08)$ & $-101 \pm 41$ & $2.1 \pm 0.2$ \\
         & High & 16 & 49 & $1.53\,(1.12$--$2.11)$ & $-140 \pm 11$ & $2.7 \pm 0.1$ \\
         \hline
        log(sSFR) & Low & 16 & 37 & $-9.16\,((-10.8)$--$(-8.78))$ & $-142 \pm 13$ & $2.1 \pm 0.2$ \\
         & High & 16 & 30 & $-8.08\,((-8.75)$--$(-7.17))$ & $-92\pm 20$ & $2.5 \pm 0.1$ \\
         \hline
        $\log(\Sigma_{\rm SFR})$ & Low & 16 & 34 & $-0.02\,((-1.03)$--$0.98)$ & $-95 \pm 11$ & $2.7 \pm 0.1$ \\
         & High & 16 & 30 & $1.65\,(1.07$--$2.41)$ & $-126 \pm 26$ & $2.0 \pm 0.2$ \\
         \hline
        $E(B-V)$ & Low & 16 & 27 & $0.14\, (0.00$--$0.47)$ & $-115 \pm 18$ & $2.5 \pm 0.2$ \\
         & High & 16 & 47 & $0.82\,(0.47$--$1.55)$ & $-134 \pm 16$ & $2.2 \pm 0.2$ \\
         \hline
        $A_{\rm V}$ & Low & 16 & 25 & $0.27\,(0.00$--$0.51)$ & $-137 \pm 17$ & $2.2 \pm 0.2$ \\
         & High & 16 & 58 & $1.27\,(0.53$--$3.6)$ & $-120 \pm 15$ & $2.7 \pm 0.1$ \\
         \hline
        $\beta$ & Low & 16 & 27 & $-1.66\,((-2.42)$--$(-1.13))$ & $-65 \pm 17$ & $2.6\pm 0.1$ \\
         & High & 16 & 34 & $-0.54\,((-1.09)$--$0.43)$ & $-139 \pm 16$ & $1.8 \pm 0.2$ \\
         \hline
        inclination & Low & 16 & 41 & $11\,(10$--$53)$ & $-129 \pm 20$ & $2.2 \pm 0.1$ \\
         & High & 16 & 33 & $313\,(53$--$84)$ & $-133 \pm 13$ & $2.3 \pm 0.1$ \\
        \hline
        EW$_{\rm H\alpha}$ & Low & 16 & 56 & $98\,(16$--$151)$ & $-147 \pm 7$ & $2.2 \pm 0.1$ \\
         & High & 16 & 23 & $343\,(169$--$652)$ & $-88\pm 23$ & $2.5\pm 0.2$ \\
        \hline
        EW$_{\rm [O\,{III}]}$ & Low & 16 & 56 & $19\,(0$--$45)$ & $-149 \pm 5$ & $2.0 \pm 0.1$ \\
         & High & 16 & 22 & $200\,(71$--$542)$ & $-39 \pm 44$ & $2.4 \pm 0.2$ \\
        \enddata
        \tablenotetext{a}{Number of galaxies included in each composite spectrum.}
        \tablenotetext{b}{Centroid velocity offset for Na\,{\sc d}.}
        \tablenotetext{c}{Rest-frame absorption equivalent width.}
        \end{deluxetable*}
        
        We also analyzed composite spectra to probe the average outflow properties across bins of galaxy parameters. These stacked profiles reveal how outflow velocities and absorption strengths vary with stellar, dust, and geometric properties, providing complementary measurements from individual galaxy spectra. In Figures \ref{fig:comp_stellar_NUV} and \ref{fig:comp_dust_NUV} we compare the composite spectra for Fe\,{\sc ii} and Mg\,{\sc ii} across the different galaxy properties.\footnote{Although not the focus of the current study, we note that Fe\,{\sc ii}$^*\lambda2612$ and $\lambda 2626$ fine-structure emission is not detected in the composite spectra shown in Figures~\ref{fig:comp_stellar_NUV} and \ref{fig:comp_dust_NUV}. The lack of detected Fe\,{\sc ii}$^*$ emission stands in contrast to the fine-structure emission observed by, e.g.,  \citet{Kornei2013} and \citet{2012Erb} in star-forming galaxies at $z\sim 1-2$. Fe\,{\sc ii}$^*$ emission is also absent in local starburst galaxy spectra \citep{Leitherer2011}, which is explained by \citet{Giavalisco2011} as being due to slit-losses. Accordingly, the physical size subtended by the $z\sim 0$ slit observations ($\sim 100$~pc) may not encompass a more extended Fe\,{\sc ii}$^*$-emitting halo. Likewise, the small (0.2") NIRSpec slit may also miss Fe\,{\sc ii}$^*$ emission if it originates in an extended ($>$~few-kpc-scale) distribution. } To quantify outflow velocities, we used $v_{80}$, defined as the velocity where the absorption reaches 80\% of the continuum on the blue side of the profile. We used $v_{80}$ instead of centroid measurements because resonant emission can partially fill in absorption features, biasing centroids towards larger blueshifts \citep{2011Prochaska,2012Erb, Martin2012, Kornei2013}. 
        
        We measured $v_{80}$ by identifying the absorption minimum and stepping toward bluer wavelengths until the sum of the flux and its uncertainty exceeded 0.8 of the continuum. Uncertainties were estimated by perturbing the spectra according to their error spectra and repeating the measurement 1000 times. For the Fe\,{\sc ii} and Mg\,{\sc ii} composites, the blue wings of the absorption profiles are consistent across most properties (Figures \ref{fig:comp_stellar_NUV}, \ref{fig:comp_dust_NUV}, Table \ref{tab:comp_NUV}), yielding comparable velocities within the uncertainties. The lack of strong monotonic trends suggests that any correlations in outflow kinematics are either weak or obscured by measurement uncertainties and spectral resolution. We find a significant correlation between $v_{80}$ and inclination, where $v_{80}$ increases in magnitude (more blueshifted) at smaller inclination (i.e., more face-on), suggesting that outflows may be collimated \citep{2011Bordoloi, Kornei2012} and that true outflow velocities may be underestimated in many cases because we are observing projected velocities. A more detailed discussion of this trend is presented in Section \ref{sec:out_geo}.

        For Na\,{\sc d}, we compare the composites of the two bins in Figure \ref{fig:comp_NaD} and Table \ref{tab:comp_NaD}. The Na\,{\sc d} composites are based on a smaller subsample than the Fe\,{\sc ii} and Mg\,{\sc ii} stacks, since we only include galaxies with individual Na\,{\sc d} detections. We adopted centroid velocities rather than $v_{80}$ because the nearby He\,{\sc i} emission line contaminates the blue wing of the Na\,{\sc d} absorption profile. The centroid therefore provides a more stable estimate of the absorption kinematics for Na\,{\sc d}. A clear trend is seen with $M_*$ and H$\alpha$ and [O\,{\sc iii}] EW, where galaxies with higher $M_*$ and lower emission-line EWs exhibit more blueshifted absorption (Table \ref{tab:comp_NaD}). A possible correlation is also seen with $\beta$, with redder (more positive $\beta$) galaxies exhibiting more blueshifted Na\,{\sc d} absorption, although the significance of this trend is unclear given the lack of similar trends in other dust properties. Unlike the NUV tracers, the inclination bins in the Na\,{\sc d} composites show consistent velocities within the uncertainties, indicating no strong geometric dependence, though selection effects in the Na\,{\sc d} sample may obscure subtler underlying trends.

        \begin{figure*}
            \centering
            \includegraphics[width=\linewidth]{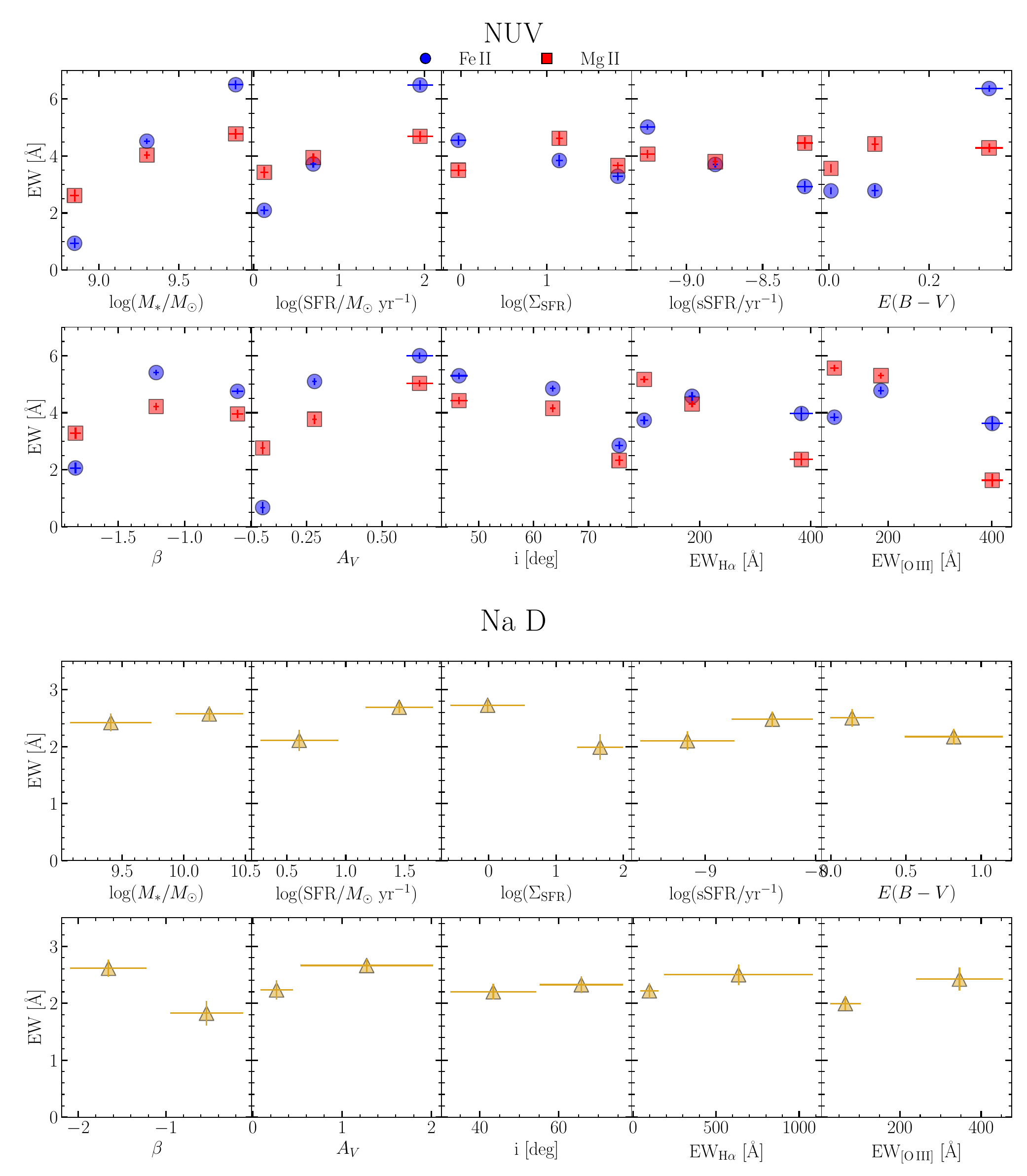}
            \caption{EWs of NUV absorption lines (top) (i.e, Fe\,{\sc ii} $\lambda2587$, $\lambda2600$, summed over both lines (blue), and Mg\,{\sc ii} $\lambda\lambda2796,2803$ (red)) and the Na\,{\sc d} $\lambda\lambda 5891,5897$ doublet (bottom; gold) measured from the composite spectra binned by galaxy properties. For the NUV lines, the EW increases with stellar mass, SFR, $A_V$, and decreases with galaxy inclination. For Na\,{\sc d}, the EW increases with SFR, and sSFR, and $A_V$, and decreases with $\Sigma_{\rm SFR}$, and $\beta$.}
            \label{fig:all_EW}
        \end{figure*}

        Fe\,{\sc ii}, Mg\,{\sc ii}, and Na\,{\sc d} absorption EWs provide an additional probe of the absorbing gas and trace the column density and the amount of outflowing material. We therefore examine how these EWs vary with galaxies properties for both the NUV tracers and Na\,{\sc d} (Figure \ref{fig:all_EW}). For the Fe\,{\sc ii} and Mg\,{\sc ii} transitions, the EWs show clear trends with increasing $M_*$, SFR, nebular $E(B-V)$, stellar $A_V$, and decreasing inclination. For Na\,{\sc d} EW, there are similar trends with increasing $M_*$ and SFR. However, the dust related trends are less uniform, with $A_V$ showing a positive correlation with EW, while $\beta$ shows the opposite. This inconsistency is likely due to the composite sample not being fully representative of the parent sample. The galaxies contributing to the Na\,{\sc d} stacks are biased toward more massive, higher SFR, and dustier systems, which shifts the effective dust properties in the composites. Recovering the trends in the composites will require a larger and higher S/N data in the Na\,{\sc d} region to reliably recover the doublet in stacked spectra using the full sample. We discuss the relationship between outflow velocity and strength and galaxy properties further in Section \ref{sec:dis_outflows}.

    \subsection{Galaxy Inflows}
        Gas inflows are a fundamental component of galaxy evolution as they replenish the supply of cold gas within galaxies. Observations have shown that, without continued gas accretion, the existing cold gas reservoirs would be depleted on relatively short timescales, preventing galaxies from sustaining their SFRs \citep{Kennicutt1989, Prochaska2009,Bauermeister2010,Leroy2008, Saintonge2017, Tacconi2018}. Gas accretion is commonly identified through inflow signatures, such as redshifted absorption features relative to the systemic velocity \citep{Rubin2012,2022Calabro,Weldon2023,Bevacqua2026}. However, such signatures are difficult to detect, as they can be weak, have low covering fractions, are metal-poor, and are often sensitive to the viewing geometry \citep{2010Steidel,Faucher2011, Fumagalli2011, 2011Kimm, Peroux2020}.

        \begin{figure}
            \centering
            \includegraphics[width = 3 in]{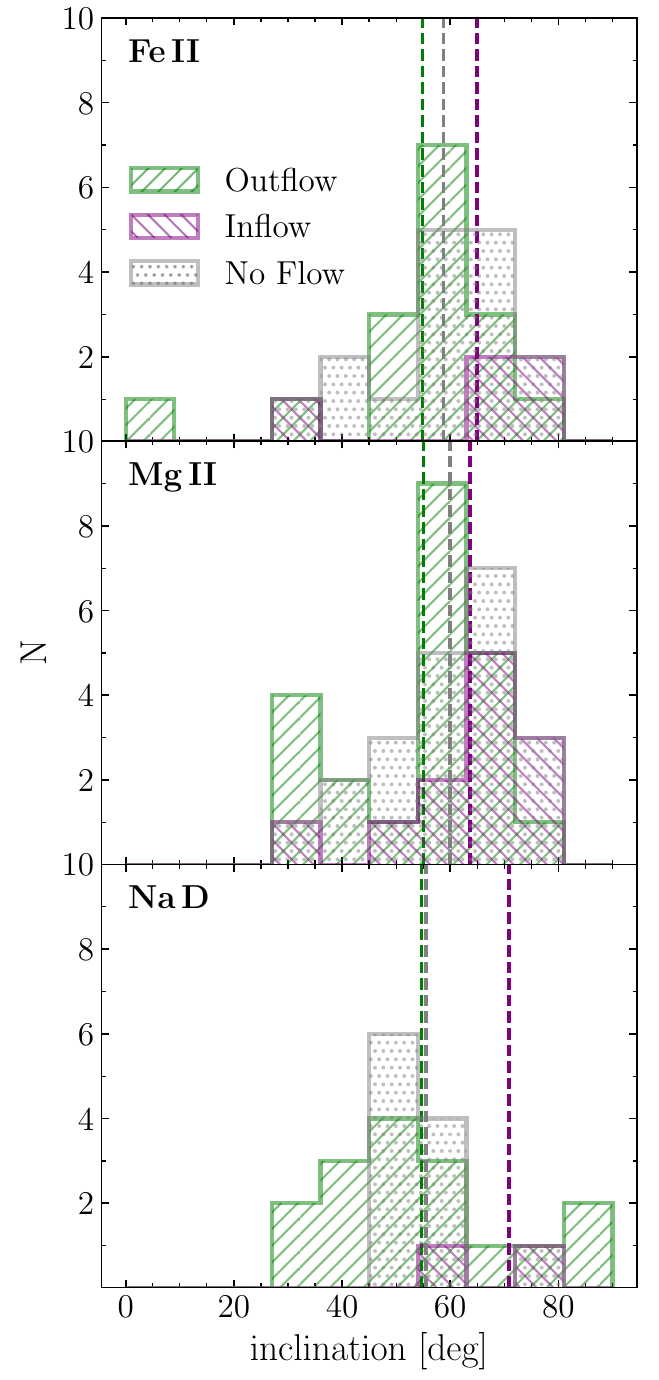}
            \caption{Histograms of galaxy inclinations for Fe\,{\sc ii} (top), Mg\,{\sc ii} (middle), and Na\,{\sc d} (bottom). Galaxies with detected outflows (green; $\Delta v <0$ and $\left|\Delta v\right| >\sigma_{\Delta v}$), inflows (purple; $\Delta v >0$ and $\left|\Delta v\right| >\sigma_{\Delta v}$), and no significant flows (gray; absorption fitting but $\left|\Delta v\right| \leq\sigma_{\Delta v}$) are shown. Vertical lines indicate the average inclination for each subsample. In all tracers, inflows are more commonly found in higher inclination (more edge-on) galaxies.}
            \label{fig:inc_hist}
        \end{figure}
        
        Defining a subsample of galaxies that exhibit inflow signatures allows us to directly examine how gas accretion relates to various galaxy properties, helping constrain the physical conditions under which inflows occur. Applying our inflow selection criteria ($\Delta v > 0$ and $\left|\Delta v\right| > \sigma$), we identify a subsample of 15 galaxies with evidence for inflows. Of these 15 inflows, 1 is traced by Fe\,{\sc ii}, 8 are traced by Mg\,{\sc ii}, 4 are traced by both Fe\,{\sc ii} and Mg\,{\sc ii}, and 2 are traced by Na\,{\sc d}. In all three tracers, inclination is the only galaxy property for which we observe a clear trend in inflow incidence. Combining the inflow detections from all tracers and comparing them to galaxies without detected inflows, we find a statistically significant difference in their inclination ($p\approx 0.03)$, corresponding to a $>$$2\sigma$ significance level. We find that inflows are preferentially detected in higher inclination (more edge-on) galaxies (Figure \ref{fig:inc_hist}), suggesting that viewing geometry plays an important role in our ability to detect accreting gas. We return to the interpretation of this trend in Section \ref{sec:in_geo}.

    \subsection{Mg\,{\sc ii} Emitters}\label{sec:result_mgiiem}
        A subset of galaxies in our sample exhibit significant Mg\,{\sc ii} emission. The Mg\,{\sc ii} line profiles can vary from strong absorption to emission-dominated, reflecting the resonant nature of the transition. In this process, absorbed photons are re-emitted and scattered by outflowing gas, producing emission near the systemic velocity and modifying the overall profile shape \citep{2009Weiner, 2010Rubin, Martin2013, 2011Prochaska, 2018Henry, Chisholm2020, Xu2023}.

        Detection rates of Mg\,{\sc ii} emission reported in the literature span a broad range ($\sim 1$--70\%) at similar redshifts, driven primarily by difference in he properties of the galaxy samples and identification methods and the quality of data \citep{2009Weiner, 2010Rubin, 2012Erb, 2013Guseva, 2018Feltre}. Methods based on flux excess redward of the Mg\,{\sc ii} transitions tend to be more conservative \citep{2009Weiner, 2010Rubin}, while pixel-based \citep{2012Erb} or EW criteria are more sensitive to weaker emission features \citep{Kornei2012, 2018Feltre}.

        \begin{figure*}
            \centering
            \includegraphics[width=\linewidth]{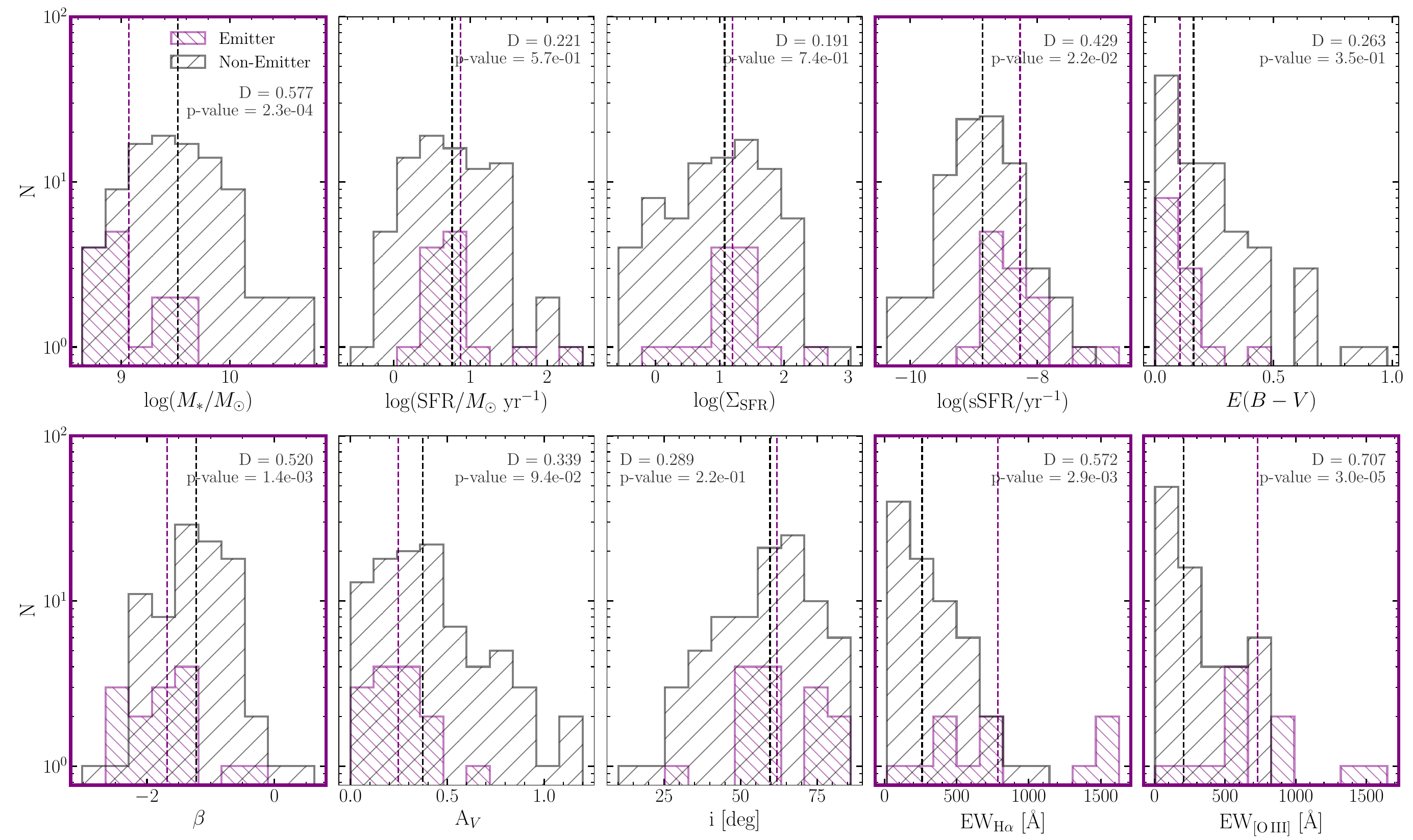}
            \caption{Histograms showing the distributions of various galaxy properties for the Mg\,{\sc ii} non-emitter (gray) and Mg\,{\sc ii} emitter (purple) samples. Dashed gray and purple lines indicate the average values for non-emitters and emitters, respectively. The KS $D$ statistic and $p$-values are reported in the upper corners of each panel. Figures with significant results are highlighted in bold purple. On average, Mg\,{\sc ii} emitters have lower stellar mass and $\beta$, and higher sSFR and H$\alpha$ and [O\,{\sc iii}] emission EW than non-emitters.}
            \label{fig:mgii_em_hist}
        \end{figure*}

        To maintain consistency with our previous analysis \citep{Kehoe2025}, we adopt the same conservative selection method based on detecting flux above the continuum redward of the Mg\,{\sc ii} lines \citep{2009Weiner, 2010Rubin}. This approach isolates the strongest Mg\,{\sc ii} emitters in our sample. We find 14 galaxies have strong Mg\,{\sc ii} emission, corresponding to 8\% of galaxies with Mg\,{\sc ii} spectral coverage. In Figure \ref{fig:mgii_em_hist}, we compare these ``Mg\,{\sc ii} emitters'' to galaxies without significant Mg\,{\sc ii} emission, restricting the comparison to objects with S/N$>3$ in the region surrounding the Mg\,{\sc ii} profile to minimize biases from low-quality spectra. Applying a KS test, we find that Mg\,{\sc ii} emitters have lower stellar masses and bluer $\beta$, and higher sSFRs relative to galaxies without significant Mg\,{\sc ii} emission. The connection between Mg\,{\sc ii} emission and galaxy properties is explored in more detail in Section \ref{sec:dis_mgii_em}.
        
        We also examine the contribution of Mg\,{\sc ii} emitters in the composite spectra binned by galaxy properties (Figure \ref{fig:comp_stellar_NUV} and Figure \ref{fig:comp_dust_NUV}). Emission filling is visible in the Mg\,{\sc ii} profiles across several bins, particularly where the underlying absorption is weaker, such as in the low stellar mass, low $E(B-V)$, low (more negative) $\beta$, and low $A_V$ bins. In these composites, the emission partially fills in the absorption features, illustrating how even a small number of strong emitters (Table \ref{tab:comp_NUV}) can influence the average line profiles.

\section{Discussion}\label{sec:Discussion}
    \subsection{Comparison to Other Outflow Studies}\label{sec:dis_outflows}
        \subsubsection{NUV Absorption}
            Fe\,{\sc ii} and Mg\,{\sc ii} trace low-ionization gas associated with cool, partially ionized material in galaxies. While we find no significant correlations between NUV outflow velocities and any galaxy property, we do find that galaxies with detected NUV outflows are biased toward higher $M_*$, SFR and $\Sigma_{\mathrm{SFR}}$ (Figure \ref{fig:prop_hist_all}). These trends are consistent with our previous analysis using only the AURORA survey \citep{Kehoe2025} and are in agreement with other observations as we discuss below. 

            Outflow velocities traced by NUV absorption at $z\sim1$ have been shown to increase with SFR and $\Sigma_{\mathrm{SFR}}$ \citep{Martin2012, Kornei2012, 2013Bradshaw, 2014Bordoloi, 2015Heckman, 2016Heckman, 2021Prusinski, 2022Xu_Classy}. In particular, when analyzing NUV observations of 1400 galaxies at $z\sim1.4$ \cite{2009Weiner} found a scaling relation $v_{10\%} \propto \mathrm{SFR}^{0.3}$. Here, $v_{10\%}$ is where the absorption reaches 90\% of the continuum on the blue side of the profile. Other studies, however, have found no significant correlation \citep{2010Steidel, 2012Talia, 2022Calabro}, due to a limited dynamic range, with SFRs spanning only 10--100$\,M_{\odot}\,\mathrm{yr}^{-1}$. In our previous analysis \citep{Kehoe2025}, we found no correlation as the sample spans a comparably narrow dynamic range. Similarly, with our expanded sample, we did not observe a clear dependence of outflow velocity on SFR as our $v_{80}$ values in the low- and high-SFR composites are consistent within the uncertainties (Table \ref{tab:comp_NUV}). A clear correlation between outflow velocity and SFR generally requires including galaxies with much lower SFRs of SFR$<1\,M_{\odot}\,\mathrm{yr}^{-1}$ \citep{2005Martin, 2022Xu_Classy}. While our expanded sample includes more galaxies in this low-SFR regime, a larger number of low-SFR galaxies is still needed to find a significant correlation with either the composites or individual measurements. However, we observed that galaxies with detected outflows are preferentially detected in systems with higher SFR and $\Sigma_{\rm SFR}$ (Figure \ref{fig:prop_hist_all}), which is broadly consistent with the picture in which strong star formation activity is associated with galactic outflows.
            
            Trends with $M_*$ in NUV observations are also variable across samples, with some studies reporting positive correlations ($v_{10\%} \propto M_*^{0.17}$) \citep{2009Weiner, 2010Rubin, 2012Erb} and others finding no dependence \citep{Martin2012, 2011Bordoloi, 2021Prusinski}. This difference can also be explained by a narrow dynamic range, as shown by \cite{2021Prusinski}, with masses $9.3\leq\log(M/M_{\odot})\leq10.8$. 
            We do not observe a clear correlation with mass as the lowest bins have large uncertainties and the middle and high bins show comparable blueshifts (Table \ref{tab:comp_NUV}). Although our sample spans a larger mass range, the lack of a robust trend may be due to a combination of selection effects, observational uncertainties, and differences in galaxy properties such as SFR, gas content, and outflow geometry that can influence measured outflow velocities \citep{2011Bordoloi, Martin2012, 2021Prusinski}. Consistent with previously reported trends \citep{2009Weiner, 2010Rubin, 2012Erb}, we found that galaxies with detected outflows are biased toward higher stellar masses (Figure \ref{fig:prop_hist_all}), suggesting that while a clear scaling relation is not evident in our NUV sample, outflow incidence may still increase with mass.
            
            We also found that the EW of absorption lines increases with $M_*$, SFR, nebular $E(B-V)$, and stellar $A_V$ (Figure \ref{fig:all_EW}), suggesting that the variations in the NUV absorption strength may be driven by changes in the column density and covering fraction of low-ionization gas rather than outflow velocity. Similar correlations between Fe\,{\sc ii} and Mg\,{\sc ii} absorption EWs and galaxy properties have been widely reported in $z\sim1$ samples \citep{2009Weiner, 2010Rubin, 2014Rubin, Martin2012, 2014Bordoloi, 2021Prusinski}. As these transitions are optically thick, the absorption EW is largely governed by the covering fraction and velocity dispersion of the absorbing material, while also being sensitive to the coupling between dust and metal-enriched neutral gas, where higher dust content is associated with  strong low-ionization absorption \citep{2021Du}. This trend may also be influenced by emission line filling, as discussed in Section \ref{sec:dis_mgii_em}. The observed increase in absorption EW with $M_*$, SFR, nebular $E(B-V)$, and stellar $A_V$, and  supports a picture in which more massive, actively star-forming, and dust-rich galaxies host outflows that cover a larger fraction of the background continuum and span a wider range of velocities \citep{2022Reddy}.

            These results indicate that while outflow velocity does not depend on any galaxy properties in our sample, the incidence of detected outflows increases with $M_*$, SFR, and $\Sigma_{\mathrm{SFR}}$, and the strength of the absorption increases with $M_*$, SFR, and $A_V$. These trends are consistent with NUV studies at $z\sim1$, which probe galaxies at significantly lower redshift than our $z\gtrsim2.5$ sample, highlighting that similar outflow trends may persist across cosmic time.

        \subsubsection{Na\,{\sc d} Absorption}
            Na\,{\sc d} absorption traces cooler, neutral gas that is often associated with dusty outflowing material \citep{2000Heckman, 2005Veilleux, 2010Chen, 2022Avery, 2024Davies}. We find no significant correlations between Na\,{\sc d} outflow velocity and galaxy properties in the individual measurements. However, galaxies with detected Na\,{\sc d} outflows tend to have higher $M_*$, SFR, nebular $E(B-V)$, and stellar $A_V$ and $\beta$, and lower [O\,{\sc iii}] EW, indicating Na\,{\sc d} outflows are more easily detected in more massive, actively star-forming, dustier galaxies. The difference between the SFR and emission-line EW trends reflects the fact that the EWs most closely trace the sSFR as opposed to total SFR. These results agree with our previous work based on only 10 galaxies \citep{Kehoe2025}, now expanded to a total of 32 detections of Na\,{\sc d}. 

            Previous work in the local universe found consistent results, where Na\,{\sc d} outflows are found in galaxies with higher masses, SFRs, and dust content \citep{2000Heckman, 2005Martin, 2010Chen, 2022Avery}. \textit{JWST} observations allow for Na\,{\sc d} to be observed at much higher redshifts ($z\sim2$) \citep{2024Belli, 2024Davies, 2024DEugenio, 2025Sun, 2026Moretti}. Using 113 galaxies at ($1\lesssim z \lesssim 3$), \cite{2024Davies} found that Na\,{\sc d} outflows are typically detected in more massive galaxies ($\log M_*/M_\odot>10$), in agreement with our mass trend. These higher redshift studies also suggest that while star formation likely dominates the driving mechanism of neutral outflows in the local Universe, AGN activity at higher redshift may contribute to or enhance Na\,{\sc d} outflows \citep{2024Belli, 2024DEugenio, 2025Sun, 2026Taylor}.
            
            We find that the composite spectra show a general increase in Na\,{\sc d} absorption EW and blueshift with $M_*$ and SFR, and less consistent behavior with sSFR and dust properties. However, these results should be interpreted with caution. The composites only include galaxies with significant individual Na\,{\sc d} detections and are therefore biased toward higher $M_*$, SFR, nebular $E(B-V)$, and stellar $A_V$ and $\beta$, and lower [O\,{\sc iii}] EW relative to the full sample. After continuum  normalizing the spectra by dividing by the best-fit stellar continuum from SED fitting, the residual Na\,{\sc d} absorption is significantly reduced, indicating that the outflow component is weak and difficult to isolate in the stacked spectra. A larger, higher signal-to-noise sample is required to robustly isolate the outflowing component of Na\,{\sc d} absorption in composite spectra and better quantify its dependence on the global properties of star-forming galaxies at high redshift.

    \subsection{Geometry of Outflowing Gas}\label{sec:out_geo}
        \begin{figure}
            \centering
            \includegraphics[width=\linewidth]{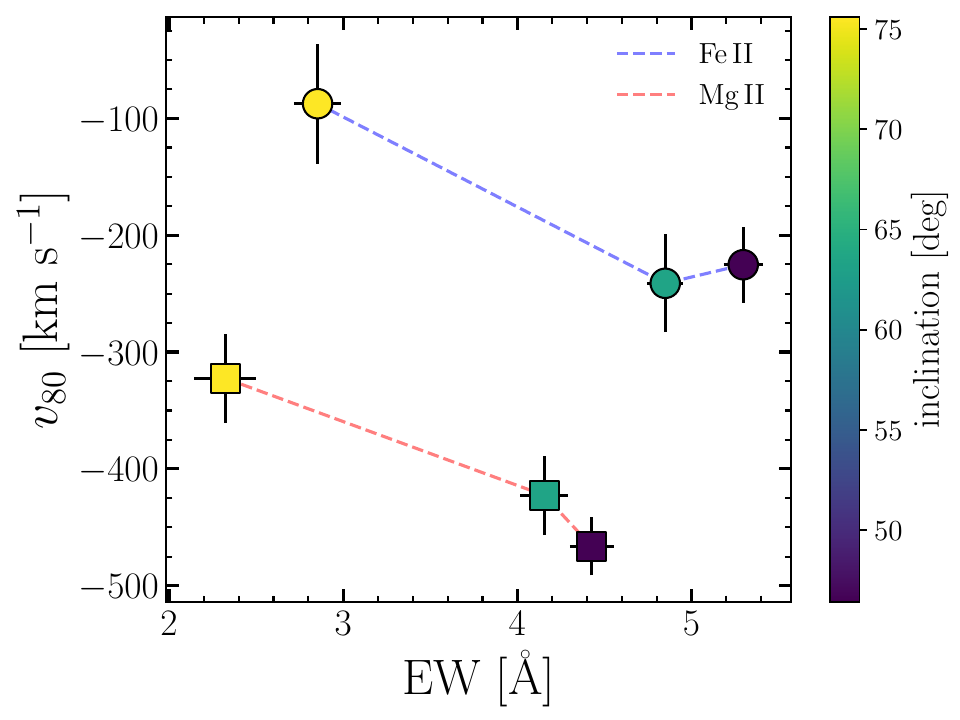}
            \caption{$v_{80}$ vs. EW from the composite spectra for Fe\,{\sc ii} (circles) and Mg\,{\sc ii} (squares) absorption. Points are color-coded by galaxy inclination. Dashed lines connect the the Fe\,{\sc ii} (blue) and Mg\,{\sc ii} (red) measurements. For both tracers, lower inclination (more face-on) galaxies exhibit both larger EWs and more blueshifted velocities, compared to higher inclination (more edge-on) systems. The Low- and High-inclination bins differ by 2.3$\sigma$ for Fe\,{\sc ii} and 3.2$\sigma$ for Mg\,{\sc ii}. The observed inclination dependence suggests that outflows are collimated and preferentially launched perpendiculat to the galactic plane.}
            \label{fig:outflow_geo}
        \end{figure}
        
        The outflow kinematics of the composite spectra exhibit a clear dependence on inclination. As shown in Figure \ref{fig:outflow_geo}, the lower inclination (more face-on) composites have higher Fe\,{\sc ii} and Mg\,{\sc ii} absorption EWs and higher maximum outflow velocities compared to the higher inclination (more edge-on) bins. The Low- and High-inclination bins differ in $v_{80}$ by 2.3$\sigma$ for Fe\,{\sc ii} and 3.2$\sigma$ for Mg\,{\sc ii}. This trend indicates that the outflow signal is stronger when galaxies are viewed more face-on, suggesting that the outflowing material is not isotropic. Instead, the inclination dependence points to a more structured outflow geometry, where the detectability and measured strength of the outflow varies with the viewing angle, consistent with a collimated geometry. 

        This picture is reproduced in simulations, in which outflows are preferentially launched perpendicular to the disk \citep{2011Brook, Glascow2013,  Fielding2017, 2019Nelson}. These simulations also show that outflowing gas is concentrated along the minor axis, while inflowing material is along the major axis \citep{Peroux2020}. In this model, feedback produces winds that escape along directions normal to the disk, leading to a bipolar structure.

        Observations across a wide range of redshifts further support this interpretation. In the local universe ($z\sim 0.1$), this geometry is strongly suggested by Na\,{\sc d} absorption measurements, where galaxies viewed more face-on exhibit larger blueshifts than more edge-on systems \citep{2000Heckman, 2010Chen, 2019Roberts}. Similar inclination trends have been reported in Mg\,{\sc ii} and Fe\,{\sc ii} absorption observations at $0.3\lesssim z \lesssim 1.4$, which find decreasing outflow strength and detection fraction with increasing inclination \citep{2011Bordoloi, Kornei2012, 2012Kacprzak, Rubin2014, Guo2023}. A comparable trend is also observed at higher redshift ($z\sim 2$), where outflow strength and detection fraction increase with decreasing inclination \citep{2012Newman}. However, this trend is not consistently recovered at $z\geq 2$, with many studies reporting no significant inclination dependence \citep{Law2012, 2019Schreiber, 2022Weldon}, likely due to the more irregular morphologies of high-redshift galaxies, which make it more difficult to robustly measure inclinations.

        In general, inclination trends are consistent with collimated outflows, where the observed outflow becomes stronger as the galaxy is viewed more face-on. The clear trend measured in our sample (Figure \ref{fig:outflow_geo}), where lower inclination (more face-on) galaxies exhibit larger EWs and more blueshifted outflow velocities, provides direct evidence for collimated outflows at earlier times.

    \subsection{Geometry of Inflowing Gas}\label{sec:in_geo}
        Our result that inflows are preferentially detected in highly inclined galaxies (Figure \ref{fig:inc_hist}) indicates that geometry influences the detectability of inflowing gas. This trend is evident across all three tracers, with 4 out of 5 (80\%) of Fe\,{\sc ii} inflows, 11 out of 12 (92\%) of Mg\,{\sc ii} inflows, and 2 out of 2 (100\%) of Na\,{\sc d} inflows detected in galaxies with inclinations greater than $50^\circ$. In support of this picture, the galaxy in our sample with the fastest inflow, SSA22-20013 (Figure \ref{fig:inflow_outlier}), is also highly inclined ($79^\circ$). The increased incidence of inflows in higher inclination (more edge-on) galaxies suggests that the gas producing the redshifted absorption is likely concentrated near the galactic plane rather than distributed isotropically throughout the system. With this geometry, inflow velocities are more easily detected when the galaxy is being viewed edge-on because gas moving within or close to the plane produces a larger line of sight velocity component. When the galaxy is closer to being viewed face-on (lower inclinations), these inflows project weakly along the line of sight, making them more difficult to detect even when they are present. This implies that part of the low detection rate of inflows may be driven by the orientation of the galaxy with respect to the observer rather than by an intrinsic lack of accretion as already suggested by lower-redshift studies.
    
        This interpretation is supported by cosmological simulations, which show that inflow rates depend strongly on the azimuthal angle and are stronger along the major axis, where gas is funneled inward toward galaxies \citep{Peroux2020}. In these simulations, galaxies are supplied by narrow, dense streams of cold gas that deliver material along the major axis, often closely aligned with the disk plane \citep{Fumagalli2011}. Cosmological simulations further show that accreting cool gas carries substantial angular momentum and often forms extended, warped, co-rotating structures that flow around the galactic disk \citep{Stewart2011}. This geometry produces stronger inflow velocities in edge-on systems, consistent with the inclination dependence observed in our sample.
    
        Observations also support this geometry of inflowing gas. At $z\leq1$, redshifted Fe\,{\sc ii} and Mg\,{\sc ii} absorption is more often detected in higher inclined galaxies. For example, \citet{Rubin2012} identified six galaxies with inflows and found that five out of the six were among the most edge-on galaxies in their sample. Similarly, \cite{Ho2017} examined Mg\,{\sc ii} absorption for galaxies at $0.15<z<0.3$ and found that the absorbing gas around highly inclined galaxies frequently shares the rotational direction of the galactic disk, suggesting that inflowing material may rotate with the galaxy before spiraling inward. A similar inclination dependence is observed with Na\,{\sc d} for local galaxies ($z\sim 0.1$). In these galaxies, the line center shifts toward redder wavelengths as inclination increases, with inflows only being detected at $i\gtrsim50^\circ$ \citep{2010Chen,2019Roberts}. These observations show that the inclination dependence observed in our sample is not unique, but instead reflects a broader trend across multiple tracers and redshift ranges.
        
        Both simulations and observations indicate that the redshifted absorption detected in our sample traces inflowing gas that is distributed near the plane of the galactic disk. The consistency of this inclination dependence across multiple tracers and redshifts suggests that accretion along the disk may be a common characteristic of cool gas inflows onto galaxies.

    \subsection{Physical Drivers of Mg\, {\sc ii} Emission}\label{sec:dis_mgii_em}
        We observe significant Mg\,{\sc ii} emission in 8\% of our sample and find that Mg\,{\sc ii} emitters have lower $M_*$, lower (more negative) $\beta$, and higher sSFRs and H$\alpha$ and [O\,{\sc iii}] emission EWs compared to non-emitters \citep[Figure \ref{fig:mgii_em_hist};][]{2025Hayes}. Our results also show that the Mg\, {\sc ii} absorption EW decreases with decreasing $M_*$, SFR, and $A_V$ (Figure \ref{fig:all_EW}). These trends indicate that Mg\,{\sc ii} emission is favored in low-mass, actively star-forming, dust-poor galaxies, whereas stronger absorption is associated with more massive and dustier galaxies. Our findings build directly on the results presented in our previous work \citep{Kehoe2025}, where we showed that Mg\,{\sc ii} emitters at $z\gtrsim 2$ preferentially exhibit lower $M_*$, bluer $\beta$, lower $A_V$ and higher sSFRs compared to non-emitters. Our current results additionally show that Mg\,{\sc ii} emitters exhibit higher H$\alpha$ and [O\,{\sc iii}] emission EWs, linking Mg\,{\sc ii} emission to intense recent star formation. These trends are also consistent with other studies, which find that Mg\,{\sc ii} emitters are often found among galaxies with $M_* \lesssim 10^{10}\,M_\odot$, low $A_V$, and high sSFR \citep[e.g.,][]{2009Weiner, 2010Rubin, 2012Erb, Kornei2012, 2013Guseva, 2019Guseva, 2015Zhu, 2017Finley, 2018Feltre}.

        One of the strongest trends for Mg\,{\sc ii} emission is the dependence on both $M_*$ and $A_V$. Observations at $1\lesssim z \lesssim 2$ show that Mg\,{\sc ii} emission strength decreases in more massive and dustier galaxies \citep{Kornei2012, 2012Erb, 2018Feltre}. This behavior is generally attributed to the increase in ISM column density and dust content in more massive systems, which alters the escape of resonantly scattered photons. In this regime -- as is the case for Ly$\alpha$ \citep[e.g.,][]{2003Shapley} -- Mg\,{\sc ii} photons undergo multiple scatterings, which increases their effective path length through the ISM, leading to a higher probability of dust absorption \citep{2011Prochaska, Scarlata2015, 2015Zhu, 2018Feltre, 2018Henry, 2022Katz, Xu2023, 2024Garel}. Consequently, Mg\, {\sc ii} is strongly attenuated in high $A_V$ objects, making Mg\,{\sc ii} emission more likely to be detected in low mass, dust-poor galaxies. This interpretation is also consistent with the observed $\beta$ trend, where bluer slopes in Mg\,{\sc ii} emitters indicate lower $A_V$ and more efficient escape of resonantly scattered photons.

        The connection between Mg\,{\sc ii} emission and sSFR further supports this picture. In our sample, Mg\,{\sc ii} emitters exhibit elevated sSFRs, indicating that strong emission is generally associated with galaxies undergoing intense, recent star formation. Earlier work at $z\sim1$ similarly finds that Mg\,{\sc ii} emission strength increases with sSFR \citep{Kornei2012, Rubin2014}. This suggests that Mg\,{\sc ii} emission arises from nebular photon production in actively star-forming regions, where lower $A_V$ further enhances the escape of the emitted photons. We also find that the Mg\,{\sc ii} absorption EW decreases with decreasing SFR, consistent with previous work at $1<z<2$, showing that stronger Mg\,{\sc ii} emission was found in galaxies with lower SFRs \citep{2009Weiner,2012Erb}. \citet{2012Erb} further note that the SFR dependence is weaker than the trends found with $M_*$ and $A_V$, suggesting that the observed SFR trend may be secondary and driven by underlying correlations with $M_*$ and $A_V$.

        The physical conditions inferred for the Mg\,{\sc ii} emitters also resemble those commonly associated with LyC-leaking galaxies \citep{2013Jaskot, 2016Izotov, 2017Finley, 2018Henry, Chisholm2020, 2022Katz, 2022Flury_a, 2022Flury_b}. These galaxies are typically low-mass, metal-poor galaxies with elevated specific star formation rates and extreme nebular emission properties. This similarity arises because Mg\,{\sc ii} and Ly$\alpha$ are governed by resonant scattering in the neutral ISM and CGM, making them sensitive to the same underlying gas distribution. Ly$\alpha$ escape has been linked to the LyC escape fraction ($f_{\rm esc}$) in both observations and theoretical studies \citep{2017Verhamme, 2020Gazagnes,2021Izotov, Pahl2021, 2022Flury_b,Choustikov2024, 2025LeReste}. However, at high redshift ($z\geq 6$), corresponding to the epoch of reionization, Ly$\alpha$ becomes increasingly attenuated by the neutral IGM, making it difficult to use as a tracer of $f_{\rm esc}$. 
        
       Mg\,{\sc ii} can be used as an alternative probe, as an indirect indicator of $f_{\rm esc}$, because its ionization potential is close to that of neutral hydrogen and it traces the neutral gas conditions that regulate $f_{\rm esc}$ \citep{2018Henry, Chisholm2020, Chang2024}. Observational studies at low redshift ($z<0.5$) have demonstrated that Mg\,{\sc ii} can be used as an indirect tracer of the LyC escape fraction \citep{Chisholm2020, 2022Izotov,Xu2022, Xu2023, 2024Leclercq}. Recently, this has been extended to the epoch of reionization, where Mg\,{\sc ii} was used to place constraints on $f_{\rm esc}$ at $z\sim7.5$ \citep{Gazagnes2025}. Comparisons with simulations \citep{2022Katz} suggested that these Mg\,{\sc ii}-based estimates overpredict $f_{\rm esc}$ and should only be used as upper limits, implying only weak LyC leakage for these high redshift galaxies \citep{Gazagnes2025}. Our results, showing a clear preference for Mg\,{\sc ii} emission in  low-mass, dust-poor, high sSFR galaxies with strong nebular emission, supports the use of Mg\,{\sc ii} as an indirect tracer of $f_{\rm esc}$, providing an alternative and more robust tracer for high redshift galaxies in the epoch of reionization.
    
\section{Conclusions}
    We have presented a joint analysis of NUV Fe\,{\sc ii} and Mg\,{\sc ii} and optical Na\,{\sc d} absorption kinematics using a combined \textit{JWST}/NIRSpec sample drawn from the LyC22, EXCELS, and AURORA surveys at $z_{\rm med}=3.07$. This work builds on the results of \citet{Kehoe2025}, which was based solely on the AURORA survey, by significantly expanding the sample size through the inclusion of additional NIRSpec programs. By combining these datasets, we have constructed a large and diverse sample of 321 star-forming galaxies with measurements of low-ionization absorption tracers, enabling a detailed study of outflow and inflow properties across a wide range of galaxy masses, star formation rates, and dust content. Our analysis focuses on how gas kinematics traced by Fe\,{\sc ii}, Mg\,{\sc ii}, and Na\,{\sc d} absorption depends on galaxy properties and viewing geometry. Our main results are as follows:
    \begin{enumerate}
        \item We find no statistically significant correlation between individual or composite NUV outflow velocity measurements and any galaxy property. However, galaxies with detected NUV outflows are preferentially located at higher $M_*$, SFR, and $\Sigma_{\rm SFR}$, suggesting that more massive and stronger star-forming galaxies are more likely to host detectable outflows.
        \item The absorption EWs of Fe\,{\sc ii} and Mg\,{\sc ii} absorption increases with $M_*$, SFR, nebular $E(B-V)$, amd stellar $A_V$, indicating that variations in NUV absorption strength are primarily driven by changes in column density and covering fraction of low-ionization gas rather than changes in outflow velocity.
        \item Na\,{\sc d} absorption shows no correlation between outflow velocity and any galaxy properties, but does exhibit higher absorption EWs and a higher incidence of outflow detections in more massive, dusty, and actively star-forming galaxies.
        \item The composite spectra reveal a strong dependence between outflow kinematics and galaxy inclination, with more face-on systems exhibiting larger absorption EWs and higher maximum outflow velocities in the NUV absorption lines. This result provides direct observational evidence that star-formation driven outflows at $z\sim 3$ are collimated, and are preferentially directed along the minor axis.
        \item The inflowing gas traced by redshifted absorption is more biased toward highly inclined (more edge-on) galaxies, suggesting that accretion primarily occurs in or near the galactic disk plane, rather than isotropically distributed throughout the halo. 
        \item Mg\,{\sc ii} emission is predominantly observed in galaxies with low $M_*$, low dust content, and high sSFR, consistent with efficient escape of resonantly scattered photons in objects with lower neutral gas content and dust attenuation. These conditions are similar to those associated with elevated ionizing $f_{\rm esc}$, suggesting that Mg\,{\sc ii} may trace conditions favorable for ionizing photon escape. Mg\,{\sc ii} emitters also exhibit higher H$\alpha$ and [O\,{\sc iii}] emission EWs, indicating that Mg\,{\sc ii} emission may be linked to young, high sSFR objects with hard ionizing conditions.
    \end{enumerate}

    Our results demonstrate that while scaling relations between outflow velocity and galaxy properties are weak or absent in our sample, both the detectability and strength of both inflows and outflows are strongly governed by galaxy orientation. We find the first observational evidence at $z\sim 3$ for a structured, directionally dependent gas flow geometry, with bipolar outflows emerging perpendicular to the disk and inflowing gas co-rotating along the galactic plane. This geometric interpretation explains the dependence of absorption strength and detection rates on viewing angle and highlights the importance of orientation in shaping low-ionization gas kinematics. We find that these trends are consistent with simulations \citep[e.g.][]{Fumagalli2011, 2019Nelson, Peroux2020} and low-redshift ($z\lesssim1$) studies \citep[e.g.][]{2010Chen, 2012Kacprzak, Ho2017}, which find that outflows are bipolar and perpendicular to the galactic disk, while inflows are co-rotating along the major axis.

    The lack of strong correlations between outflow velocity and galaxy properties likely reflects the limited dynamic range of our sample in M$_*$ and SFR, as well as the large fraction of low S/N spectra, which may hide weaker absorption features. Future {\it JWST} surveys with larger and more diverse samples, particularly extending to lower-mass and lower-SFR regimes and improved S/N for individual spectra will be essential to search for correlations more robustly. Extending such analyses to higher redshift, where both rest-frame FUV and NUV tracers can be covered within the same spectra \citep{Kehoe2025}, will also enable more direct comparisons of gas flow kinematics across multiple transitions of the same gas phase.

\begin{acknowledgments}
This work is based on observations made with the NASA/ESA/CSA James Webb Space Telescope. The data were
obtained from the Mikulski Archive for Space Telescopes at
the Space Telescope Science Institute, which is operated by the
Association of Universities for Research in Astronomy, Inc.,
under NASA contract NAS5-03127 for JWST.  The specific observations analyzed can be accessed via \dataset[DOI: 10.17909/6mza-5q55]{https://archive.stsci.edu/doi/resolve/resolve.html?doi=10.17909/6mza-5q55}, \dataset[DOI: 10.17909/4r31-j678]{https://archive.stsci.edu/doi/resolve/resolve.html?doi=10.17909/4r31-j678}, and
\dataset[DOI: 10.17909/sqde-1112]{https://archive.stsci.edu/doi/resolve/resolve.html?doi=10.17909/sqde-1112}. 
We also acknowledge support from NASA grants JWST-GO-01914, GO-01869, and GO-03543. ACC, LT, and HL acknowledge support from a UKRI Frontier Research Grantee Grant (PI Carnall; grant reference EP/Y037065/1). DJM acknowledge the support of the Royal Society through the award of a Royal Society University Research Professorship to Prof. James Dunlop. AKI acknowledge support from JSPS KAKENHI 26H02069, 25K00020, 24H00002, and 23H00131. YI and NGG acknowledge support from the National Academy of Sciences of Ukraine by its project no. 0126U000353. SDS acknowledges support from a UKRI Frontier Research Guarantee Grant (PI Carnall; grant reference EP/Y037065/1). ASL acknowledges support from Knut and Alice Wallenberg Foundation. MS acknowledges support for Program number JWST-GO-03543.014 that was provided through a grant from the STScI under NASA contract NAS5-03127.

\end{acknowledgments}

%


\bibliography{ms}{}

@ARTICLE{Kehoe2024,
       author = {{Kehoe}, Emily and {Shapley}, Alice E. and {Schreiber}, N.~M. F{\"o}rster and {Pahl}, Anthony J. and {Topping}, Michael W. and {Reddy}, Naveen A. and {Genzel}, Reinhard and {Price}, Sedona H. and {Tacconi}, L.~J.},
        title = "{The First Combined H{\ensuremath{\alpha}} and Rest-UV Spectroscopic Probe of Galactic Outflows at High Redshift}",
      journal = {\apj},
         year = 2024,
        month = nov,
       volume = {976},
       number = {1},
          eid = {28},
        pages = {28},
          doi = {10.3847/1538-4357/ad7ebb},
archivePrefix = {arXiv},
       eprint = {2406.07621},
 primaryClass = {astro-ph.GA},
       adsurl = {https://ui.adsabs.harvard.edu/abs/2024ApJ...976...28K}
}

@ARTICLE{Martin2012,
       author = {{Martin}, Crystal L. and {Shapley}, Alice E. and {Coil}, Alison L. and {Kornei}, Katherine A. and {Bundy}, Kevin and {Weiner}, Benjamin J. and {Noeske}, Kai G. and {Schiminovich}, David},
        title = "{Demographics and Physical Properties of Gas Outflows/Inflows at 0.4 < z < 1.4}",
      journal = {\apj},
         year = 2012,
        month = dec,
       volume = {760},
       number = {2},
          eid = {127},
        pages = {127},
          doi = {10.1088/0004-637X/760/2/127},
archivePrefix = {arXiv},
       eprint = {1206.5552},
 primaryClass = {astro-ph.CO},
       adsurl = {https://ui.adsabs.harvard.edu/abs/2012ApJ...760..127M}
}

@ARTICLE{Chisholm2020,
       author = {{Chisholm}, J. and {Prochaska}, J.~X. and {Schaerer}, D. and {Gazagnes}, S. and {Henry}, A.},
        title = "{Optically thin spatially resolved Mg II emission maps the escape of ionizing photons}",
      journal = {\mnras},
         year = 2020,
        month = oct,
       volume = {498},
       number = {2},
        pages = {2554-2574},
          doi = {10.1093/mnras/staa2470},
archivePrefix = {arXiv},
       eprint = {2008.06059},
 primaryClass = {astro-ph.GA},
       adsurl = {https://ui.adsabs.harvard.edu/abs/2020MNRAS.498.2554C}
}

@ARTICLE{Kornei2013,
       author = {{Kornei}, Katherine A. and {Shapley}, Alice E. and {Martin}, Crystal L. and {Coil}, Alison L. and {Lotz}, Jennifer M. and {Weiner}, Benjamin J.},
        title = "{Fine-structure Fe II* Emission and Resonant Mg II Emission in z \raisebox{-0.5ex}\textasciitilde 1 Star-forming Galaxies}",
      journal = {\apj},
         year = 2013,
        month = sep,
       volume = {774},
       number = {1},
          eid = {50},
        pages = {50},
          doi = {10.1088/0004-637X/774/1/50},
archivePrefix = {arXiv},
       eprint = {1302.6997},
 primaryClass = {astro-ph.CO},
       adsurl = {https://ui.adsabs.harvard.edu/abs/2013ApJ...774...50K}
}

@ARTICLE{Martin2013,
       author = {{Martin}, Crystal L. and {Shapley}, Alice E. and {Coil}, Alison L. and {Kornei}, Katherine A. and {Murray}, Norman and {Pancoast}, Anna},
        title = "{Scattered Emission from z \raisebox{-0.5ex}\textasciitilde 1 Galactic Outflows}",
      journal = {\apj},
         year = 2013,
        month = jun,
       volume = {770},
       number = {1},
          eid = {41},
        pages = {41},
          doi = {10.1088/0004-637X/770/1/41},
archivePrefix = {arXiv},
       eprint = {1304.6405},
 primaryClass = {astro-ph.CO},
       adsurl = {https://ui.adsabs.harvard.edu/abs/2013ApJ...770...41M}
}

@ARTICLE{Kornei2012,
       author = {{Kornei}, Katherine A. and {Shapley}, Alice E. and {Martin}, Crystal L. and {Coil}, Alison L. and {Lotz}, Jennifer M. and {Schiminovich}, David and {Bundy}, Kevin and {Noeske}, Kai G.},
        title = "{The Properties and Prevalence of Galactic Outflows at z \raisebox{-0.5ex}\textasciitilde 1 in the Extended Groth Strip}",
      journal = {\apj},
         year = 2012,
        month = oct,
       volume = {758},
       number = {2},
          eid = {135},
        pages = {135},
          doi = {10.1088/0004-637X/758/2/135},
archivePrefix = {arXiv},
       eprint = {1205.0812},
 primaryClass = {astro-ph.CO},
       adsurl = {https://ui.adsabs.harvard.edu/abs/2012ApJ...758..135K}
}

@article{2009Weiner,
doi = {10.1088/0004-637X/692/1/187},
url = {https://dx.doi.org/10.1088/0004-637X/692/1/187},
year = {2009},
month = {feb},
publisher = {The American Astronomical Society},
volume = {692},
number = {1},
pages = {187},
author = {Weiner, Benjamin J. and Coil, Alison L. and Prochaska, Jason X. and Newman, Jeffrey A. and Cooper, Michael C. and Bundy, Kevin and Conselice, Christopher J. and Dutton, Aaron A. and Faber, S. M. and Koo, David C. and Lotz, Jennifer M. and Rieke, G. H. and Rubin, K. H. R.},
title = {UBIQUITOUS OUTFLOWS IN DEEP2 SPECTRA OF STAR-FORMING GALAXIES AT z = 1.4},
journal = {The Astrophysical Journal}
}

@ARTICLE{2003Shapley,
       author = {{Shapley}, Alice E. and {Steidel}, Charles C. and {Pettini}, Max and {Adelberger}, Kurt L.},
        title = "{Rest-Frame Ultraviolet Spectra of z\raisebox{-0.5ex}\textasciitilde3 Lyman Break Galaxies}",
      journal = {\apj},
         year = 2003,
        month = may,
       volume = {588},
       number = {1},
        pages = {65-89},
          doi = {10.1086/373922},
archivePrefix = {arXiv},
       eprint = {astro-ph/0301230},
 primaryClass = {astro-ph},
       adsurl = {https://ui.adsabs.harvard.edu/abs/2003ApJ...588...65S}
}

@ARTICLE{2010Steidel,
       author = {{Steidel}, Charles C. and {Erb}, Dawn K. and {Shapley}, Alice E. and {Pettini}, Max and {Reddy}, Naveen and {Bogosavljevi{\'c}}, Milan and {Rudie}, Gwen C. and {Rakic}, Olivera},
        title = "{The Structure and Kinematics of the Circumgalactic Medium from Far-ultraviolet Spectra of z \raisebox{-0.5ex}\textasciitilde= 2-3 Galaxies}",
      journal = {\apj},
         year = 2010,
        month = jul,
       volume = {717},
       number = {1},
        pages = {289-322},
          doi = {10.1088/0004-637X/717/1/289},
archivePrefix = {arXiv},
       eprint = {1003.0679},
 primaryClass = {astro-ph.CO},
       adsurl = {https://ui.adsabs.harvard.edu/abs/2010ApJ...717..289S}
}

@ARTICLE{2022Weldon,
       author = {{Weldon}, Andrew and {Reddy}, Naveen A. and {Topping}, Michael W. and {Shapley}, Alice E. and {Sanders}, Ryan L. and {Du}, Xinnan and {Price}, Sedona H. and {Kriek}, Mariska and {Coil}, Alison L. and {Siana}, Brian and {Mobasher}, Bahram and {Fetherolf}, Tara and {Shivaei}, Irene and {Rezaee}, Saeed},
        title = "{The MOSDEF-LRIS survey: connection between galactic-scale outflows and the properties of z   2 star-forming galaxies}",
      journal = {\mnras},
         year = 2022,
        month = sep,
       volume = {515},
       number = {1},
        pages = {841-856},
          doi = {10.1093/mnras/stac1822},
archivePrefix = {arXiv},
       eprint = {2203.09539},
 primaryClass = {astro-ph.GA},
       adsurl = {https://ui.adsabs.harvard.edu/abs/2022MNRAS.515..841W}
}

@ARTICLE{2022Calabro,
       author = {{Calabr{\`o}}, A. and {Pentericci}, L. and {Talia}, M. and {Cresci}, G. and {Castellano}, M. and {Belfiori}, D. and {Mascia}, S. and {Zamorani}, G. and {Amor{\'\i}n}, R. and {Fynbo}, J.~P.~U. and {Ginolfi}, M. and {Guaita}, L. and {Hathi}, N.~P. and {Koekemoer}, A. and {Llerena}, M. and {Mannucci}, F. and {Santini}, P. and {Saxena}, A. and {Schaerer}, D.},
        title = "{Properties of the interstellar medium in star-forming galaxies at redshifts 2 {\ensuremath{\leq}} z {\ensuremath{\leq}} 5 from the VANDELS survey}",
      journal = {\aap},
         year = 2022,
        month = nov,
       volume = {667},
          eid = {A117},
        pages = {A117},
          doi = {10.1051/0004-6361/202244364},
archivePrefix = {arXiv},
       eprint = {2206.14918},
 primaryClass = {astro-ph.GA},
       adsurl = {https://ui.adsabs.harvard.edu/abs/2022A&A...667A.117C}
}

@ARTICLE{2012Talia,
       author = {{Talia}, M. and {Mignoli}, M. and {Cimatti}, A. and {Kurk}, J. and {Berta}, S. and {Bolzonella}, M. and {Cassata}, P. and {Daddi}, E. and {Dickinson}, M. and {Franceschini}, A. and {Halliday}, C. and {Pozzetti}, L. and {Renzini}, A. and {Rodighiero}, G. and {Rosati}, P. and {Zamorani}, G.},
        title = "{GMASS ultradeep spectroscopy of galaxies at z \raisebox{-0.5ex}\textasciitilde 2. VI. Star formation, extinction, and gas outflows from UV spectra}",
      journal = {\aap},
         year = 2012,
        month = mar,
       volume = {539},
          eid = {A61},
        pages = {A61},
          doi = {10.1051/0004-6361/201117683},
archivePrefix = {arXiv},
       eprint = {1111.4402},
 primaryClass = {astro-ph.CO},
       adsurl = {https://ui.adsabs.harvard.edu/abs/2012A&A...539A..61T}
}

@article{2012Erb,
doi = {10.1088/0004-637X/759/1/26},
url = {https://dx.doi.org/10.1088/0004-637X/759/1/26},
year = {2012},
month = {oct},
publisher = {The American Astronomical Society},
volume = {759},
number = {1},
pages = {26},
author = {Erb, Dawn K. and Quider, Anna M. and Henry, Alaina L. and Martin, Crystal L.},
title = {GALACTIC OUTFLOWS IN ABSORPTION AND EMISSION: NEAR-ULTRAVIOLET SPECTROSCOPY OF GALAXIES AT 1 &lt; z &lt; 2*},
journal = {The Astrophysical Journal}
}

@article{2010Rubin,
doi = {10.1088/0004-637X/719/2/1503},
url = {https://dx.doi.org/10.1088/0004-637X/719/2/1503},
year = {2010},
month = {jul},
publisher = {The American Astronomical Society},
volume = {719},
number = {2},
pages = {1503},
author = {Rubin, Kate H. R. and Weiner, Benjamin J. and Koo, David C. and Martin, Crystal L. and Prochaska, J. Xavier and Coil, Alison L. and Newman, Jeffrey A.},
title = {THE PERSISTENCE OF COOL GALACTIC WINDS IN HIGH STELLAR MASS GALAXIES BETWEEN z ∼ 1.4 AND ∼1*},
journal = {The Astrophysical Journal}
}

@ARTICLE{2018Feltre,
       author = {{Feltre}, Anna and {Bacon}, Roland and {Tresse}, Laurence and {Finley}, Hayley and {Carton}, David and {Blaizot}, J{\'e}r{\'e}my and {Bouch{\'e}}, Nicolas and {Garel}, Thibault and {Inami}, Hanae and {Boogaard}, Leindert A. and {Brinchmann}, Jarle and {Charlot}, St{\'e}phane and {Chevallard}, Jacopo and {Contini}, Thierry and {Michel-Dansac}, Leo and {Mahler}, Guillaume and {Marino}, Raffaella A. and {Maseda}, Michael V. and {Richard}, Johan and {Schmidt}, Kasper B. and {Verhamme}, Anne},
        title = "{The MUSE Hubble Ultra Deep Field Survey. XII. Mg II emission and absorption in star-forming galaxies}",
      journal = {\aap},
         year = 2018,
        month = sep,
       volume = {617},
          eid = {A62},
        pages = {A62},
          doi = {10.1051/0004-6361/201833281},
archivePrefix = {arXiv},
       eprint = {1806.01864},
 primaryClass = {astro-ph.GA},
       adsurl = {https://ui.adsabs.harvard.edu/abs/2018A&A...617A..62F}
}

@ARTICLE{1989Cardelli,
       author = {{Cardelli}, Jason A. and {Clayton}, Geoffrey C. and {Mathis}, John S.},
        title = "{The Relationship between Infrared, Optical, and Ultraviolet Extinction}",
      journal = {\apj},
         year = 1989,
        month = oct,
       volume = {345},
        pages = {245},
          doi = {10.1086/167900},
       adsurl = {https://ui.adsabs.harvard.edu/abs/1989ApJ...345..245C}
}

@ARTICLE{2024Clarke,
       author = {{Clarke}, Leonardo and {Shapley}, Alice E. and {Sanders}, Ryan L. and {Topping}, Michael W. and {Brammer}, Gabriel B. and {Bento}, Trinity and {Reddy}, Naveen A. and {Kehoe}, Emily},
        title = "{The Star-forming Main Sequence in JADES and CEERS at z > 1.4: Investigating the Burstiness of Star Formation}",
      journal = {\apj},
         year = 2024,
        month = dec,
       volume = {977},
       number = {1},
          eid = {133},
        pages = {133},
          doi = {10.3847/1538-4357/ad8ba4},
archivePrefix = {arXiv},
       eprint = {2406.05178},
 primaryClass = {astro-ph.GA},
       adsurl = {https://ui.adsabs.harvard.edu/abs/2024ApJ...977..133C}
}

@ARTICLE{2018Stanway,
    author = {Stanway, E R and Eldridge, J J},
    title = {Re-evaluating old stellar populations},
    journal = {Monthly Notices of the Royal Astronomical Society},
    volume = {479},
    number = {1},
    pages = {75-93},
    year = {2018},
    month = {05},
    issn = {0035-8711},
    doi = {10.1093/mnras/sty1353},
    url = {https://doi.org/10.1093/mnras/sty1353},
    eprint = {https://academic.oup.com/mnras/article-pdf/479/1/75/25111406/sty1353.pdf},
}

@ARTICLE{2022Reddy,
       author = {{Reddy}, Naveen A. and {Topping}, Michael W. and {Shapley}, Alice E. and {Steidel}, Charles C. and {Sanders}, Ryan L. and {Du}, Xinnan and {Coil}, Alison L. and {Mobasher}, Bahram and {Price}, Sedona H. and {Shivaei}, Irene},
        title = "{The Effects of Stellar Population and Gas Covering Fraction on the Emergent Ly{\ensuremath{\alpha}} Emission of High-redshift Galaxies}",
      journal = {\apj},
         year = 2022,
        month = feb,
       volume = {926},
       number = {1},
          eid = {31},
        pages = {31},
          doi = {10.3847/1538-4357/ac3b4c},
archivePrefix = {arXiv},
       eprint = {2108.05363},
 primaryClass = {astro-ph.GA},
       adsurl = {https://ui.adsabs.harvard.edu/abs/2022ApJ...926...31R}
}

@ARTICLE{2009Conroy,
       author = {{Conroy}, Charlie and {Gunn}, James E. and {White}, Martin},
        title = "{The Propagation of Uncertainties in Stellar Population Synthesis Modeling. I. The Relevance of Uncertain Aspects of Stellar Evolution and the Initial Mass Function to the Derived Physical Properties of Galaxies}",
      journal = {\apj},
         year = 2009,
        month = jul,
       volume = {699},
       number = {1},
        pages = {486-506},
          doi = {10.1088/0004-637X/699/1/486},
archivePrefix = {arXiv},
       eprint = {0809.4261},
 primaryClass = {astro-ph},
       adsurl = {https://ui.adsabs.harvard.edu/abs/2009ApJ...699..486C}
}

@ARTICLE{2003Chabrier,
       author = {{Chabrier}, Gilles},
        title = "{Galactic Stellar and Substellar Initial Mass Function}",
      journal = {\pasp},
         year = 2003,
        month = jul,
       volume = {115},
       number = {809},
        pages = {763-795},
          doi = {10.1086/376392},
archivePrefix = {arXiv},
       eprint = {astro-ph/0304382},
 primaryClass = {astro-ph},
       adsurl = {https://ui.adsabs.harvard.edu/abs/2003PASP..115..763C}
}

@ARTICLE{2000Calzetti,
       author = {{Calzetti}, Daniela and {Armus}, Lee and {Bohlin}, Ralph C. and {Kinney}, Anne L. and {Koornneef}, Jan and {Storchi-Bergmann}, Thaisa},
        title = "{The Dust Content and Opacity of Actively Star-forming Galaxies}",
      journal = {\apj},
         year = 2000,
        month = apr,
       volume = {533},
       number = {2},
        pages = {682-695},
          doi = {10.1086/308692},
archivePrefix = {arXiv},
       eprint = {astro-ph/9911459},
 primaryClass = {astro-ph},
       adsurl = {https://ui.adsabs.harvard.edu/abs/2000ApJ...533..682C}
}

@ARTICLE{2003Gordon,
doi = {10.1086/376774},
url = {https://dx.doi.org/10.1086/376774},
year = {2003},
month = {sep},
publisher = {},
volume = {594},
number = {1},
pages = {279},
author = {Gordon, Karl D. and Clayton, Geoffrey C. and Misselt, K. A. and Landolt, Arlo U. and Wolff, Michael J.},
title = {A Quantitative Comparison of the Small Magellanic Cloud, Large Magellanic Cloud, and Milky Way Ultraviolet to Near-Infrared Extinction Curves*},
journal = {The Astrophysical Journal}
}

@ARTICLE{2013Bradshaw,
       author = {{Bradshaw}, E.~J. and {Almaini}, O. and {Hartley}, W.~G. and {Smith}, K.~T. and {Conselice}, C.~J. and {Dunlop}, J.~S. and {Simpson}, C. and {Chuter}, R.~W. and {Cirasuolo}, M. and {Foucaud}, S. and {McLure}, R.~J. and {Mortlock}, A. and {Pearce}, H.},
        title = "{High-velocity outflows from young star-forming galaxies in the UKIDSS Ultra-Deep Survey}",
      journal = {\mnras},
         year = 2013,
        month = jul,
       volume = {433},
       number = {1},
        pages = {194-208},
          doi = {10.1093/mnras/stt715},
archivePrefix = {arXiv},
       eprint = {1304.7276},
 primaryClass = {astro-ph.CO},
       adsurl = {https://ui.adsabs.harvard.edu/abs/2013MNRAS.433..194B}
}

@ARTICLE{2024Carniani,
       author = {{Carniani}, Stefano and {Venturi}, Giacomo and {Parlanti}, Eleonora and {de Graaff}, Anna and {Maiolino}, Roberto and {Arribas}, Santiago and {Bonaventura}, Nina and {Boyett}, Kristan and {Bunker}, Andrew J. and {Cameron}, Alex J. and {Charlot}, Stephane and {Chevallard}, Jacopo and {Curti}, Mirko and {Curtis-Lake}, Emma and {Eisenstein}, Daniel J. and {Giardino}, Giovanna and {Hausen}, Ryan and {Kumari}, Nimisha and {Maseda}, Michael V. and {Nelson}, Erica and {Perna}, Michele and {Rix}, Hans-Walter and {Robertson}, Brant and {Del Pino}, Bruno Rodr{\'\i}guez and {Sandles}, Lester and {Scholtz}, Jan and {Simmonds}, Charlotte and {Smit}, Renske and {Tacchella}, Sandro and {{\"U}bler}, Hannah and {Williams}, Christina C. and {Willott}, Chris and {Witstok}, Joris},
        title = "{JADES: The incidence rate and properties of galactic outflows in low-mass galaxies across 3 < z < 9}",
      journal = {\aap},
         year = 2024,
        month = may,
       volume = {685},
          eid = {A99},
        pages = {A99},
          doi = {10.1051/0004-6361/202347230},
archivePrefix = {arXiv},
       eprint = {2306.11801},
 primaryClass = {astro-ph.GA},
       adsurl = {https://ui.adsabs.harvard.edu/abs/2024A&A...685A..99C}
}

@article{2005Rupke,
doi = {10.1086/432886},
url = {https://doi.org/10.1086/432886},
year = {2005},
month = {sep},
publisher = {},
volume = {160},
number = {1},
pages = {87},
author = {Rupke, David S. and Veilleux, Sylvain and Sanders, D. B.},
title = {Outflows in Infrared-Luminous Starbursts at z &lt; 0.5. I. Sample, Na I D Spectra, and Profile Fitting* ** ***},
journal = {The Astrophysical Journal Supplement Series}
}

@ARTICLE{2010Chen,
       author = {{Chen}, Yan-Mei and {Tremonti}, Christy A. and {Heckman}, Timothy M. and {Kauffmann}, Guinevere and {Weiner}, Benjamin J. and {Brinchmann}, Jarle and {Wang}, Jing},
        title = "{Absorption-line Probes of the Prevalence and Properties of Outflows in Present-day Star-forming Galaxies}",
      journal = {\aj},
         year = 2010,
        month = aug,
       volume = {140},
       number = {2},
        pages = {445-461},
          doi = {10.1088/0004-6256/140/2/445},
archivePrefix = {arXiv},
       eprint = {1003.5425},
 primaryClass = {astro-ph.GA},
       adsurl = {https://ui.adsabs.harvard.edu/abs/2010AJ....140..445C}
}

@ARTICLE{2021Prusinski,
       author = {{Prusinski}, Nikolaus Z. and {Erb}, Dawn K. and {Martin}, Crystal L.},
        title = "{Connecting Galactic Outflows and Star Formation: Inferences from H{\ensuremath{\alpha}} Maps and Absorption-line Spectroscopy at 1 {\ensuremath{\lesssim}} z {\ensuremath{\lesssim}} 1.5}",
      journal = {\aj},
         year = 2021,
        month = may,
       volume = {161},
       number = {5},
          eid = {212},
        pages = {212},
          doi = {10.3847/1538-3881/abe85b},
archivePrefix = {arXiv},
       eprint = {2102.10187},
 primaryClass = {astro-ph.GA},
       adsurl = {https://ui.adsabs.harvard.edu/abs/2021AJ....161..212P}
}

@article{2000Heckman,
doi = {10.1086/313421},
url = {https://dx.doi.org/10.1086/313421},
year = {2000},
month = {aug},
publisher = {},
volume = {129},
number = {2},
pages = {493},
author = {Timothy M. Heckman and Matthew D. Lehnert and David K. Strickland and Lee Armus},
title = {Absorption-Line Probes of Gas and Dust in Galactic
Superwinds},
journal = {The Astrophysical Journal Supplement Series}
}

@ARTICLE{2014Bordoloi,
       author = {{Bordoloi}, R. and {Lilly}, S.~J. and {Hardmeier}, E. and {Contini}, T. and {Kneib}, J. -P. and {Le Fevre}, O. and {Mainieri}, V. and {Renzini}, A. and {Scodeggio}, M. and {Zamorani}, G. and {Bardelli}, S. and {Bolzonella}, M. and {Bongiorno}, A. and {Caputi}, K. and {Carollo}, C.~M. and {Cucciati}, O. and {de la Torre}, S. and {de Ravel}, L. and {Garilli}, B. and {Iovino}, A. and {Kampczyk}, P. and {Kova{\v{c}}}, K. and {Knobel}, C. and {Lamareille}, F. and {Le Borgne}, J. -F. and {Le Brun}, V. and {Maier}, C. and {Mignoli}, M. and {Oesch}, P. and {Pello}, R. and {Peng}, Y. and {Perez Montero}, E. and {Presotto}, V. and {Silverman}, J. and {Tanaka}, M. and {Tasca}, L. and {Tresse}, L. and {Vergani}, D. and {Zucca}, E. and {Cappi}, A. and {Cimatti}, A. and {Coppa}, G. and {Franzetti}, P. and {Koekemoer}, A. and {Moresco}, M. and {Nair}, P. and {Pozzetti}, L.},
        title = "{The Dependence of Galactic Outflows on the Properties and Orientation of zCOSMOS Galaxies at z \raisebox{-0.5ex}\textasciitilde 1}",
      journal = {\apj},
         year = 2014,
        month = oct,
       volume = {794},
       number = {2},
          eid = {130},
        pages = {130},
          doi = {10.1088/0004-637X/794/2/130},
archivePrefix = {arXiv},
       eprint = {1307.6553},
 primaryClass = {astro-ph.CO},
       adsurl = {https://ui.adsabs.harvard.edu/abs/2014ApJ...794..130B}
}

@ARTICLE{2017Finley,
       author = {{Finley}, Hayley and {Bouch{\'e}}, Nicolas and {Contini}, Thierry and {Paalvast}, Mieke and {Boogaard}, Leindert and {Maseda}, Michael and {Bacon}, Roland and {Blaizot}, J{\'e}r{\'e}my and {Brinchmann}, Jarle and {Epinat}, Beno{\^\i}t and {Feltre}, Anna and {Marino}, Raffaella Anna and {Muzahid}, Sowgat and {Richard}, Johan and {Schaye}, Joop and {Verhamme}, Anne and {Weilbacher}, Peter M. and {Wisotzki}, Lutz},
        title = "{The MUSE Hubble Ultra Deep Field Survey. VII. Fe II* emission in star-forming galaxies}",
      journal = {\aap},
         year = 2017,
        month = dec,
       volume = {608},
          eid = {A7},
        pages = {A7},
          doi = {10.1051/0004-6361/201731499},
archivePrefix = {arXiv},
       eprint = {1710.09195},
 primaryClass = {astro-ph.GA},
       adsurl = {https://ui.adsabs.harvard.edu/abs/2017A&A...608A...7F}
}

@ARTICLE{2018Henry,
       author = {{Henry}, Alaina and {Berg}, Danielle A. and {Scarlata}, Claudia and {Verhamme}, Anne and {Erb}, Dawn},
        title = "{A Close Relationship between Ly{\ensuremath{\alpha}} and Mg II in Green Pea Galaxies}",
      journal = {\apj},
         year = 2018,
        month = mar,
       volume = {855},
       number = {2},
          eid = {96},
        pages = {96},
          doi = {10.3847/1538-4357/aab099},
archivePrefix = {arXiv},
       eprint = {1803.10243},
 primaryClass = {astro-ph.GA},
       adsurl = {https://ui.adsabs.harvard.edu/abs/2018ApJ...855...96H}
}

@ARTICLE{2011Prochaska,
doi = {10.1088/0004-637X/734/1/24},
url = {https://dx.doi.org/10.1088/0004-637X/734/1/24},
year = {2011},
month = {may},
publisher = {The American Astronomical Society},
volume = {734},
number = {1},
pages = {24},
author = {Prochaska, J. Xavier and Kasen, Daniel and Rubin, Kate},
title = {SIMPLE MODELS OF METAL-LINE ABSORPTION AND EMISSION FROM COOL GAS OUTFLOWS},
journal = {The Astrophysical Journal}
}

@ARTICLE{2005Martin,
       author = {{Martin}, Crystal L.},
        title = "{Mapping Large-Scale Gaseous Outflows in Ultraluminous Galaxies with Keck II ESI Spectra: Variations in Outflow Velocity with Galactic Mass}",
      journal = {\apj},
         year = 2005,
        month = mar,
       volume = {621},
       number = {1},
        pages = {227-245},
          doi = {10.1086/427277},
archivePrefix = {arXiv},
       eprint = {astro-ph/0410247},
 primaryClass = {astro-ph},
       adsurl = {https://ui.adsabs.harvard.edu/abs/2005ApJ...621..227M}
}

@ARTICLE{2014Rubin,
       author = {{Rubin}, Kate H.~R. and {Prochaska}, J. Xavier and {Koo}, David C. and {Phillips}, Andrew C. and {Martin}, Crystal L. and {Winstrom}, Lucas O.},
        title = "{Evidence for Ubiquitous Collimated Galactic-scale Outflows along the Star-forming Sequence at z \raisebox{-0.5ex}\textasciitilde 0.5}",
      journal = {\apj},
         year = 2014,
        month = oct,
       volume = {794},
       number = {2},
          eid = {156},
        pages = {156},
          doi = {10.1088/0004-637X/794/2/156},
archivePrefix = {arXiv},
       eprint = {1307.1476},
 primaryClass = {astro-ph.CO},
       adsurl = {https://ui.adsabs.harvard.edu/abs/2014ApJ...794..156R}
}

@ARTICLE{2019Schreiber,
doi = {10.3847/1538-4357/ab0ca2},
url = {https://dx.doi.org/10.3847/1538-4357/ab0ca2},
year = {2019},
month = {apr},
publisher = {The American Astronomical Society},
volume = {875},
number = {1},
pages = {21},
author = {N. M. {Förster Schreiber} and H. Übler and R. L. Davies and R. Genzel and E. Wisnioski and S. Belli and T. Shimizu and D. Lutz and M. Fossati and R. Herrera-Camus and J. T. Mendel and L. J. Tacconi and D. Wilman and A. Beifiori and G. B. Brammer and A. Burkert and C. M. Carollo and R. I. Davies and F. Eisenhauer and M. Fabricius and S. J. Lilly and I. Momcheva and T. Naab and E. J. Nelson and S. H. Price and A. Renzini and R. Saglia and A. Sternberg and P. van Dokkum and S. Wuyts},
title = {The KMOS3D Survey: Demographics and Properties of Galactic Outflows at z=0.6–2.7*},
journal = {The Astrophysical Journal},
}

@article{2011Murray,
doi = {10.1088/0004-637X/735/1/66},
url = {https://dx.doi.org/10.1088/0004-637X/735/1/66},
year = {2011},
month = {jun},
publisher = {The American Astronomical Society},
volume = {735},
number = {1},
pages = {66},
author = {Murray, Norman and Ménard, Brice and Thompson, Todd A.},
title = {RADIATION PRESSURE FROM MASSIVE STAR CLUSTERS AS A LAUNCHING MECHANISM FOR SUPER-GALACTIC WINDS},
journal = {The Astrophysical Journal}
}

@ARTICLE{Topping2025,
       author = {{Topping}, Michael W. and {Sanders}, Ryan L. and {Shapley}, Alice E. and {Pahl}, Anthony J. and {Reddy}, Naveen A. and {Stark}, Daniel P. and {Berg}, Danielle A. and {Clarke}, Leonardo and {Cullen}, Fergus and {Dunlop}, James S. and {Ellis}, Richard S. and {F{\"o}rster Schreiber}, N.~M. and {Illingworth}, Garth D. and {Jones}, Tucker and {Narayanan}, Desika and {Pettini}, Max and {Schaerer}, Daniel},
     title = {The AURORA survey: the evolution of multiphase electron densities at high redshift},
    journal = {Monthly Notices of the Royal Astronomical Society},
    volume = {541},
    number = {2},
    pages = {1707-1721},
    year = {2025},
    month = {07},
    issn = {0035-8711},
    doi = {10.1093/mnras/staf903},
    url = {https://doi.org/10.1093/mnras/staf903},
    eprint = {https://academic.oup.com/mnras/article-pdf/541/2/1707/63734437/staf903.pdf},
}

@ARTICLE{2022Katz,
    author = {Katz, Harley and Garel, Thibault and Rosdahl, Joakim and Mauerhofer, Valentin and Kimm, Taysun and Blaizot, Jérémy and Michel-Dansac, Léo and Devriendt, Julien and Slyz, Adrianne and Haehnelt, Martin},
    title = {Mg ii in the JWST era: a probe of Lyman continuum escape?},
    journal = {Monthly Notices of the Royal Astronomical Society},
    volume = {515},
    number = {3},
    pages = {4265-4286},
    year = {2022},
    month = {05},
    issn = {0035-8711},
    doi = {10.1093/mnras/stac1437},
    url = {https://doi.org/10.1093/mnras/stac1437},
    eprint = {https://academic.oup.com/mnras/article-pdf/515/3/4265/47846595/stac1437.pdf},
}

@article{2005Veilleux,
   author = "Veilleux, Sylvain and Cecil, Gerald and Bland-Hawthorn, Joss",
   title = "Galactic Winds", 
   journal= "Annual Review of Astronomy and Astrophysics",
   year = "2005",
   volume = "43",
   number = "Volume 43, 2005",
   pages = "769-826",
   doi = "https://doi.org/10.1146/annurev.astro.43.072103.150610",
   url = "https://www.annualreviews.org/content/journals/10.1146/annurev.astro.43.072103.150610",
   publisher = "Annual Reviews",
   issn = "1545-4282",
   type = "Journal Article",
  }

@ARTICLE{1998Silk,
       author = {{Silk}, Joseph and {Rees}, Martin J.},
        title = "{Quasars and galaxy formation}",
      journal = {\aap},
         year = 1998,
        month = mar,
       volume = {331},
        pages = {L1-L4},
          doi = {10.48550/arXiv.astro-ph/9801013},
archivePrefix = {arXiv},
       eprint = {astro-ph/9801013},
 primaryClass = {astro-ph},
       adsurl = {https://ui.adsabs.harvard.edu/abs/1998A&A...331L...1S}
}

@ARTICLE{2017Tumlinson,
       author = {{Tumlinson}, Jason and {Peeples}, Molly S. and {Werk}, Jessica K.},
        title = "{The Circumgalactic Medium}",
      journal = {\araa},
         year = 2017,
        month = aug,
       volume = {55},
       number = {1},
        pages = {389-432},
          doi = {10.1146/annurev-astro-091916-055240},
archivePrefix = {arXiv},
       eprint = {1709.09180},
 primaryClass = {astro-ph.GA},
       adsurl = {https://ui.adsabs.harvard.edu/abs/2017ARA&A..55..389T}
}

@ARTICLE{2005Scannapieco,
       author = {{Scannapieco}, Evan and {Silk}, Joseph and {Bouwens}, Rychard},
        title = "{AGN Feedback Causes Downsizing}",
      journal = {\apjl},
         year = 2005,
        month = dec,
       volume = {635},
       number = {1},
        pages = {L13-L16},
          doi = {10.1086/499271},
archivePrefix = {arXiv},
       eprint = {astro-ph/0511116},
 primaryClass = {astro-ph},
       adsurl = {https://ui.adsabs.harvard.edu/abs/2005ApJ...635L..13S}
}

@ARTICLE{2005NDiMatteo,
       author = {{Di Matteo}, Tiziana and {Springel}, Volker and {Hernquist}, Lars},
        title = "{Energy input from quasars regulates the growth and activity of black holes and their host galaxies}",
      journal = {\nat},
         year = 2005,
        month = feb,
       volume = {433},
       number = {7026},
        pages = {604-607},
          doi = {10.1038/nature03335},
archivePrefix = {arXiv},
       eprint = {astro-ph/0502199},
 primaryClass = {astro-ph},
       adsurl = {https://ui.adsabs.harvard.edu/abs/2005Natur.433..604D}
}

@ARTICLE{2008Somerville,
       author = {{Somerville}, Rachel S. and {Hopkins}, Philip F. and {Cox}, Thomas J. and {Robertson}, Brant E. and {Hernquist}, Lars},
        title = "{A semi-analytic model for the co-evolution of galaxies, black holes and active galactic nuclei}",
      journal = {\mnras},
         year = 2008,
        month = dec,
       volume = {391},
       number = {2},
        pages = {481-506},
          doi = {10.1111/j.1365-2966.2008.13805.x},
archivePrefix = {arXiv},
       eprint = {0808.1227},
 primaryClass = {astro-ph},
       adsurl = {https://ui.adsabs.harvard.edu/abs/2008MNRAS.391..481S}
}

@ARTICLE{2015Erb,
       author = {{Erb}, Dawn K.},
        title = "{Feedback in low-mass galaxies in the early Universe}",
      journal = {\nat},
         year = 2015,
        month = jul,
       volume = {523},
       number = {7559},
        pages = {169-176},
          doi = {10.1038/nature14454},
archivePrefix = {arXiv},
       eprint = {1507.02374},
 primaryClass = {astro-ph.GA},
       adsurl = {https://ui.adsabs.harvard.edu/abs/2015Natur.523..169E}
}

@article{2008Hopkins,
doi = {10.1086/524363},
url = {https://dx.doi.org/10.1086/524363},
year = {2008},
month = {apr},
publisher = {},
volume = {175},
number = {2},
pages = {390},
author = {Hopkins, Philip F. and Cox, Thomas J. and Kereš, Dušan and Hernquist, Lars},
title = {A Cosmological Framework for the Co-Evolution of Quasars, Supermassive Black Holes, and Elliptical Galaxies. II. Formation of Red Ellipticals},
journal = {The Astrophysical Journal Supplement Series}
}

@ARTICLE{1990Heckman,
       author = {{Heckman}, Timothy M. and {Armus}, Lee and {Miley}, George K.},
        title = "{On the Nature and Implications of Starburst-driven Galactic Superwinds}",
      journal = {\apjs},
         year = 1990,
        month = dec,
       volume = {74},
        pages = {833},
          doi = {10.1086/191522},
       adsurl = {https://ui.adsabs.harvard.edu/abs/1990ApJS...74..833H}
}

@article{2017Chisholm,
    author = {Chisholm, John and Tremonti, Christy A. and Leitherer, Claus and Chen, Yanmei},
    title = "{The mass and momentum outflow rates of photoionized galactic outflows}",
    journal = {Monthly Notices of the Royal Astronomical Society},
    volume = {469},
    number = {4},
    pages = {4831-4849},
    year = {2017},
    month = {05},
    issn = {0035-8711},
    doi = {10.1093/mnras/stx1164},
    url = {https://doi.org/10.1093/mnras/stx1164},
    eprint = {https://academic.oup.com/mnras/article-pdf/469/4/4831/17729336/stx1164.pdf},
}

@article{2015Zhu,
doi = {10.1088/0004-637X/815/1/48},
url = {https://dx.doi.org/10.1088/0004-637X/815/1/48},
year = {2015},
month = {dec},
publisher = {The American Astronomical Society},
volume = {815},
number = {1},
pages = {48},
author = {Zhu, Guangtun Ben and Comparat, Johan and Kneib, Jean-Paul and Delubac, Timothée and Raichoor, Anand and Dawson, Kyle S. and Newman, Jeffrey and Yèche, Christophe and Zhou, Xu and Schneider, Donald P.},
title = {NEAR-ULTRAVIOLET SPECTROSCOPY OF STAR-FORMING GALAXIES FROM eBOSS: SIGNATURES OF UBIQUITOUS GALACTIC-SCALE OUTFLOWS},
journal = {The Astrophysical Journal}
}

@article{2012Newman,
doi = {10.1088/0004-637X/761/1/43},
url = {https://dx.doi.org/10.1088/0004-637X/761/1/43},
year = {2012},
month = {nov},
publisher = {The American Astronomical Society},
volume = {761},
number = {1},
pages = {43},
author = {Newman, Sarah F. and Genzel, Reinhard and Förster-Schreiber, Natascha M. and Griffin, Kristen Shapiro and Mancini, Chiara and Lilly, Simon J. and Renzini, Alvio and Bouché, Nicolas and Burkert, Andreas and Buschkamp, Peter and Carollo, C. Marcella and Cresci, Giovanni and Davies, Ric and Eisenhauer, Frank and Genel, Shy and Hicks, Erin K. S. and Kurk, Jaron and Lutz, Dieter and Naab, Thorsten and Peng, Yingjie and Sternberg, Amiel and Tacconi, Linda J. and Vergani, Daniela and Wuyts, Stijn and Zamorani, Gianni},
title = {THE SINS/zC-SINF SURVEY of z ∼ 2 GALAXY KINEMATICS: OUTFLOW PROPERTIES*},
journal = {The Astrophysical Journal}
}

@ARTICLE{2009Strickland,
       author = {{Strickland}, David K. and {Heckman}, Timothy M.},
        title = "{Supernova Feedback Efficiency and Mass Loading in the Starburst and Galactic Superwind Exemplar M82}",
      journal = {\apj},
         year = 2009,
        month = jun,
       volume = {697},
       number = {2},
        pages = {2030-2056},
          doi = {10.1088/0004-637X/697/2/2030},
archivePrefix = {arXiv},
       eprint = {0903.4175},
 primaryClass = {astro-ph.CO},
       adsurl = {https://ui.adsabs.harvard.edu/abs/2009ApJ...697.2030S}
}

@ARTICLE{2010Tombesi,
       author = {{Tombesi}, F. and {Cappi}, M. and {Reeves}, J.~N. and {Palumbo}, G.~G.~C. and {Yaqoob}, T. and {Braito}, V. and {Dadina}, M.},
        title = "{Evidence for ultra-fast outflows in radio-quiet AGNs. I. Detection and statistical incidence of Fe K-shell absorption lines}",
      journal = {\aap},
         year = 2010,
        month = oct,
       volume = {521},
          eid = {A57},
        pages = {A57},
          doi = {10.1051/0004-6361/200913440},
archivePrefix = {arXiv},
       eprint = {1006.2858},
 primaryClass = {astro-ph.HE},
       adsurl = {https://ui.adsabs.harvard.edu/abs/2010A&A...521A..57T}
}

@article{Sanders2025,
doi = {10.3847/1538-4357/adf066},
url = {https://doi.org/10.3847/1538-4357/adf066},
year = {2025},
month = {aug},
publisher = {The American Astronomical Society},
volume = {989},
number = {2},
pages = {209},
author = {Sanders, Ryan L. and Shapley, Alice E. and Topping, Michael W. and Reddy, Naveen A. and Berg, Danielle A. and Bouwens, Rychard J. and Brammer, Gabriel and Carnall, Adam C. and Cullen, Fergus and Davé, Romeel and Dunlop, James S. and Ellis, Richard S. and Förster Schreiber, N. M. and Furlanetto, Steven R. and Glazebrook, Karl and Illingworth, Garth D. and Jones, Tucker and Kriek, Mariska and McLeod, Derek J. and McLure, Ross J. and Narayanan, Desika and Oesch, Pascal A. and Pahl, Anthony J. and Pettini, Max and Schaerer, Daniel and Stark, Daniel P. and Steidel, Charles C. and Tang, Mengtao and Clarke, Leonardo and Donnan, Callum T. and Kehoe, Emily},
title = {The AURORA Survey: The Nebular Attenuation Curve of a Galaxy at z = 4.41 from Ultraviolet to Near-infrared Wavelengths},
journal = {The Astrophysical Journal}
}

@article{Valentino2023,
doi = {10.3847/1538-4357/acbefa},
url = {https://doi.org/10.3847/1538-4357/acbefa},
year = {2023},
month = {apr},
publisher = {The American Astronomical Society},
volume = {947},
number = {1},
pages = {20},
author = {Valentino, Francesco and Brammer, Gabriel and Gould, Katriona M. L. and Kokorev, Vasily and Fujimoto, Seiji and Jespersen, Christian Kragh and Vijayan, Aswin P. and Weaver, John R. and Ito, Kei and Tanaka, Masayuki and Ilbert, Olivier and Magdis, Georgios E. and Whitaker, Katherine E. and Faisst, Andreas L. and Gallazzi, Anna and Gillman, Steven and Giménez-Arteaga, Clara and Gómez-Guijarro, Carlos and Kubo, Mariko and Heintz, Kasper E. and Hirschmann, Michaela and Oesch, Pascal and Onodera, Masato and Rizzo, Francesca and Lee, Minju and Strait, Victoria and Toft, Sune},
title = {An Atlas of Color-selected Quiescent Galaxies at z &gt; 3 in Public JWST Fields },
journal = {The Astrophysical Journal}
}

@article{Kriek2009,
doi = {10.1088/0004-637X/700/1/221},
url = {https://doi.org/10.1088/0004-637X/700/1/221},
year = {2009},
month = {jul},
publisher = {The American Astronomical Society},
volume = {700},
number = {1},
pages = {221},
author = {Kriek, Mariska and van Dokkum, Pieter G. and Labbé, Ivo and Franx, Marijn and Illingworth, Garth D. and Marchesini, Danilo and Quadri, Ryan F.},
title = {AN ULTRA-DEEP NEAR-INFRARED SPECTRUM OF A COMPACT QUIESCENT GALAXY AT z = 2.2},
journal = {The Astrophysical Journal}
}

@ARTICLE{Eisenstein2026,
       author = {{Eisenstein}, Daniel J. and {Willott}, Chris and {Alberts}, Stacey and {Arribas}, Santiago and {Bonaventura}, Nina and {Bunker}, Andrew J. and {Cameron}, Alex J. and {Carniani}, Stefano and {Charlot}, Stephane and {Curtis-Lake}, Emma and {D'Eugenio}, Francesco and {Ferruit}, Pierre and {Giardino}, Giovanna and {Hainline}, Kevin and {Hausen}, Ryan and {Jakobsen}, Peter and {Johnson}, Benjamin D. and {Maiolino}, Roberto and {Rauscher}, Bernard J. and {Rieke}, Marcia and {Rieke}, George and {Rix}, Hans-Walter and {Robertson}, Brant and {Stark}, Daniel P. and {Tacchella}, Sandro and {Williams}, Christina C. and {Willmer}, Christopher N.~A. and {Baker}, William M. and {Baum}, Stefi and {Bhatawdekar}, Rachana and {Boyett}, Kristan and {Chen}, Zuyi and {Chevallard}, Jacopo and {Circosta}, Chiara and {Curti}, Mirko and {Danhaive}, A. Lola and {DeCoursey}, Christa and {Endsley}, Ryan and {de Graaff}, Anna and {Dressler}, Alan and {Egami}, Eiichi and {Helton}, Jakob M. and {Hviding}, Raphael E. and {Ji}, Zhiyuan and {Jones}, Gareth C. and {Kumari}, Nimisha and {L{\"u}tzgendorf}, Nora and {Laseter}, Isaac and {Looser}, Tobias J. and {Lyu}, Jianwei and {Maseda}, Michael V. and {Nelson}, Erica and {Parlanti}, Eleonora and {Perna}, Michele and {Pusk{\'a}s}, D{\'a}vid and {Rawle}, Tim and {Rodr{\'\i}guez Del Pino}, Bruno and {Rujopakarn}, Wiphu and {Sandles}, Lester and {Saxena}, Aayush and {Scholtz}, Jan and {Sharpe}, Katherine and {Shivaei}, Irene and {Silcock}, Maddie S. and {Simmonds}, Charlotte and {Skarbinski}, Maya and {Smit}, Renske and {Stone}, Meredith and {Suess}, Katherine A. and {Sun}, Fengwu and {Tang}, Mengtao and {Topping}, Michael W. and {{\"U}bler}, Hannah and {Villanueva}, Natalia C. and {Wallace}, Imaan E.~B. and {Whitler}, Lily and {Witstok}, Joris and {Woodrum}, Charity},
        title = "{Overview of the JWST Advanced Deep Extragalactic Survey (JADES)}",
      journal = {\apjs},
         year = 2026,
        month = mar,
       volume = {283},
       number = {1},
          eid = {6},
        pages = {6},
          doi = {10.3847/1538-4365/ae3163},
archivePrefix = {arXiv},
       eprint = {2306.02465},
 primaryClass = {astro-ph.GA},
       adsurl = {https://ui.adsabs.harvard.edu/abs/2026ApJS..283....6E}
}

@article{Carnall2018,
    author = {Carnall, A C and McLure, R J and Dunlop, J S and Davé, R},
    title = {Inferring the star formation histories of massive quiescent galaxies with bagpipes: evidence for multiple quenching mechanisms},
    journal = {Monthly Notices of the Royal Astronomical Society},
    volume = {480},
    number = {4},
    pages = {4379-4401},
    year = {2018},
    month = {08},
    issn = {0035-8711},
    doi = {10.1093/mnras/sty2169},
    url = {https://doi.org/10.1093/mnras/sty2169},
    eprint = {https://academic.oup.com/mnras/article-pdf/480/4/4379/25539546/sty2169.pdf},
}

@article{Carnall2019,
doi = {10.3847/1538-4357/ab04a2},
url = {https://doi.org/10.3847/1538-4357/ab04a2},
year = {2019},
month = {mar},
publisher = {The American Astronomical Society},
volume = {873},
number = {1},
pages = {44},
author = {Carnall, Adam C. and Leja, Joel and Johnson, Benjamin D. and McLure, Ross J. and Dunlop, James S. and Conroy, Charlie},
title = {How to Measure Galaxy Star Formation Histories. I. Parametric Models},
journal = {The Astrophysical Journal}
}

@article{Eldridge_2017, title={Binary Population and Spectral Synthesis Version 2.1: Construction, Observational Verification, and New Results}, volume={34}, DOI={10.1017/pasa.2017.51}, journal={Publications of the Astronomical Society of Australia}, author={Eldridge, J. J. and Stanway, E. R. and Xiao, L. and McClelland, L. A. S. and Taylor, G. and Ng, M. and Greis, S. M. L. and Bray, J. C.}, year={2017}, pages={e058}}

@article{Kroupa2001,
    author = {Kroupa, Pavel},
    title = {On the variation of the initial mass function},
    journal = {Monthly Notices of the Royal Astronomical Society},
    volume = {322},
    number = {2},
    pages = {231-246},
    year = {2001},
    month = {04},
    issn = {0035-8711},
    doi = {10.1046/j.1365-8711.2001.04022.x},
    url = {https://doi.org/10.1046/j.1365-8711.2001.04022.x},
    eprint = {https://academic.oup.com/mnras/article-pdf/322/2/231/2852412/322-2-231.pdf},
}

@ARTICLE{2025Genin,
       author = {{Genin}, Aur{\'e}lien and {Shuntov}, Marko and {Brammer}, Gabe and {Allen}, Natalie and {Ito}, Kei and {Magdis}, Georgios and {Matharu}, Jasleen and {Oesch}, Pascal A. and {Toft}, Sune and {Valentino}, Francesco},
        title = "{DAWN JWST Archive: Morphology from profile fitting of over 340 000 galaxies in major JWST fields: Morphology evolution with redshift and galaxy type}",
      journal = {\aap},
         year = 2025,
        month = jul,
       volume = {699},
          eid = {A343},
        pages = {A343},
          doi = {10.1051/0004-6361/202555504},
archivePrefix = {arXiv},
       eprint = {2505.21622},
 primaryClass = {astro-ph.GA},
       adsurl = {https://ui.adsabs.harvard.edu/abs/2025A&A...699A.343G}
}

@article{Stanton2026,
    author = {Stanton, T M and Cullen, F and Carnall, A C and Scholte, D and Arellano-Córdova, K Z and Shapley, A E and McLeod, D J and Donnan, C T and Begley, R and Davé, R and Dunlop, J S and McLure, R J and Rowlands, K and Bondestam, C and Hamadouche, M L and Leung, H-H and Stevenson, S D and Taylor, E},
    title = {The JWST EXCELS Survey: gas-phase metallicity evolution at 2 \&lt; z \&lt; 8},
    journal = {Monthly Notices of the Royal Astronomical Society},
    pages = {stag449},
    year = {2026},
    month = {03},
    issn = {0035-8711},
    doi = {10.1093/mnras/stag449},
    url = {https://doi.org/10.1093/mnras/stag449},
    eprint = {https://academic.oup.com/mnras/advance-article-pdf/doi/10.1093/mnras/stag449/67243273/stag449.pdf},
}

@article{Salim2018,
doi = {10.3847/1538-4357/aabf3c},
url = {https://doi.org/10.3847/1538-4357/aabf3c},
year = {2018},
month = {may},
publisher = {The American Astronomical Society},
volume = {859},
number = {1},
pages = {11},
author = {Salim, Samir and Boquien, Médéric and Lee, Janice C.},
title = {Dust Attenuation Curves in the Local Universe: Demographics and New Laws for Star-forming Galaxies and High-redshift Analogs},
journal = {The Astrophysical Journal}
}

@ARTICLE{2024Davies,
       author = {{Davies}, Rebecca L. and {Belli}, Sirio and {Park}, Minjung and {Mendel}, J. Trevor and {Johnson}, Benjamin D. and {Conroy}, Charlie and {Benton}, Chlo{\"e} and {Bugiani}, Letizia and {Emami}, Razieh and {Leja}, Joel and {Li}, Yijia and {Maheson}, Gabriel and {Mathews}, Elijah P. and {Naidu}, Rohan P. and {Nelson}, Erica J. and {Tacchella}, Sandro and {Terrazas}, Bryan A. and {Weinberger}, Rainer},
        title = "{JWST reveals widespread AGN-driven neutral gas outflows in massive z   2 galaxies}",
      journal = {\mnras},
         year = 2024,
        month = mar,
       volume = {528},
       number = {3},
        pages = {4976-4992},
          doi = {10.1093/mnras/stae327},
archivePrefix = {arXiv},
       eprint = {2310.17939},
 primaryClass = {astro-ph.GA},
       adsurl = {https://ui.adsabs.harvard.edu/abs/2024MNRAS.528.4976D}
}

@ARTICLE{2024Belli,
       author = {{Belli}, Sirio and {Park}, Minjung and {Davies}, Rebecca L. and {Mendel}, J. Trevor and {Johnson}, Benjamin D. and {Conroy}, Charlie and {Benton}, Chlo{\"e} and {Bugiani}, Letizia and {Emami}, Razieh and {Leja}, Joel and {Li}, Yijia and {Maheson}, Gabriel and {Mathews}, Elijah P. and {Naidu}, Rohan P. and {Nelson}, Erica J. and {Tacchella}, Sandro and {Terrazas}, Bryan A. and {Weinberger}, Rainer},
        title = "{Star formation shut down by multiphase gas outflow in a galaxy at a redshift of 2.45}",
      journal = {\nat},
         year = 2024,
        month = jun,
       volume = {630},
       number = {8015},
        pages = {54-58},
          doi = {10.1038/s41586-024-07412-1},
archivePrefix = {arXiv},
       eprint = {2308.05795},
 primaryClass = {astro-ph.GA},
       adsurl = {https://ui.adsabs.harvard.edu/abs/2024Natur.630...54B}
}

@ARTICLE{2024DEugenio,
       author = {{D'Eugenio}, Francesco and {P{\'e}rez-Gonz{\'a}lez}, Pablo G. and {Maiolino}, Roberto and {Scholtz}, Jan and {Perna}, Michele and {Circosta}, Chiara and {{\"U}bler}, Hannah and {Arribas}, Santiago and {B{\"o}ker}, Torsten and {Bunker}, Andrew J. and {Carniani}, Stefano and {Charlot}, Stephane and {Chevallard}, Jacopo and {Cresci}, Giovanni and {Curtis-Lake}, Emma and {Jones}, Gareth C. and {Kumari}, Nimisha and {Lamperti}, Isabella and {Looser}, Tobias J. and {Parlanti}, Eleonora and {Rix}, Hans-Walter and {Robertson}, Brant and {Rodr{\'\i}guez Del Pino}, Bruno and {Tacchella}, Sandro and {Venturi}, Giacomo and {Willott}, Chris J.},
        title = "{A fast-rotator post-starburst galaxy quenched by supermassive black-hole feedback at z = 3}",
      journal = {Nature Astronomy},
         year = 2024,
        month = nov,
       volume = {8},
        pages = {1443-1456},
          doi = {10.1038/s41550-024-02345-1},
archivePrefix = {arXiv},
       eprint = {2308.06317},
 primaryClass = {astro-ph.GA},
       adsurl = {https://ui.adsabs.harvard.edu/abs/2024NatAs...8.1443D}
}

@ARTICLE{2025Sun,
       author = {{Sun}, Yang and {Ji}, Zhiyuan and {Rieke}, George H. and {D'Eugenio}, Francesco and {Zhu}, Yongda and {Sun}, Fengwu and {Lin}, Xiaojing and {Bunker}, Andrew J. and {Lyu}, Jianwei and {Rinaldi}, Pierluigi and {Willmer}, Christopher N.~A.},
        title = "{Extreme Neutral Outflow in an Inactive Quenching Galaxy at z$\sim$1.3}",
      journal = {arXiv e-prints},
         year = 2025,
        month = apr,
          eid = {arXiv:2504.14682},
        pages = {arXiv:2504.14682},
          doi = {10.48550/arXiv.2504.14682},
archivePrefix = {arXiv},
       eprint = {2504.14682},
 primaryClass = {astro-ph.GA},
       adsurl = {https://ui.adsabs.harvard.edu/abs/2025arXiv250414682S}
}

@ARTICLE{2013Foreman,
       author = {{Foreman-Mackey}, Daniel and {Hogg}, David W. and {Lang}, Dustin and {Goodman}, Jonathan},
        title = "{emcee: The MCMC Hammer}",
      journal = {\pasp},
         year = 2013,
        month = mar,
       volume = {125},
       number = {925},
        pages = {306},
          doi = {10.1086/670067},
archivePrefix = {arXiv},
       eprint = {1202.3665},
 primaryClass = {astro-ph.IM},
       adsurl = {https://ui.adsabs.harvard.edu/abs/2013PASP..125..306F}
}

@article{Pahl2022,
    author = {Pahl, Anthony J and Shapley, Alice and Steidel, Charles C and Reddy, Naveen A and Chen, Yuguang},
    title = {Searching for the connection between ionizing-photon escape and the surface density of star formation at z ∼ 3},
    journal = {Monthly Notices of the Royal Astronomical Society},
    volume = {516},
    number = {2},
    pages = {2062-2073},
    year = {2022},
    month = {06},
    issn = {0035-8711},
    doi = {10.1093/mnras/stac1767},
    url = {https://doi.org/10.1093/mnras/stac1767},
    eprint = {https://academic.oup.com/mnras/article-pdf/516/2/2062/45787644/stac1767.pdf},
}

@article{2022Avery,
    author = {Avery, Charlotte R and Wuyts, Stijn and Förster Schreiber, Natascha M and Villforth, Carolin and Bertemes, Caroline and Hamer, Stephen L and Sharma, Raman and Toshikawa, Jun and Zhang, Junkai},
    title = {Cool outflows in MaNGA: a systematic study and comparison to the warm phase},
    journal = {Monthly Notices of the Royal Astronomical Society},
    volume = {511},
    number = {3},
    pages = {4223-4237},
    year = {2022},
    month = {01},
    issn = {0035-8711},
    doi = {10.1093/mnras/stac190},
    url = {https://doi.org/10.1093/mnras/stac190},
    eprint = {https://academic.oup.com/mnras/article-pdf/511/3/4223/48230138/stac190.pdf},
}

@article{Kehoe2025,
doi = {10.3847/1538-4357/ae10b3},
url = {https://doi.org/10.3847/1538-4357/ae10b3},
year = {2025},
month = {nov},
publisher = {The American Astronomical Society},
volume = {994},
number = {2},
pages = {170},
author = {Kehoe, Emily and Shapley, Alice E. and Sanders, Ryan L. and Reddy, Naveen A. and Topping, Michael W. and Lam, Natalie and Clarke, Leonardo and Cullen, Fergus and Ellis, Richard S. and Förster Schreiber, N. M. and Jones, Tucker and Khostovan, Ali Ahmad and McLeod, Derek J. and McLure, Ross J. and Narayanan, Desika and Oesch, Pascal and Pahl, Anthony J.},
title = {The AURORA Survey: Tracing Galactic Outflows at z ≳ 2.5 with JWST/NIRSpec Near-ultraviolet Absorption Lines},
journal = {The Astrophysical Journal}
}

@article{Weldon2023,
    author = {Weldon, Andrew and Reddy, Naveen A and Topping, Michael W and Shapley, Alice E and Du, Xinnan and Price, Sedona H and Sanders, Ryan L and Coil, Alison L and Mobasher, Bahram and Kriek, Mariska and Siana, Brian and Rezaee, Saeed},
    title = {The MOSDEF-LRIS survey: detection of inflowing gas towards three star-forming galaxies at z ∼ 2},
    journal = {Monthly Notices of the Royal Astronomical Society},
    volume = {523},
    number = {4},
    pages = {5624-5634},
    year = {2023},
    month = {06},
    issn = {0035-8711},
    doi = {10.1093/mnras/stad1615},
    url = {https://doi.org/10.1093/mnras/stad1615},
    eprint = {https://academic.oup.com/mnras/article-pdf/523/4/5624/50740269/stad1615.pdf},
}

@article{Bevacqua2026,
doi = {10.3847/1538-4357/ae247c},
url = {https://doi.org/10.3847/1538-4357/ae247c},
year = {2026},
month = {jan},
publisher = {The American Astronomical Society},
volume = {997},
number = {2},
pages = {189},
author = {Bevacqua, Davide and Marchesini, Danilo and Saracco, Paolo and La Barbera, Francesco and Pan, Richard and Belli, Sirio and Brammer, Gabriel and De Marchi, Guido and Ditrani, Fabio R. and Giardino, Giovanna and Glazebrook, Karl and La Torre, Valentina and Lin, Jamie and Muzzin, Adam and Roy, Namrata and Santini, Paola and Vulcani, Benedetta and Watson, Peter J. and Wang, Xin},
title = {Feeding the Dead: Neutral Gas Inflow in a Long-quenched Ancient Massive Galaxy at z&nbsp;∼&nbsp;2.7 Observed with JWST/NIRSpec},
journal = {The Astrophysical Journal}
}

@article{Leroy2008,
doi = {10.1088/0004-6256/136/6/2782},
url = {https://doi.org/10.1088/0004-6256/136/6/2782},
year = {2008},
month = {nov},
publisher = {The American Astronomical Society},
volume = {136},
number = {6},
pages = {2782},
author = {Leroy, Adam K. and Walter, Fabian and Brinks, Elias and Bigiel, Frank and de Blok, W. J. G. and Madore, Barry and Thornley, M. D.},
title = {THE STAR FORMATION EFFICIENCY IN NEARBY GALAXIES: MEASURING WHERE GAS FORMS STARS EFFECTIVELY},
journal = {The Astronomical Journal}
}

@article{Saintonge2017,
doi = {10.3847/1538-4365/aa97e0},
url = {https://doi.org/10.3847/1538-4365/aa97e0},
year = {2017},
month = {dec},
publisher = {The American Astronomical Society},
volume = {233},
number = {2},
pages = {22},
author = {Saintonge, Amélie and Catinella, Barbara and Tacconi, Linda J. and Kauffmann, Guinevere and Genzel, Reinhard and Cortese, Luca and Davé, Romeel and Fletcher, Thomas J. and Graciá-Carpio, Javier and Kramer, Carsten and Heckman, Timothy M. and Janowiecki, Steven and Lutz, Katharina and Rosario, David and Schiminovich, David and Schuster, Karl and Wang, Jing and Wuyts, Stijn and Borthakur, Sanchayeeta and Lamperti, Isabella and Roberts-Borsani, Guido W.},
title = {xCOLD GASS: The Complete IRAM 30 m Legacy Survey of Molecular Gas for Galaxy Evolution Studies},
journal = {The Astrophysical Journal Supplement Series}
}

@article{Tacconi2018,
doi = {10.3847/1538-4357/aaa4b4},
url = {https://doi.org/10.3847/1538-4357/aaa4b4},
year = {2018},
month = {feb},
publisher = {The American Astronomical Society},
volume = {853},
number = {2},
pages = {179},
author = {Tacconi, L. J. and Genzel, R. and Saintonge, A. and Combes, F. and García-Burillo, S. and Neri, R. and Bolatto, A. and Contini, T. and Schreiber, N. M. Förster and Lilly, S. and Lutz, D. and Wuyts, S. and Accurso, G. and Boissier, J. and Boone, F. and Bouché, N. and Bournaud, F. and Burkert, A. and Carollo, M. and Cooper, M. and Cox, P. and Feruglio, C. and Freundlich, J. and Herrera-Camus, R. and Juneau, S. and Lippa, M. and Naab, T. and Renzini, A. and Salome, P. and Sternberg, A. and Tadaki, K. and Übler, H. and Walter, F. and Weiner, B. and Weiss, A.},
title = {PHIBSS: Unified Scaling Relations of Gas Depletion Time and Molecular Gas Fractions*},
journal = {The Astrophysical Journal}
}

@ARTICLE{Kennicutt1989,
       author = {{Kennicutt}, Jr., Robert C.},
        title = "{The Star Formation Law in Galactic Disks}",
      journal = {\apj},
         year = 1989,
        month = sep,
       volume = {344},
        pages = {685},
          doi = {10.1086/167834},
       adsurl = {https://ui.adsabs.harvard.edu/abs/1989ApJ...344..685K}
}

@article{Prochaska2009,
doi = {10.1088/0004-637X/696/2/1543},
url = {https://doi.org/10.1088/0004-637X/696/2/1543},
year = {2009},
month = {apr},
publisher = {The American Astronomical Society},
volume = {696},
number = {2},
pages = {1543},
author = {Prochaska, J. Xavier and Wolfe, Arthur M.},
title = {ON THE (NON)EVOLUTION OF H i GAS IN GALAXIES OVER COSMIC TIME},
journal = {The Astrophysical Journal}
}

@article{Bauermeister2010,
doi = {10.1088/0004-637X/717/1/323},
url = {https://doi.org/10.1088/0004-637X/717/1/323},
year = {2010},
month = {jun},
publisher = {The American Astronomical Society},
volume = {717},
number = {1},
pages = {323},
author = {Bauermeister, Amber and Blitz, Leo and Ma, Chung-Pei},
title = {THE GAS CONSUMPTION HISTORY TO REDSHIFT 4},
journal = {The Astrophysical Journal}
}

@article{Faucher2011,
    author = {Faucher-Giguère, Claude-André and Kereš, Dušan},
    title = {The small covering factor of cold accretion streams},
    journal = {Monthly Notices of the Royal Astronomical Society: Letters},
    volume = {412},
    number = {1},
    pages = {L118-L122},
    year = {2011},
    month = {03},
    issn = {1745-3925},
    doi = {10.1111/j.1745-3933.2011.01018.x},
    url = {https://doi.org/10.1111/j.1745-3933.2011.01018.x},
    eprint = {https://academic.oup.com/mnrasl/article-pdf/412/1/L118/54670333/mnrasl_412_1_l118.pdf},
}

@article{Peroux2020,
    author = {Péroux, Céline and Nelson, Dylan and van de Voort, Freeke and Pillepich, Annalisa and Marinacci, Federico and Vogelsberger, Mark and Hernquist, Lars},
    title = {Predictions for the angular dependence of gas mass flow rate and metallicity in the circumgalactic medium},
    journal = {Monthly Notices of the Royal Astronomical Society},
    volume = {499},
    number = {2},
    pages = {2462-2473},
    year = {2020},
    month = {09},
    issn = {0035-8711},
    doi = {10.1093/mnras/staa2888},
    url = {https://doi.org/10.1093/mnras/staa2888},
    eprint = {https://academic.oup.com/mnras/article-pdf/499/2/2462/33975642/staa2888.pdf},
}

@ARTICLE{Fumagalli2011,
       author = {{Fumagalli}, Michele and {Prochaska}, J. Xavier and {Kasen}, Daniel and {Dekel}, Avishai and {Ceverino}, Daniel and {Primack}, Joel R.},
        title = "{Absorption-line systems in simulated galaxies fed by cold streams}",
      journal = {\mnras},
         year = 2011,
        month = dec,
       volume = {418},
       number = {3},
        pages = {1796-1821},
          doi = {10.1111/j.1365-2966.2011.19599.x},
archivePrefix = {arXiv},
       eprint = {1103.2130},
 primaryClass = {astro-ph.CO},
       adsurl = {https://ui.adsabs.harvard.edu/abs/2011MNRAS.418.1796F}
}

@article{Rubin2012,
doi = {10.1088/2041-8205/747/2/L26},
url = {https://doi.org/10.1088/2041-8205/747/2/L26},
year = {2012},
month = {feb},
publisher = {The American Astronomical Society},
volume = {747},
number = {2},
pages = {L26},
author = {Rubin, Kate H. R. and Xavier Prochaska, J. and Koo, David C. and Phillips, Andrew C.},
title = {THE DIRECT DETECTION OF COOL, METAL-ENRICHED GAS ACCRETION ONTO GALAXIES AT z ∼ 0.5},
journal = {The Astrophysical Journal Letters}
}

@article{Ho2017,
doi = {10.3847/1538-4357/835/2/267},
url = {https://doi.org/10.3847/1538-4357/835/2/267},
year = {2017},
month = {feb},
publisher = {The American Astronomical Society},
volume = {835},
number = {2},
pages = {267},
author = {Ho, Stephanie H. and Martin, Crystal L. and Kacprzak, Glenn G. and Churchill, Christopher W.},
title = {Quasars Probing Galaxies. I. Signatures of Gas Accretion at Redshift z ≈ 0.2∗

†},
journal = {The Astrophysical Journal}
}

@ARTICLE{2019Roberts,
       author = {{Roberts-Borsani}, G.~W. and {Saintonge}, A.},
        title = "{The prevalence and properties of cold gas inflows and outflows around galaxies in the local Universe}",
      journal = {\mnras},
         year = 2019,
        month = jan,
       volume = {482},
       number = {3},
        pages = {4111-4145},
          doi = {10.1093/mnras/sty2824},
archivePrefix = {arXiv},
       eprint = {1807.07575},
 primaryClass = {astro-ph.GA},
       adsurl = {https://ui.adsabs.harvard.edu/abs/2019MNRAS.482.4111R}
}

@ARTICLE{2011Bordoloi,
       author = {{Bordoloi}, R. and {Lilly}, S.~J. and {Knobel}, C. and {Bolzonella}, M. and {Kampczyk}, P. and {Carollo}, C.~M. and {Iovino}, A. and {Zucca}, E. and {Contini}, T. and {Kneib}, J. -P. and {Le Fevre}, O. and {Mainieri}, V. and {Renzini}, A. and {Scodeggio}, M. and {Zamorani}, G. and {Balestra}, I. and {Bardelli}, S. and {Bongiorno}, A. and {Caputi}, K. and {Cucciati}, O. and {de la Torre}, S. and {de Ravel}, L. and {Garilli}, B. and {Kova{\v{c}}}, K. and {Lamareille}, F. and {Le Borgne}, J. -F. and {Le Brun}, V. and {Maier}, C. and {Mignoli}, M. and {Pello}, R. and {Peng}, Y. and {Perez Montero}, E. and {Presotto}, V. and {Scarlata}, C. and {Silverman}, J. and {Tanaka}, M. and {Tasca}, L. and {Tresse}, L. and {Vergani}, D. and {Barnes}, L. and {Cappi}, A. and {Cimatti}, A. and {Coppa}, G. and {Diener}, C. and {Franzetti}, P. and {Koekemoer}, A. and {L{\'o}pez-Sanjuan}, C. and {McCracken}, H.~J. and {Moresco}, M. and {Nair}, P. and {Oesch}, P. and {Pozzetti}, L. and {Welikala}, N.},
        title = "{The Radial and Azimuthal Profiles of Mg II Absorption around 0.5 < z < 0.9 zCOSMOS Galaxies of Different Colors, Masses, and Environments}",
      journal = {\apj},
         year = 2011,
        month = dec,
       volume = {743},
       number = {1},
          eid = {10},
        pages = {10},
          doi = {10.1088/0004-637X/743/1/10},
archivePrefix = {arXiv},
       eprint = {1106.0616},
 primaryClass = {astro-ph.CO},
       adsurl = {https://ui.adsabs.harvard.edu/abs/2011ApJ...743...10B}
}

@article{Glascow2013,
    author = {Glasow, W. von and Krause, M. G. H. and Sommer-Larsen, J. and Burkert, A.},
    title = {Galactic winds – how to launch galactic outflows in typical Lyman-break galaxies},
    journal = {Monthly Notices of the Royal Astronomical Society},
    volume = {434},
    number = {2},
    pages = {1151-1170},
    year = {2013},
    month = {09},
    issn = {0035-8711},
    doi = {10.1093/mnras/stt1060},
    url = {https://doi.org/10.1093/mnras/stt1060},
    eprint = {https://academic.oup.com/mnras/article-pdf/434/2/1151/18751669/stt1060.pdf},
}

@article{Fielding2017,
    author = {Fielding, Drummond and Quataert, Eliot and Martizzi, Davide and Faucher-Giguère, Claude-André},
    title = {How supernovae launch galactic winds?},
    journal = {Monthly Notices of the Royal Astronomical Society: Letters},
    volume = {470},
    number = {1},
    pages = {L39-L43},
    year = {2017},
    month = {09},
    issn = {1745-3925},
    doi = {10.1093/mnrasl/slx072},
    url = {https://doi.org/10.1093/mnrasl/slx072},
    eprint = {https://academic.oup.com/mnrasl/article-pdf/470/1/L39/56928682/mnrasl_470_1_l39.pdf},
}

@article{Rubin2014,
doi = {10.1088/0004-637X/794/2/156},
url = {https://doi.org/10.1088/0004-637X/794/2/156},
year = {2014},
month = {oct},
publisher = {The American Astronomical Society},
volume = {794},
number = {2},
pages = {156},
author = {Rubin, Kate H. R. and Prochaska, J. Xavier and Koo, David C. and Phillips, Andrew C. and Martin, Crystal L. and Winstrom, Lucas O.},
title = {EVIDENCE FOR UBIQUITOUS COLLIMATED GALACTIC-SCALE OUTFLOWS ALONG THE STAR-FORMING SEQUENCE AT z ∼ 0.5},
journal = {The Astrophysical Journal}
}

@article{Law2012,
doi = {10.1088/0004-637X/759/1/29},
url = {https://doi.org/10.1088/0004-637X/759/1/29},
year = {2012},
month = {oct},
publisher = {The American Astronomical Society},
volume = {759},
number = {1},
pages = {29},
author = {Law, David R. and Steidel, Charles C. and Shapley, Alice E. and Nagy, Sarah R. and Reddy, Naveen A. and Erb, Dawn K.},
title = {A HST/WFC3-IR MORPHOLOGICAL SURVEY OF GALAXIES AT z = 1.5–3.6. II. THE RELATION BETWEEN MORPHOLOGY AND GAS-PHASE KINEMATICS*},
journal = {The Astrophysical Journal}
}

@article{Scarlata2015,
doi = {10.1088/0004-637X/801/1/43},
url = {https://doi.org/10.1088/0004-637X/801/1/43},
year = {2015},
month = {mar},
publisher = {The American Astronomical Society},
volume = {801},
number = {1},
pages = {43},
author = {Scarlata, C. and Panagia, N.},
title = {A SEMI-ANALYTICAL LINE TRANSFER MODEL TO INTERPRET THE SPECTRA OF GALAXY OUTFLOWS},
journal = {The Astrophysical Journal}
}

@ARTICLE{2026Moretti,
       author = {{Moretti}, Lorenzo and {Belli}, Sirio and {Rudie}, Gwen C. and {Newman}, Andrew B. and {Park}, Minjung and {Khoram}, Amir H. and {Chartab}, Nima and {Donevski}, Darko},
        title = "{Empirical calibration of Na I D and other absorption lines as tracers of high-redshift neutral outflows}",
      journal = {\aap},
         year = 2026,
        month = mar,
       volume = {707},
          eid = {A146},
        pages = {A146},
          doi = {10.1051/0004-6361/202556336},
archivePrefix = {arXiv},
       eprint = {2507.07160},
 primaryClass = {astro-ph.GA},
       adsurl = {https://ui.adsabs.harvard.edu/abs/2026A&A...707A.146M}
}

@ARTICLE{Schaerer2026,
       author = {{Schaerer}, D. and {Izotov}, Y.~I. and {Marques-Chaves}, R. and {Steidel}, C.~C. and {Reddy}, N. and {Shapley}, A.~E. and {Mascia}, S. and {Chisholm}, J. and {Flury}, S.~R. and {Guseva}, N. and {Heckman}, T. and {Henry}, A. and {Inoue}, A.~K. and {Jung}, I. and {Kusakabe}, H. and {Mawatari}, K. and {Oesch}, P. and {{\"O}stlin}, G. and {Pentericci}, L. and {Roy}, N. and {Saldana-Lopez}, A. and {Sato}, R. and {Vanzella}, E. and {Verhamme}, A. and {Wang}, B.},
        title = "{Nitrogen abundances in star-forming galaxies 2.2 Gyr after the Big Bang are not elevated}",
      journal = {\aap},
         year = 2026,
        month = apr,
       volume = {708},
          eid = {A242},
        pages = {A242},
          doi = {10.1051/0004-6361/202556832},
archivePrefix = {arXiv},
       eprint = {2601.06968},
 primaryClass = {astro-ph.GA},
       adsurl = {https://ui.adsabs.harvard.edu/abs/2026A&A...708A.242S}
}

@ARTICLE{Carnall2024,
       author = {{Carnall}, A.~C. and {Cullen}, F. and {McLure}, R.~J. and {McLeod}, D.~J. and {Begley}, R. and {Donnan}, C.~T. and {Dunlop}, J.~S. and {Shapley}, A.~E. and {Rowlands}, K. and {Almaini}, O. and {Arellano-C{\'o}rdova}, K.~Z. and {Barrufet}, L. and {Cimatti}, A. and {Ellis}, R.~S. and {Grogin}, N.~A. and {Hamadouche}, M.~L. and {Illingworth}, G.~D. and {Koekemoer}, A.~M. and {Leung}, H.-H. and {Lovell}, C.~C. and {P{\'e}rez-Gonz{\'a}lez}, P.~G. and {Santini}, P. and {Stanton}, T.~M. and {Wild}, V.},
        title = "{The JWST EXCELS survey: too much, too young, too fast? Ultra-massive quiescent galaxies at 3 < z < 5}",
      journal = {\mnras},
         year = 2024,
        month = oct,
       volume = {534},
       number = {1},
        pages = {325-348},
          doi = {10.1093/mnras/stae2092},
archivePrefix = {arXiv},
       eprint = {2405.02242},
 primaryClass = {astro-ph.GA},
       adsurl = {https://ui.adsabs.harvard.edu/abs/2024MNRAS.534..325C}
}

@article{Shapley2025,
doi = {10.3847/1538-4357/adad68},
url = {https://doi.org/10.3847/1538-4357/adad68},
year = {2025},
month = {feb},
publisher = {The American Astronomical Society},
volume = {980},
number = {2},
pages = {242},
author = {Shapley, Alice E. and Sanders, Ryan L. and Topping, Michael W. and Reddy, Naveen A. and Berg, Danielle A. and Bouwens, Rychard J. and Brammer, Gabriel and Carnall, Adam C. and Cullen, Fergus and Davé, Romeel and Dunlop, James S. and Ellis, Richard S. and Förster Schreiber, N. M. and Furlanetto, Steven R. and Glazebrook, Karl and Illingworth, Garth D. and Jones, Tucker and Kriek, Mariska and McLeod, Derek J. and McLure, Ross J. and Narayanan, Desika and Oesch, Pascal and Pahl, Anthony J. and Pettini, Max and Schaerer, Daniel and Stark, Daniel P. and Steidel, Charles C. and Tang, Mengtao and Clarke, Leonardo and Donnan, Callum T. and Kehoe, Emily},
title = {The AURORA Survey: A New Era of Emission-line Diagrams with JWST/NIRSpec},
journal = {The Astrophysical Journal}
}

@article{Steidel2018,
  title = {The {{Keck Lyman Continuum Spectroscopic Survey}} ({{KLCS}}): {{The Emergent Ionizing Spectrum}} of {{Galaxies}} at {\emph{z}} {$\sim$} 3},
  author = {Steidel, Charles C. and Bogosavljevi{\'c}, Milan and Shapley, Alice E. and Reddy, Naveen A. and Rudie, Gwen C. and Pettini, Max and Trainor, Ryan F. and Strom, Allison L.},
  year = {2018},
  month = dec,
  journal = {The Astrophysical Journal},
  volume = {869},
  number = {2},
  pages = {123},
  publisher = {IOP Publishing},
  issn = {1538-4357},
  doi = {10.3847/1538-4357/aaed28},
  urldate = {2019-03-26}
}

@ARTICLE{2021Du,
       author = {{Du}, Xinnan and {Shapley}, Alice E. and {Topping}, Michael W. and {Reddy}, Naveen A. and {Sanders}, Ryan L. and {Coil}, Alison L. and {Kriek}, Mariska and {Mobasher}, Bahram and {Siana}, Brian},
        title = "{The MOSDEF-LRIS Survey: Probing the ISM/CGM Structure of Star-forming Galaxies at z   2 Using Rest-UV Spectroscopy}",
      journal = {\apj},
         year = 2021,
        month = oct,
       volume = {920},
       number = {2},
          eid = {95},
        pages = {95},
          doi = {10.3847/1538-4357/ac1273},
archivePrefix = {arXiv},
       eprint = {2103.15824},
 primaryClass = {astro-ph.GA},
       adsurl = {https://ui.adsabs.harvard.edu/abs/2021ApJ...920...95D}
}

@ARTICLE{2011Kimm,
       author = {{Kimm}, Taysun and {Slyz}, Adrianne and {Devriendt}, Julien and {Pichon}, Christophe},
        title = "{Are cold flows detectable with metal absorption lines?}",
      journal = {\mnras},
         year = 2011,
        month = may,
       volume = {413},
       number = {1},
        pages = {L51-L55},
          doi = {10.1111/j.1745-3933.2011.01031.x},
archivePrefix = {arXiv},
       eprint = {1012.0059},
 primaryClass = {astro-ph.CO},
       adsurl = {https://ui.adsabs.harvard.edu/abs/2011MNRAS.413L..51K}
}

@ARTICLE{2012Kacprzak,
       author = {{Kacprzak}, Glenn G. and {Churchill}, Christopher W. and {Nielsen}, Nikole M.},
        title = "{Tracing Outflows and Accretion: A Bimodal Azimuthal Dependence of Mg II Absorption}",
      journal = {\apjl},
         year = 2012,
        month = nov,
       volume = {760},
       number = {1},
          eid = {L7},
        pages = {L7},
          doi = {10.1088/2041-8205/760/1/L7},
archivePrefix = {arXiv},
       eprint = {1205.0245},
 primaryClass = {astro-ph.CO},
       adsurl = {https://ui.adsabs.harvard.edu/abs/2012ApJ...760L...7K}
}

@article{Stewart2011,
doi = {10.1088/0004-637X/738/1/39},
url = {https://doi.org/10.1088/0004-637X/738/1/39},
year = {2011},
month = {aug},
publisher = {The American Astronomical Society},
volume = {738},
number = {1},
pages = {39},
author = {Stewart, Kyle R. and Kaufmann, Tobias and Bullock, James S. and Barton, Elizabeth J. and Maller, Ariyeh H. and Diemand, Jürg and Wadsley, James},
title = {ORBITING CIRCUMGALACTIC GAS AS A SIGNATURE OF COSMOLOGICAL ACCRETION},
journal = {The Astrophysical Journal}
}

@ARTICLE{2019Nelson,
       author = {{Nelson}, Dylan and {Pillepich}, Annalisa and {Springel}, Volker and {Pakmor}, R{\"u}diger and {Weinberger}, Rainer and {Genel}, Shy and {Torrey}, Paul and {Vogelsberger}, Mark and {Marinacci}, Federico and {Hernquist}, Lars},
        title = "{First results from the TNG50 simulation: galactic outflows driven by supernovae and black hole feedback}",
      journal = {\mnras},
         year = 2019,
        month = dec,
       volume = {490},
       number = {3},
        pages = {3234-3261},
          doi = {10.1093/mnras/stz2306},
archivePrefix = {arXiv},
       eprint = {1902.05554},
 primaryClass = {astro-ph.GA},
       adsurl = {https://ui.adsabs.harvard.edu/abs/2019MNRAS.490.3234N}
}

@article{2011Brook,
    author = {Brook, C. B. and Governato, F. and Roškar, R. and Stinson, G. and Brooks, A. M. and Wadsley, J. and Quinn, T. and Gibson, B. K. and Snaith, O. and Pilkington, K. and House, E. and Pontzen, A.},
    title = {Hierarchical formation of bulgeless galaxies: why outflows have low angular momentum},
    journal = {Monthly Notices of the Royal Astronomical Society},
    volume = {415},
    number = {2},
    pages = {1051-1060},
    year = {2011},
    month = {08},
    issn = {0035-8711},
    doi = {10.1111/j.1365-2966.2011.18545.x},
    url = {https://doi.org/10.1111/j.1365-2966.2011.18545.x},
    eprint = {https://academic.oup.com/mnras/article-pdf/415/2/1051/3464763/mnras0415-1051.pdf},
}

@ARTICLE{2013Jaskot,
       author = {{Jaskot}, A.~E. and {Oey}, M.~S.},
        title = "{The Origin and Optical Depth of Ionizing Radiation in the ``Green Pea'' Galaxies}",
      journal = {\apj},
         year = 2013,
        month = apr,
       volume = {766},
       number = {2},
          eid = {91},
        pages = {91},
          doi = {10.1088/0004-637X/766/2/91},
archivePrefix = {arXiv},
       eprint = {1301.0530},
 primaryClass = {astro-ph.CO},
       adsurl = {https://ui.adsabs.harvard.edu/abs/2013ApJ...766...91J}
}

@ARTICLE{2016Izotov,
       author = {{Izotov}, Y.~I. and {Schaerer}, D. and {Thuan}, T.~X. and {Worseck}, G. and {Guseva}, N.~G. and {Orlitov{\'a}}, I. and {Verhamme}, A.},
        title = "{Detection of high Lyman continuum leakage from four low-redshift compact star-forming galaxies}",
      journal = {\mnras},
         year = 2016,
        month = oct,
       volume = {461},
       number = {4},
        pages = {3683-3701},
          doi = {10.1093/mnras/stw1205},
archivePrefix = {arXiv},
       eprint = {1605.05160},
 primaryClass = {astro-ph.GA},
       adsurl = {https://ui.adsabs.harvard.edu/abs/2016MNRAS.461.3683I}
}

@article{Choustikov2024,
    author = {Choustikov, Nicholas and Katz, Harley and Saxena, Aayush and Garel, Thibault and Devriendt, Julien and Slyz, Adrianne and Kimm, Taysun and Blaizot, Jeremy and Rosdahl, Joki},
    title = {The great escape: understanding the connection between Ly α emission and LyC escape in simulated JWST analogues},
    journal = {Monthly Notices of the Royal Astronomical Society},
    volume = {532},
    number = {2},
    pages = {2463-2484},
    year = {2024},
    month = {08},
    issn = {0035-8711},
    doi = {10.1093/mnras/stae1586},
    url = {https://doi.org/10.1093/mnras/stae1586},
    eprint = {https://academic.oup.com/mnras/article-pdf/532/2/2463/58558860/stae1586.pdf},
}

@article{Pahl2021,
    author = {Pahl, Anthony J and Shapley, Alice and Steidel, Charles C and Chen, Yuguang and Reddy, Naveen A},
    title = {An uncontaminated measurement of the escaping Lyman continuum at z ∼ 3},
    journal = {Monthly Notices of the Royal Astronomical Society},
    volume = {505},
    number = {2},
    pages = {2447-2467},
    year = {2021},
    month = {08},
    issn = {0035-8711},
    doi = {10.1093/mnras/stab1374},
    url = {https://doi.org/10.1093/mnras/stab1374},
    eprint = {https://academic.oup.com/mnras/article-pdf/505/2/2447/38576884/stab1374.pdf},
}

@article{2021Izotov,
    author = {Izotov, Y I and Worseck, G and Schaerer, D and Guseva, N G and Chisholm, J and Thuan, T X and Fricke, K J and Verhamme, A},
    title = {Lyman continuum leakage from low-mass galaxies with M⋆ \&lt; 108 M⊙},
    journal = {Monthly Notices of the Royal Astronomical Society},
    volume = {503},
    number = {2},
    pages = {1734-1752},
    year = {2021},
    month = {05},
    issn = {0035-8711},
    doi = {10.1093/mnras/stab612},
    url = {https://doi.org/10.1093/mnras/stab612},
    eprint = {https://academic.oup.com/mnras/article-pdf/503/2/1734/36653768/stab612.pdf},
}

@ARTICLE{2017Verhamme,
       author = {{Verhamme}, A. and {Orlitov{\'a}}, I. and {Schaerer}, D. and {Izotov}, Y. and {Worseck}, G. and {Thuan}, T.~X. and {Guseva}, N.},
        title = "{Lyman-{\ensuremath{\alpha}} spectral properties of five newly discovered Lyman continuum emitters}",
      journal = {\aap},
         year = 2017,
        month = jan,
       volume = {597},
          eid = {A13},
        pages = {A13},
          doi = {10.1051/0004-6361/201629264},
archivePrefix = {arXiv},
       eprint = {1609.03477},
 primaryClass = {astro-ph.GA},
       adsurl = {https://ui.adsabs.harvard.edu/abs/2017A&A...597A..13V}
}

@article{Gazagnes2025,
    author = {Gazagnes, S and Chisholm, J and Endsley, R and Berg, D A and Leclercq, F and Jurlin, N and Saldana-Lopez, A and Finkelstein, S L and Flury, S R and Guseva, N G and Henry, A and Izotov, Y I and Jung, I and Matthee, J and Schaerer, D},
    title = {A negligible contribution of two luminous z ∼ 7.5 galaxies to the ionizing photon budget of reionization},
    journal = {Monthly Notices of the Royal Astronomical Society},
    volume = {540},
    number = {3},
    pages = {2331-2348},
    year = {2025},
    month = {07},
    issn = {0035-8711},
    doi = {10.1093/mnras/staf768},
    url = {https://doi.org/10.1093/mnras/staf768},
    eprint = {https://academic.oup.com/mnras/article-pdf/540/3/2331/63140707/staf768.pdf},
}

@article{Xu2022,
doi = {10.3847/1538-4357/ac7225},
url = {https://doi.org/10.3847/1538-4357/ac7225},
year = {2022},
month = {jul},
publisher = {The American Astronomical Society},
volume = {933},
number = {2},
pages = {202},
author = {Xu, Xinfeng and Henry, Alaina and Heckman, Timothy and Chisholm, John and Worseck, Gábor and Gronke, Max and Jaskot, Anne and McCandliss, Stephan R. and Flury, Sophia R. and Giavalisco, Mauro and Ji, Zhiyuan and Amorín, Ricardo O. and Berg, Danielle A. and Borthakur, Sanchayeeta and Bouche, Nicolas and Carr, Cody and Erb, Dawn K. and Ferguson, Harry and Garel, Thibault and Hayes, Matthew and Makan, Kirill and Marques-Chaves, Rui and Rutkowski, Michael and Östlin, Göran and Rafelski, Marc and Saldana-Lopez, Alberto and Scarlata, Claudia and Schaerer, Daniel and Trebitsch, Maxime and Tremonti, Christy and Verhamme, Anne and Wang, Bingjie},
title = {Tracing Lyα and LyC Escape in Galaxies with Mg ii Emission},
journal = {The Astrophysical Journal}
}

@article{Xu2023,
doi = {10.3847/1538-4357/aca89a},
url = {https://doi.org/10.3847/1538-4357/aca89a},
year = {2023},
month = {jan},
publisher = {The American Astronomical Society},
volume = {943},
number = {2},
pages = {94},
author = {Xu, Xinfeng and Henry, Alaina and Heckman, Timothy and Chisholm, John and Marques-Chaves, Rui and Leclercq, Floriane and Berg, Danielle A. and Jaskot, Anne and Schaerer, Daniel and Worseck, Gábor and Amorín, Ricardo O. and Atek, Hakim and Hayes, Matthew and Ji, Zhiyuan and Östlin, Göran and Saldana-Lopez, Alberto and Thuan, Trinh},
title = {The Low-redshift Lyman Continuum Survey: Optically Thin and Thick Mg ii Lines as Probes of Lyman Continuum Escape},
journal = {The Astrophysical Journal}
}

@ARTICLE{2024Leclercq,
       author = {{Leclercq}, Floriane and {Chisholm}, John and {King}, Wichahpi and {Zeimann}, Greg and {Jaskot}, Anne E. and {Henry}, Alaina and {Hayes}, Matthew and {Flury}, Sophia R. and {Izotov}, Yuri and {Prochaska}, Xavier J. and {Verhamme}, Anne and {Amor{\'\i}n}, Ricardo O. and {Atek}, Hakim and {Bait}, Omkar and {Blaizot}, J{\'e}r{\'e}my and {Carr}, Cody and {Ji}, Zhiyuan and {Le Reste}, Alexandra and {Ferguson}, Harry C. and {Gazagnes}, Simon and {Heckman}, Timothy and {Komarova}, Lena and {Marques-Chaves}, Rui and {{\"O}stlin}, G{\"o}ran and {Saldana-Lopez}, Alberto and {Scarlata}, Claudia and {Schaerer}, Daniel and {Thuan}, Trinh X. and {Trebitsch}, Maxime and {Worseck}, G{\'a}bor and {Wang}, Bingjie and {Xu}, Xinfeng},
        title = "{Linking Mg II and [O II] spatial distribution to ionizing photon escape in confirmed LyC leakers and non-leakers}",
      journal = {\aap},
         year = 2024,
        month = jul,
       volume = {687},
          eid = {A73},
        pages = {A73},
          doi = {10.1051/0004-6361/202449362},
archivePrefix = {arXiv},
       eprint = {2401.14981},
 primaryClass = {astro-ph.GA},
       adsurl = {https://ui.adsabs.harvard.edu/abs/2024A&A...687A..73L}
}

@article{Chang2024,
    author = {Chang, Seok-Jun and Gronke, Max},
    title = {Probing cold gas with Mg ii and Ly α radiative transfer},
    journal = {Monthly Notices of the Royal Astronomical Society},
    volume = {532},
    number = {3},
    pages = {3526-3555},
    year = {2024},
    month = {08},
    issn = {0035-8711},
    doi = {10.1093/mnras/stae1664},
    url = {https://doi.org/10.1093/mnras/stae1664},
    eprint = {https://academic.oup.com/mnras/article-pdf/532/3/3526/58652917/stae1664.pdf},
}

@ARTICLE{Pasha2023,
       author = {{Pasha}, Imad and {Miller}, Tim B.},
        title = "{pysersic: A Python package for determining galaxy structural properties via Bayesian inference, accelerated with jax}",
      journal = {The Journal of Open Source Software},
         year = 2023,
        month = sep,
       volume = {8},
       number = {89},
          eid = {5703},
        pages = {5703},
          doi = {10.21105/joss.05703},
archivePrefix = {arXiv},
       eprint = {2306.05454},
 primaryClass = {astro-ph.GA},
       adsurl = {https://ui.adsabs.harvard.edu/abs/2023JOSS....8.5703P}
}

@ARTICLE{Birrer2021,
       author = {{Birrer}, Simon},
        title = "{Gravitational Lensing Formalism in a Curved Arc Basis: A Continuous Description of Observables and Degeneracies from the Weak to the Strong Lensing Regime}",
      journal = {\apj},
         year = 2021,
        month = sep,
       volume = {919},
       number = {1},
          eid = {38},
        pages = {38},
          doi = {10.3847/1538-4357/ac1108},
archivePrefix = {arXiv},
       eprint = {2104.09522},
 primaryClass = {astro-ph.CO},
       adsurl = {https://ui.adsabs.harvard.edu/abs/2021ApJ...919...38B}
}

@software{Birrer2022,
       author = {{Birrer}, Simon and {Bhamre}, Vikram and {Nierenberg}, Anna and {Yang}, Lilan and {Van de Vyvere}, Lyne},
        title = "{PSFr: Point Spread Function reconstruction}",
 howpublished = {Astrophysics Source Code Library, record ascl:2210.005},
         year = 2022,
        month = oct,
          eid = {ascl:2210.005},
archivePrefix = {ascl},
       eprint = {2210.005},
       adsurl = {https://ui.adsabs.harvard.edu/abs/2022ascl.soft10005B}
}

@INPROCEEDINGS{Kacprzak2017,
       author = {{Kacprzak}, Glenn G.},
        title = "{Gas Accretion in Star-Forming Galaxies}",
    booktitle = {Gas Accretion onto Galaxies},
         year = 2017,
       editor = {{Fox}, Andrew and {Dav{\'e}}, Romeel},
       series = {Astrophysics and Space Science Library},
       volume = {430},
        month = jan,
        pages = {145},
          doi = {10.1007/978-3-319-52512-9_7},
archivePrefix = {arXiv},
       eprint = {1612.00451},
 primaryClass = {astro-ph.GA},
       adsurl = {https://ui.adsabs.harvard.edu/abs/2017ASSL..430..145K}
}

@ARTICLE{2026Taylor,
       author = {{Taylor}, Elizabeth and {Carnall}, Adam C. and {Maltby}, David and {Almaini}, Omar and {Leung}, Ho-Hin and {Stevenson}, Struan D. and {Negri}, Andrea and {Cullen}, Fergus and {Wild}, Vivienne and {McLure}, Ross J. and {Shapley}, Alice E. and {Arellano-C{\'o}rdova}, Karla Z. and {Begley}, Ryan and {Bondestam}, Cecilia and {de Lisle}, Thomas and {Donnan}, Callum T. and {Dunlop}, James S. and {Ellis}, Richard and {Hewitt}, Guillaume and {Koekemoer}, Anton M. and {Frey Liu}, Feng-Yuan and {McLeod}, Derek J. and {Rowlands}, Kate and {Sanders}, Ryan L. and {Scholte}, Dirk and {Skarbinski}, Maya and {Stanton}, Thomas M.},
        title = "{The JWST EXCELS survey: Outflows in 1.5 < z < 5 quiescent and recently quenched galaxies are likely relics from episodic AGN activity}",
      journal = {arXiv e-prints},
         year = 2026,
        month = jan,
          eid = {arXiv:2601.02269},
        pages = {arXiv:2601.02269},
          doi = {10.48550/arXiv.2601.02269},
archivePrefix = {arXiv},
       eprint = {2601.02269},
 primaryClass = {astro-ph.GA},
       adsurl = {https://ui.adsabs.harvard.edu/abs/2026arXiv260102269T}
}

@ARTICLE{2015Wisnioski,
       author = {{Wisnioski}, E. and {F{\"o}rster Schreiber}, N.~M. and {Wuyts}, S. and {Wuyts}, E. and {Bandara}, K. and {Wilman}, D. and {Genzel}, R. and {Bender}, R. and {Davies}, R. and {Fossati}, M. and {Lang}, P. and {Mendel}, J.~T. and {Beifiori}, A. and {Brammer}, G. and {Chan}, J. and {Fabricius}, M. and {Fudamoto}, Y. and {Kulkarni}, S. and {Kurk}, J. and {Lutz}, D. and {Nelson}, E.~J. and {Momcheva}, I. and {Rosario}, D. and {Saglia}, R. and {Seitz}, S. and {Tacconi}, L.~J. and {van Dokkum}, P.~G.},
        title = "{The KMOS$^{3D}$ Survey: Design, First Results, and the Evolution of Galaxy Kinematics from 0.7 <= z <= 2.7}",
      journal = {\apj},
         year = 2015,
        month = feb,
       volume = {799},
       number = {2},
          eid = {209},
        pages = {209},
          doi = {10.1088/0004-637X/799/2/209},
archivePrefix = {arXiv},
       eprint = {1409.6791},
 primaryClass = {astro-ph.GA},
       adsurl = {https://ui.adsabs.harvard.edu/abs/2015ApJ...799..209W}
}

@ARTICLE{2019Wisnioski,
       author = {{Wisnioski}, E. and {F{\"o}rster Schreiber}, N.~M. and {Fossati}, M. and {Mendel}, J.~T. and {Wilman}, D. and {Genzel}, R. and {Bender}, R. and {Wuyts}, S. and {Davies}, R.~L. and {{\"U}bler}, H. and {Bandara}, K. and {Beifiori}, A. and {Belli}, S. and {Brammer}, G. and {Chan}, J. and {Davies}, R.~I. and {Fabricius}, M. and {Galametz}, A. and {Lang}, P. and {Lutz}, D. and {Nelson}, E.~J. and {Momcheva}, I. and {Price}, S. and {Rosario}, D. and {Saglia}, R. and {Seitz}, S. and {Shimizu}, T. and {Tacconi}, L.~J. and {Tadaki}, K. and {van Dokkum}, P.~G. and {Wuyts}, E.},
        title = "{The KMOS$^{3D}$ Survey: Data Release and Final Survey Paper}",
      journal = {\apj},
         year = 2019,
        month = dec,
       volume = {886},
       number = {2},
          eid = {124},
        pages = {124},
          doi = {10.3847/1538-4357/ab4db8},
archivePrefix = {arXiv},
       eprint = {1909.11096},
 primaryClass = {astro-ph.GA},
       adsurl = {https://ui.adsabs.harvard.edu/abs/2019ApJ...886..124W}
}

@ARTICLE{2025Saldana,
       author = {{Saldana-Lopez}, A. and {Chisholm}, J. and {Gazagnes}, S. and {Endsley}, R. and {Hayes}, M.~J. and {Berg}, D.~A. and {Finkelstein}, S.~L. and {Flury}, S.~R. and {Guseva}, N.~G. and {Henry}, A. and {Izotov}, Y.~I. and {Lambrides}, E. and {Marques-Chaves}, R. and {Richardson}, C.~T.},
        title = "{Feedback and dynamical masses in high-z galaxies: the advent of high-resolution NIRSpec spectroscopy}",
      journal = {\mnras},
         year = 2025,
        month = nov,
       volume = {544},
       number = {1},
        pages = {132-151},
          doi = {10.1093/mnras/staf1680},
archivePrefix = {arXiv},
       eprint = {2501.17145},
 primaryClass = {astro-ph.GA},
       adsurl = {https://ui.adsabs.harvard.edu/abs/2025MNRAS.544..132S}
}

@ARTICLE{2026Lyu,
       author = {{Lyu}, Cheqiu and {Yu}, Haoran and {Wang}, Enci and {Wang}, Junxian and {Jia}, Cheng and {Song}, Jie and {Chen}, Yangyao and {Wang}, Jinyang and {Chen}, Zeyu and {Ma}, Chengyu and {Wang}, Yifan and {Kong}, Xu},
        title = "{First Statistical Detection of Mg II-traced Cool Gas Outflows with JWST toward Cosmic Dawn}",
      journal = {\apjl},
         year = 2026,
        month = mar,
       volume = {1000},
       number = {1},
          eid = {L3},
        pages = {L3},
          doi = {10.3847/2041-8213/ae48ee},
archivePrefix = {arXiv},
       eprint = {2512.05622},
 primaryClass = {astro-ph.GA},
       adsurl = {https://ui.adsabs.harvard.edu/abs/2026ApJ..1000L...3L}
}

@INPROCEEDINGS{Bertin2020,
       author = {{Bertin}, E. and {Schefer}, M. and {Apostolakos}, N. and {{\'A}lvarez-Ayll{\'o}n}, A. and {Dubath}, P. and {K{\"u}mmel}, M.},
        title = "{The SourceXtractor++ Software}",
    booktitle = {Astronomical Data Analysis Software and Systems XXIX},
         year = 2020,
       editor = {{Pizzo}, R. and {Deul}, E.~R. and {Mol}, J.~D. and {de Plaa}, J. and {Verkouter}, H.},
       series = {Astronomical Society of the Pacific Conference Series},
       volume = {527},
        month = jan,
        pages = {461},
       adsurl = {https://ui.adsabs.harvard.edu/abs/2020ASPC..527..461B}
}

@ARTICLE{2022Xu_Classy,
       author = {{Xu}, Xinfeng and {Heckman}, Timothy and {Henry}, Alaina and {Berg}, Danielle A. and {Chisholm}, John and {James}, Bethan L. and {Martin}, Crystal L. and {Stark}, Daniel P. and {Aloisi}, Alessandra and {Amor{\'\i}n}, Ricardo O. and {Arellano-C{\'o}rdova}, Karla Z. and {Bordoloi}, Rongmon and {Charlot}, St{\'e}phane and {Chen}, Zuyi and {Hayes}, Matthew and {Mingozzi}, Matilde and {Sugahara}, Yuma and {Kewley}, Lisa J. and {Ouchi}, Masami and {Scarlata}, Claudia and {Steidel}, Charles C.},
        title = "{CLASSY III. The Properties of Starburst-driven Warm Ionized Outflows}",
      journal = {\apj},
         year = 2022,
        month = jul,
       volume = {933},
       number = {2},
          eid = {222},
        pages = {222},
          doi = {10.3847/1538-4357/ac6d56},
archivePrefix = {arXiv},
       eprint = {2204.09181},
 primaryClass = {astro-ph.GA},
       adsurl = {https://ui.adsabs.harvard.edu/abs/2022ApJ...933..222X}
}

@ARTICLE{2015Heckman,
       author = {{Heckman}, Timothy M. and {Alexandroff}, Rachel M. and {Borthakur}, Sanchayeeta and {Overzier}, Roderik and {Leitherer}, Claus},
        title = "{The Systematic Properties of the Warm Phase of Starburst-Driven Galactic Winds}",
      journal = {\apj},
         year = 2015,
        month = aug,
       volume = {809},
       number = {2},
          eid = {147},
        pages = {147},
          doi = {10.1088/0004-637X/809/2/147},
archivePrefix = {arXiv},
       eprint = {1507.05622},
 primaryClass = {astro-ph.GA},
       adsurl = {https://ui.adsabs.harvard.edu/abs/2015ApJ...809..147H}
}

@ARTICLE{2016Heckman,
       author = {{Heckman}, Timothy M. and {Borthakur}, Sanchayeeta},
        title = "{The Implications of Extreme Outflows from Extreme Starbursts}",
      journal = {\apj},
         year = 2016,
        month = may,
       volume = {822},
       number = {1},
          eid = {9},
        pages = {9},
          doi = {10.3847/0004-637X/822/1/9},
archivePrefix = {arXiv},
       eprint = {1603.03036},
 primaryClass = {astro-ph.GA},
       adsurl = {https://ui.adsabs.harvard.edu/abs/2016ApJ...822....9H}
}

@ARTICLE{2024Garel,
       author = {{Garel}, T. and {Michel-Dansac}, L. and {Verhamme}, A. and {Mauerhofer}, V. and {Katz}, H. and {Blaizot}, J. and {Leclercq}, F. and {Salvignol}, G.},
        title = "{A public grid of radiative transfer simulations for Ly{\ensuremath{\alpha}} and metal lines in idealised galactic outflows}",
      journal = {\aap},
         year = 2024,
        month = nov,
       volume = {691},
          eid = {A213},
        pages = {A213},
          doi = {10.1051/0004-6361/202450654},
archivePrefix = {arXiv},
       eprint = {2408.03605},
 primaryClass = {astro-ph.GA},
       adsurl = {https://ui.adsabs.harvard.edu/abs/2024A&A...691A.213G}
}

@ARTICLE{2020Gazagnes,
       author = {{Gazagnes}, S. and {Chisholm}, J. and {Schaerer}, D. and {Verhamme}, A. and {Izotov}, Y.},
        title = "{The origin of the escape of Lyman {\ensuremath{\alpha}} and ionizing photons in Lyman continuum emitters}",
      journal = {\aap},
         year = 2020,
        month = jul,
       volume = {639},
          eid = {A85},
        pages = {A85},
          doi = {10.1051/0004-6361/202038096},
archivePrefix = {arXiv},
       eprint = {2005.07215},
 primaryClass = {astro-ph.GA},
       adsurl = {https://ui.adsabs.harvard.edu/abs/2020A&A...639A..85G}
}

@ARTICLE{2022Izotov,
       author = {{Izotov}, Y.~I. and {Chisholm}, J. and {Worseck}, G. and {Guseva}, N.~G. and {Schaerer}, D. and {Prochaska}, J.~X.},
        title = "{Lyman alpha and Lyman continuum emission of Mg II-selected star-forming galaxies}",
      journal = {\mnras},
         year = 2022,
        month = sep,
       volume = {515},
       number = {2},
        pages = {2864-2881},
          doi = {10.1093/mnras/stac1899},
archivePrefix = {arXiv},
       eprint = {2207.04483},
 primaryClass = {astro-ph.GA},
       adsurl = {https://ui.adsabs.harvard.edu/abs/2022MNRAS.515.2864I}
}

@ARTICLE{2013Guseva,
       author = {{Guseva}, N.~G. and {Izotov}, Y.~I. and {Fricke}, K.~J. and {Henkel}, C.},
        title = "{The Mg II {\ensuremath{\lambda}}2797, {\ensuremath{\lambda}}2803 emission in low-metallicity star-forming galaxies from the SDSS}",
      journal = {\aap},
         year = 2013,
        month = jul,
       volume = {555},
          eid = {A90},
        pages = {A90},
          doi = {10.1051/0004-6361/201221010},
archivePrefix = {arXiv},
       eprint = {1306.1848},
 primaryClass = {astro-ph.CO},
       adsurl = {https://ui.adsabs.harvard.edu/abs/2013A&A...555A..90G}
}

@ARTICLE{2019Guseva,
       author = {{Guseva}, N.~G. and {Izotov}, Y.~I. and {Fricke}, K.~J. and {Henkel}, C.},
        title = "{Mg II {\ensuremath{\lambda}}2797, {\ensuremath{\lambda}}2803 emission in a large sample of low-metallicity star-forming galaxies from SDSS DR14}",
      journal = {\aap},
         year = 2019,
        month = apr,
       volume = {624},
          eid = {A21},
        pages = {A21},
          doi = {10.1051/0004-6361/201834935},
archivePrefix = {arXiv},
       eprint = {1902.11083},
 primaryClass = {astro-ph.GA},
       adsurl = {https://ui.adsabs.harvard.edu/abs/2019A&A...624A..21G}
}

@ARTICLE{Tumlinson2011,
       author = {{Tumlinson}, J. and {Thom}, C. and {Werk}, J.~K. and {Prochaska}, J.~X. and {Tripp}, T.~M. and {Weinberg}, D.~H. and {Peeples}, M.~S. and {O'Meara}, J.~M. and {Oppenheimer}, B.~D. and {Meiring}, J.~D. and {Katz}, N.~S. and {Dav{\'e}}, R. and {Ford}, A.~B. and {Sembach}, K.~R.},
        title = "{The Large, Oxygen-Rich Halos of Star-Forming Galaxies Are a Major Reservoir of Galactic Metals}",
      journal = {Science},
         year = 2011,
        month = nov,
       volume = {334},
       number = {6058},
        pages = {948},
          doi = {10.1126/science.1209840},
archivePrefix = {arXiv},
       eprint = {1111.3980},
 primaryClass = {astro-ph.CO},
       adsurl = {https://ui.adsabs.harvard.edu/abs/2011Sci...334..948T}
}

@ARTICLE{2022Flury_a,
       author = {{Flury}, Sophia R. and {Jaskot}, Anne E. and {Ferguson}, Harry C. and {Worseck}, G{\'a}bor and {Makan}, Kirill and {Chisholm}, John and {Saldana-Lopez}, Alberto and {Schaerer}, Daniel and {McCandliss}, Stephan and {Wang}, Bingjie and {Ford}, N.~M. and {Heckman}, Timothy and {Ji}, Zhiyuan and {Giavalisco}, Mauro and {Amorin}, Ricardo and {Atek}, Hakim and {Blaizot}, Jeremy and {Borthakur}, Sanchayeeta and {Carr}, Cody and {Castellano}, Marco and {Cristiani}, Stefano and {De Barros}, Stephane and {Dickinson}, Mark and {Finkelstein}, Steven L. and {Fleming}, Brian and {Fontanot}, Fabio and {Garel}, Thibault and {Grazian}, Andrea and {Hayes}, Matthew and {Henry}, Alaina and {Mauerhofer}, Valentin and {Micheva}, Genoveva and {Oey}, M.~S. and {Ostlin}, Goran and {Papovich}, Casey and {Pentericci}, Laura and {Ravindranath}, Swara and {Rosdahl}, Joakim and {Rutkowski}, Michael and {Santini}, Paola and {Scarlata}, Claudia and {Teplitz}, Harry and {Thuan}, Trinh and {Trebitsch}, Maxime and {Vanzella}, Eros and {Verhamme}, Anne and {Xu}, Xinfeng},
        title = "{The Low-redshift Lyman Continuum Survey. I. New, Diverse Local Lyman Continuum Emitters}",
      journal = {\apjs},
         year = 2022,
        month = may,
       volume = {260},
       number = {1},
          eid = {1},
        pages = {1},
          doi = {10.3847/1538-4365/ac5331},
archivePrefix = {arXiv},
       eprint = {2201.11716},
 primaryClass = {astro-ph.GA},
       adsurl = {https://ui.adsabs.harvard.edu/abs/2022ApJS..260....1F}
}

@ARTICLE{2022Flury_b,
       author = {{Flury}, Sophia R. and {Jaskot}, Anne E. and {Ferguson}, Harry C. and {Worseck}, G{\'a}bor and {Makan}, Kirill and {Chisholm}, John and {Saldana-Lopez}, Alberto and {Schaerer}, Daniel and {McCandliss}, Stephan R. and {Xu}, Xinfeng and {Wang}, Bingjie and {Oey}, M.~S. and {Ford}, N.~M. and {Heckman}, Timothy and {Ji}, Zhiyuan and {Giavalisco}, Mauro and {Amor{\'\i}n}, Ricardo and {Atek}, Hakim and {Blaizot}, Jeremy and {Borthakur}, Sanchayeeta and {Carr}, Cody and {Castellano}, Marco and {De Barros}, Stephane and {Dickinson}, Mark and {Finkelstein}, Steven L. and {Fleming}, Brian and {Fontanot}, Fabio and {Garel}, Thibault and {Grazian}, Andrea and {Hayes}, Matthew and {Henry}, Alaina and {Mauerhofer}, Valentin and {Micheva}, Genoveva and {Ostlin}, Goran and {Papovich}, Casey and {Pentericci}, Laura and {Ravindranath}, Swara and {Rosdahl}, Joakim and {Rutkowski}, Michael and {Santini}, Paola and {Scarlata}, Claudia and {Teplitz}, Harry and {Thuan}, Trinh and {Trebitsch}, Maxime and {Vanzella}, Eros and {Verhamme}, Anne},
        title = "{The Low-redshift Lyman Continuum Survey. II. New Insights into LyC Diagnostics}",
      journal = {\apj},
         year = 2022,
        month = may,
       volume = {930},
       number = {2},
          eid = {126},
        pages = {126},
          doi = {10.3847/1538-4357/ac61e4},
archivePrefix = {arXiv},
       eprint = {2203.15649},
 primaryClass = {astro-ph.GA},
       adsurl = {https://ui.adsabs.harvard.edu/abs/2022ApJ...930..126F}
}

@ARTICLE{2025LeReste,
       author = {{Le Reste}, Alexandra and {Scarlata}, Claudia and {Hayes}, Matthew J. and {Melinder}, Jens and {Saldana-Lopez}, Alberto and {Smith}, Aaron and {Runnholm}, Axel and {Lin}, Yu-Heng and {Amor{\'\i}n}, Ricardo O. and {Atek}, Hakim and {Borthakur}, Sanchayeeta and {Carr}, Cody A. and {Fleming}, Brian and {Flury}, Sophia R. and {Giavalisco}, Mauro and {Henry}, Alaina and {Jaskot}, Anne E. and {Ji}, Zhiyuan and {Jung}, Intae and {Leclercq}, Floriane and {Marques-Chaves}, Rui and {McCandliss}, Stephan R. and {Oey}, M.~S. and {{\"O}stlin}, G{\"o}ran and {Ravindranath}, Swara and {Schaerer}, Daniel and {Thuan}, Trinh X. and {Xu}, Xinfeng},
        title = "{The Ly{\ensuremath{\alpha}} and Continuum Origins Survey. I. Survey Description and Ly{\ensuremath{\alpha}} Imaging}",
      journal = {\apjs},
         year = 2025,
        month = sep,
       volume = {280},
       number = {1},
          eid = {27},
        pages = {27},
          doi = {10.3847/1538-4365/adf227},
archivePrefix = {arXiv},
       eprint = {2504.07056},
 primaryClass = {astro-ph.GA},
       adsurl = {https://ui.adsabs.harvard.edu/abs/2025ApJS..280...27L}
}

@ARTICLE{2025Hayes,
       author = {{Hayes}, Matthew J. and {Saldana-Lopez}, Alberto and {Citro}, Annalisa and {James}, Bethan L. and {Mingozzi}, Matilde and {Scarlata}, Claudia and {Martinez}, Zorayda and {Berg}, Danielle A.},
        title = "{On the Average Ultraviolet Emission-line Spectra of High-redshift Galaxies: Hot and Cold, Carbon-poor, Nitrogen Modest, and Oozing Ionizing Photons}",
      journal = {\apj},
         year = 2025,
        month = mar,
       volume = {982},
       number = {1},
          eid = {14},
        pages = {14},
          doi = {10.3847/1538-4357/adaea1},
archivePrefix = {arXiv},
       eprint = {2411.09262},
 primaryClass = {astro-ph.GA},
       adsurl = {https://ui.adsabs.harvard.edu/abs/2025ApJ...982...14H}
}

@ARTICLE{Sanders2026,
       author = {{Sanders}, Ryan L. and {Shapley}, Alice E. and {Topping}, Michael W. and {Reddy}, Naveen A. and {Berg}, Danielle A. and {Khostovan}, Ali Ahmad and {Bouwens}, Rychard J. and {Brammer}, Gabriel and {Carnall}, Adam C. and {Cullen}, Fergus and {Dav{\'e}}, Romeel and {Dunlop}, James S. and {Ellis}, Richard S. and {F{\"o}rster Schreiber}, N.~M. and {Furlanetto}, Steven R. and {Glazebrook}, Karl and {Illingworth}, Garth D. and {Jones}, Tucker and {Kriek}, Mariska and {McLeod}, Derek J. and {McLure}, Ross J. and {Narayanan}, Desika and {Oesch}, Pascal A. and {Pahl}, Anthony J. and {Pettini}, Max and {Schaerer}, Daniel and {Stark}, Daniel P. and {Steidel}, Charles C. and {Tang}, Mengtao and {Clarke}, Leonardo and {Donnan}, Callum T. and {Kehoe}, Emily},
        title = "{The AURORA Survey: High-redshift Empirical Metallicity Calibrations from Electron Temperature Measurements at z = 2─10}",
      journal = {\apj},
         year = 2026,
        month = jun,
       volume = {1003},
       number = {2},
          eid = {228},
        pages = {228},
          doi = {10.3847/1538-4357/ae66e2},
archivePrefix = {arXiv},
       eprint = {2508.10099},
 primaryClass = {astro-ph.GA},
       adsurl = {https://ui.adsabs.harvard.edu/abs/2026ApJ..1003..228S}
}

@article{Leitherer2011,
doi = {10.1088/0004-6256/141/2/37},
url = {https://doi.org/10.1088/0004-6256/141/2/37},
year = {2011},
month = {jan},
publisher = {The American Astronomical Society},
volume = {141},
number = {2},
pages = {37},
author = {Leitherer, Claus and Tremonti, Christy A. and Heckman, Timothy M. and Calzetti, Daniela},
title = {AN ULTRAVIOLET SPECTROSCOPIC ATLAS OF LOCAL STARBURSTS AND STAR-FORMING GALAXIES: THE LEGACY OF FOS AND GHRS},
journal = {The Astronomical Journal}
}

@article{Giavalisco2011,
doi = {10.1088/0004-637X/743/1/95},
url = {https://doi.org/10.1088/0004-637X/743/1/95},
year = {2011},
month = {nov},
publisher = {The American Astronomical Society},
volume = {743},
number = {1},
pages = {95},
author = {Giavalisco, Mauro and Vanzella, Eros and Salimbeni, Sara and Tripp, Todd M. and Dickinson, Mark and Cassata, Paolo and Renzini, Alvio and Guo, Yicheng and Ferguson, Henry C. and Nonino, Mario and Cimatti, Andrea and Kurk, Jaron and Mignoli, Marco and Tang, Yuping},
title = {DISCOVERY OF COLD, PRISTINE GAS POSSIBLY ACCRETING ONTO AN OVERDENSITY OF STAR-FORMING GALAXIES AT REDSHIFT z ∼ 1.6*},
journal = {The Astrophysical Journal}
}

@ARTICLE{Guo2023,
       author = {{Guo}, Yucheng and {Bacon}, Roland and {Bouch{\'e}}, Nicolas F. and {Wisotzki}, Lutz and {Schaye}, Joop and {Blaizot}, J{\'e}r{\'e}my and {Verhamme}, Anne and {Cantalupo}, Sebastiano and {Boogaard}, Leindert A. and {Brinchmann}, Jarle and {Cherrey}, Maxime and {Kusakabe}, Haruka and {Langan}, Ivanna and {Leclercq}, Floriane and {Matthee}, Jorryt and {Michel-Dansac}, L{\'e}o and {Schroetter}, Ilane and {Wendt}, Martin},
        title = "{Bipolar outflows out to 10 kpc for massive galaxies at redshift z {\ensuremath{\approx}} 1}",
      journal = {\nat},
         year = 2023,
        month = dec,
       volume = {624},
       number = {7990},
        pages = {53-56},
          doi = {10.1038/s41586-023-06718-w},
archivePrefix = {arXiv},
       eprint = {2312.05167},
 primaryClass = {astro-ph.GA},
       adsurl = {https://ui.adsabs.harvard.edu/abs/2023Natur.624...53G}
}
\bibliographystyle{aasjournal}



\end{document}